\documentclass[twocolumn]{aastex63}

\newcommand{\code}[1]{\texttt{#1}}

\graphicspath{{./}{figures/}}

\usepackage{multirow}
\usepackage{rotating}
\usepackage{amsmath}
\usepackage{graphicx}

\received{January 1, 2018}
\revised{January 7, 2018}
\accepted{\today}
\submitjournal{ApJS}

\shorttitle{Techniques for detecting quasi-periodic pulsations}
\shortauthors{Broomhall et al.}

\begin{document}

\title{A blueprint of state-of-the-art techniques for detecting quasi-periodic pulsations in solar and stellar flares}

\correspondingauthor{Anne-Marie Broomhall}
\email{a-m.broomhall@warwick.ac.uk}

\author{Anne-Marie Broomhall}
\affil{Department of Physics, University of Warwick, Coventry, CV4 7AL, UK}
\affil{Centre for Exoplanets and Habitability, University of Warwick, Coventry CV4 7AL, UK}

\author{James R.A. Davenport}
\affil{Department of Physics \& Astronomy, Western Washington University, 516 High St., Bellingham, WA 98225, USA}
\affil{Department of Astronomy, University of Washington, Seattle, WA 98195, USA}

\author{Laura A. Hayes}
\affil{School of Physics, Trinity College Dublin, Dublin 2, Ireland}
\affil{Solar Physics Laboratory, NASA Goddard Space Flight Center, Greenbelt, Maryland, 20771, USA}

\author{Andrew R. Inglis}
\affil{Solar Physics Laboratory, NASA Goddard Space Flight Center, Greenbelt, MD, 20771, USA}

\author{Dmitrii Y. Kolotkov}
\affil{Department of Physics, University of Warwick, Coventry, CV4 7AL, UK}

\author{James A. McLaughlin}
\affil{Northumbria University, Newcastle upon Tyne, NE1 8ST, UK}

\author{Tishtrya Mehta}
\affil{Department of Physics, University of Warwick, Coventry, CV4 7AL, UK}

\author{Valery M. Nakariakov}
\affil{Department of Physics, University of Warwick, Coventry, CV4 7AL, UK}
\affil{St. Petersburg Branch, Special Astrophysical Observatory, Russian Academy of Sciences, 196140, St. Petersburg, Russia}

\author{Yuta Notsu}
\altaffiliation{JSPS Overseas Research Fellow}
\affil{Laboratory for Atmospheric and Space Physics, University of Colorado Boulder, 3665 Discovery Drive, Boulder, Colorado 80303, USA}
\affil{National Solar Observatory, 3665 Discovery Drive, Boulder, CO 80303, USA}
\affil{Department of Astronomy, Kyoto University, Sakyo, Kyoto 606-8502, Japan}

\author{David J. Pascoe}
\affil{Centre for Mathematical Plasma Astrophysics, Mathematics Department, KU~Leuven, Celestijnenlaan 200B bus 2400, B-3001 Leuven, Belgium}

\author{Chloe E. Pugh}
\affil{Department of Physics, University of Warwick, Coventry, CV4 7AL, UK}

\author{Tom Van Doorsselaere}
\affil{Centre for Mathematical Plasma Astrophysics, Mathematics Department, KU~Leuven, Celestijnenlaan 200B bus 2400, B-3001 Leuven, Belgium}

\begin{abstract}
Quasi-periodic pulsations (QPPs) appear to be a common feature observed in the light curves of both solar and stellar flares. However, their quasi-periodic nature, along with the fact that they can be small in amplitude and short-lived, makes QPPs difficult to unequivocally detect. In this paper, we test the strengths and limitations of state-of-the-art methods for detecting QPPs using a series of hare-and-hounds exercises. The hare simulated a set of flares, both with and without QPPs of a variety of forms, while the hounds attempted to detect QPPs in blind tests. We use the results of these exercises to create a blueprint for anyone who wishes to detect QPPs in real solar and stellar data. We present eight clear recommendations to be kept in mind for future QPP detections, with the plethora of solar and stellar flare data from new and future satellites. These recommendations address the key pitfalls in QPP detection, including detrending, trimming data, accounting for colored noise, detecting stationary-period QPPs, detecting QPPs with nonstationary periods, and ensuring thatdetections are robust and false detections are minimized. We find that QPPs can be detected reliably and robustly by a variety of methods, which are clearly identified and described, if the appropriate care and due diligence are taken. 
\end{abstract}



\section{Introduction} \label{sec:intro}

Solar flares are multiwavelength, powerful, impulsive energy releases on the Sun. Flares are subject to intensive studies in the context of space weather, as a driver of extreme events in the heliosphere, and also of fundamental plasma astrophysics, allowing for high-resolution observations of basic plasma physics processes such as magnetic reconnection, charged particle acceleration, turbulence, and the generation of electromagnetic radiation. The appearance of a flare at different wavelengths, which is associated with different emission mechanisms occurring in different phases of the phenomenon, is rather different. Light curves of flares, measured in different observational bands, could be considered as a superposition of a rather smooth, often asymmetric trend and variations with a characteristic time scale shorter than the characteristic times of the trend. Such a short-time variability is a common feature detected in all phases of a flare, at all wavelengths, from radio to gamma-rays \citep[e.g.][]{2012ApJ...749L..16D, 2014ApJ...791...44H, 2016ApJ...833..284I,  2017ApJ...836..121K, 2017A&A...608A.101P}.  The short-time variations occur in different parameters of the emission: its intensity, polarization, spectrum, spatial characteristics, etc. Often, such variations are seen in the form of apparently quasi-periodic patterns, which are called quasi-periodic pulsations (QPPs). 

The first observational detection of QPPs in solar flares, as a well-pronounced 16\,s periodic modulation of the hard X-ray emission generated by a flare, was reported 50 years ago \citep{1969ApJ...155L.117P}. Since this discovery, QPPs have been a subject to a number of observational case studies and theoretical models \citep[see, e.g.][for comprehensive reviews]{1987SoPh..111..113A, 2009SSRv..149..119N, 2010PPCF...52l4009N, 2016SoPh..291.3143V, 2018SSRv..214...45M, 2019PPCF...61a4024N}. QPPs have been detected in flares of all intensity classes, from microflares \citep[e.g.][]{2018ApJ...859..154N} to the most powerful flares \citep[e.g.][]{2006A&A...460..865M, 2018ApJ...858L...3K}. The observed depth of the modulation of the trend signal ranges from a few percent to almost 100\%. There have been several attempts to assess statistically the prevalence of QPP patterns in solar flares, drawing a conclusion that QPPs are a common feature of the light curves associated with both nonthermal and thermal emission \citep{2010SoPh..267..329K, 2015SoPh..290.3625S, 2016ApJ...833..284I, 2017A&A...608A.101P}. In some cases, the coexistence of several QPP patterns with different periods and other properties in the same flare has been established \citep[e.g.][]{2009A&A...493..259I, 2013ApJ...778L..28S, 2015A&A...574A..53K, 2019ApJ...875...33H}.

Similar apparently quasi-periodic patterns have been detected in stellar flares too \citep[e.g.][]{2003A&A...403.1101M, 2004AstL...30..319Z, 2005A&A...436.1041M, 2015MNRAS.450..956B, 2016MNRAS.459.3659P}, including super- and megaflares \citep[e.g.][]{2013ApJ...773..156A, 2015EP&S...67...59M, 2019MNRAS.482.5553J}. Moreover, properties of QPPs in solar and stellar flares have been found to show interesting similarities \citep{2015ApJ...813L...5P, 2016ApJ...830..110C}, which may indicate similarities in the background physical processes. 

Typical periods of QPPs range from a fraction of a second to several tens of minutes. This range coincides with the range of the predicted and observed periods of magnetohydrodynamic (MHD) oscillations in the plasma nonuniformities in the vicinity of the flaring active region  \citep[e.g.][for a recent review]{2016SSRv..200...75N}. Because of that, QPPs are often considered as a manifestation of various MHD oscillatory modes. There are a number of specific mechanisms that could be responsible for the modulation of flaring emission by MHD oscillations, either preexisting or even inducing the flare, or being excited by the flare itself. Mechanisms for the excitation of QPPs can be roughly divided into three main groups: direct modulation of the emitting plasma or kinematics of nonthermal particles, periodically induced magnetic reconnection, and self-oscillations \citep[e.g.][for recent reviews]{2016SoPh..291.3143V, 2018SSRv..214...45M}. In addition, numerical simulations demonstrate spontaneous repetitive regimes of magnetic reconnection \citep[e.g.][]{2000A&A...360..715K, 2009A&A...493..227M, 2009A&A...494..329M, 2012ApJ...749...30M, 2017ApJ...844....2T, 2018A&A...611A..10S}, i.e., the magnetic dripping mechanism \citep{2010PPCF...52l4009N}. On the other hand, there are numerical simulations that show that the process of magnetic reconnection is essentially nonsteady or even turbulent, but without a built-in characteristic time or spatial scale  \citep[e.g.][]{2011ApJ...737...24B}. In particular, parameters of shedded plasmoids were shown to obey a power-law relationship with a negative slope \citep[e.g.][]{2012PhPl...19d2303L}, which could result in a red-noise-like spectrum in the frequency domain. When the shedded plasmoids impact the underlying post-flare arcade, they trigger transverse oscillations \citep{2017ApJ...847...98J}.

Mechanisms of QPPs in flares remain a subject of intensive theoretical studies \citep{2018SSRv..214...45M}. If QPPs are a prevalent feature of the solar and stellar flare phenomenon, theoretical models of flares, summarized in, e.g. \citet {2011LRSP....8....6S}, must include QPPs as one of its key ingredients, as is attempted by, for example, \citet{2016ApJ...823..150T}. QPPs offer a promising tool for the seismological probing of the plasma in the flare site and its vicinity. In addition, a comparative study of QPPs in solar and stellar flares opens up interesting perspectives for the exploitation of the solar-stellar analogy. 

In different case studies, as well as in statistical studies, QPPs have been detected with different methods. These include direct best fitting by a guessed oscillatory function, Fourier transform methods, Wigner-Ville method, wavelet transforms with different mother functions, and the empirical mode decomposition (EMD) technique. Through use of these methods, different false-alarm estimation techniques are implemented, different models for the noise are assumed, and different detection criteria are often used. Moreover, some authors have routinely made use of signal smoothing (filtering or detrending), or work with the time derivatives of the analyzed signal or it’s autocorrelation function. In some studies, the detection technique is applied directly to the raw signal. This variety of analytical techniques and methods used by authors is caused by several intrinsic features of QPPs in flares. The quasi-periodic signal often occurs on top of a time-varying trend. The QPP signal is often very different from the underlying monochromatic signal and almost always has a pronounced amplitude and period modulation, i.e. QPP signals could be referred to as nonstationary oscillations. QPP signals are often essentially anharmonic, i.e. its shape is visibly different from a sinusoid. The QPP quality-factor (QF), which is the duration of the QPPs measured in terms of the number of oscillation cycles, is often low, as it is limited by the duration of the flare itself and also by signal damping or a wave-train-like signature. 

Thus, in the research community there is an urgent need for a unification of the QPP detection criteria, understanding advantages and shortcomings of different QPP detection techniques (along with associated artifacts), and working out recommended recipes and practical guides for QPP detection, based on best-practice examples. In this paper, we perform a series of hare-and-hounds exercises where the \lq{hare}\rq{} produced a set of simulated flares, which are described in Section \ref{sec:sims}, for the \lq{hounds}\rq{} to analyze. The hounds were aiming to produce reliable and robust detections of QPPs, and the methods they used are described in Section \ref{sec:method}. The results of the hare-and-hounds exercises are given in Section \ref{sec:results}, which includes discussion of the false-alarm rates of each methodology, along with the precision of the detected QPP periods. In Section \ref{sec:blueprint} we draw together our conclusions from these results, making a series of recommendations for anyone attempting to detect QPPs in flare time series. Finally, in Section \ref{sec:future} we look to new and future observational data, yet to be explored in a QPP framework.

\section{Simulations of QPP flares} \label{sec:sims}
In this paper we will discuss three hare-and-hounds exercises that aimed to test methods for detection of QPPs. The first hare-and-hounds exercise, HH1, contained 101 flares simulated by the \lq{hare}\rq{} (Broomhall--AMB) and these were analyzed for QPPs by the \lq{hounds}\rq{} (Davenport--JRAD; Hayes--LAH; Inglis--ARI; McLaughlin--JAM; Kolotkov \& Mehta--DK and TM; Pascoe--DJP; Pugh--CEP; Van Doorsselaere--TVD). The HH1 sample was the only completely blind test performed, where the hounds did not know how any of the simulated flares had been produced. Following the initial analysis of the results of HH1, it was deemed necessary to perform further hare-and-hounds exercises to investigate issues not covered by the HH1 sample. Accordingly, two further sets of simulated flares were produced: HH2 contained 100 flares and HH3 contained 18. Flares for all exercises were simulated using the methodology described in this section and, in fact, were produced prior to the hounds' analysis of HH1. Before the hounds received HH2 and HH3, they were informed of how the simulated flares had been produced but were not aware of which of the components described below were present in each individual flare, i.e. the tests were still semi-blind. 

Each simulated flare was assigned a randomly selected ID number to make sure the different types of simulated QPP flares could not be identified prior to analysis. All simulated time series contained 300 data points and a synthetic flare. Each flare was initially simulated to be 20 time units in length and was heavily oversampled (with a time step of 0.001 fiducial time units) to prevent resolution issues upon rescaling. Once simulated, the length of the flare was rescaled to equal a length randomly chosen from a uniform distribution, $L_{\textrm{\scriptsize{flare}}}$, and further details are given in Table \ref{table[flares]}. The respective lengths of the rise and decay phases relative to $L_{\textrm{\scriptsize{flare}}}$ are described below. The flare was then interpolated onto a regular grid where data points were separated by one time unit. The simulated flare was inserted into a null array of length 300 such that the timing of the peak, $t_\textrm{\scriptsize{peak}}$, was determined by a value randomly selected from a uniform distribution (See Table \ref{table[flares]}).
%
%

\begin{figure*}[htp]
  \centering   \includegraphics[width=0.9\textwidth]{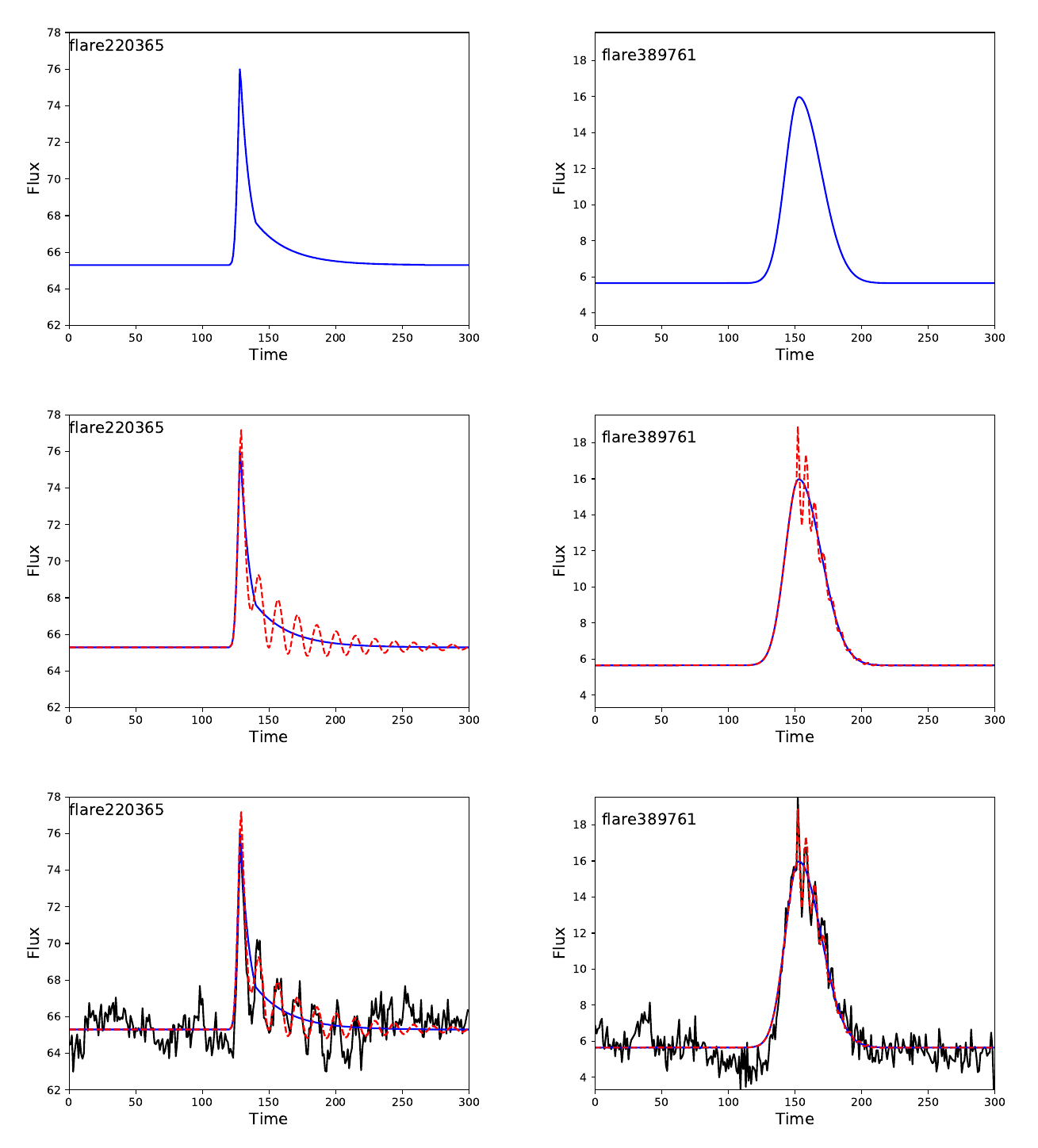}\\
   \caption{Top left: example of a simulated flare based on the flare template of \citet{2014ApJ...797..122D}, with $L_\textrm{flare}=145.1$, $t_\textrm{peak}=129.3$, and $A_\textrm{flare}=11.5$. Top right: example of a simulated flare constructed from two half-Gaussians with $L_\textrm{flare}=133.3$, $t_\textrm{peak}=157.8$, $A_\textrm{flare}=10.3$, $\sigma_\textrm{rise}=2.3$, and $\sigma_\textrm{decay}=6.3$. Middle left: example of a flare (blue solid) with a simple QPP signal (red dashed), described by equation \ref{eqn[qpp_simple]}, with $P=14.5$, $t_e=58.1$, $A_\textrm{qpp}=2.3$, and $\phi=0.5\,\rm rad$. Middle right: example of a flare (blue solid) with a simple QPP signal (red dashed), described by equation \ref{eqn[qpp_simple]}, with $P=6.7$, $t_e=13.3$, $A_\textrm{qpp}=3.1$, and $\phi=0.3\,\rm rad$. Bottom left: simulated flare including noise where the signal-to-noise of the flare was 5.0. Bottom right: simulated flare including noise where the S/N of the flare was 5.0.}
   \label{figure[flare_shape]}%
\end{figure*}

The synthetic flare shapes took two forms: The first shape was based on the results of \citet{2014ApJ...797..122D}, who produced a flare template using 885 flares observed on the active M4 star GJ 1243, which was observed by the \textit{Kepler} satellite \citep{2010Sci...327..977B}. The flare template includes a polynomial rise phase and a two-stage exponential decay. A limitation of this template is that it produces a very sharp peak. This is likely to arise in the flares observed by \citet{2014ApJ...797..122D} because of the limited time cadence of the \textit{Kepler} data. In better-resolved data, a smoother turnover at the peak is often observed \citep[e.g.][]{2019MNRAS.482.5553J}. To better replicate this, a flare shape consisting of two half-Gaussian curves was created, whereby the first half-Gaussian was used to simulate the rising phase and had smaller width than the second half-Gaussian, used to simulate the decay phase. The widths of the rising and decay Gaussian curves were determined by the standard deviations, $\sigma_\textrm{\scriptsize{rise}}$ and $\sigma_\textrm{\scriptsize{decay}}$ respectively, which were selected from uniform distributions as detailed in Table \ref{table[flares]}. For both flare shapes the amplitude of the flare, $A_{\textrm{\scriptsize{flare}}}$, was allowed to vary randomly, as determined by a normal distribution centered on 10, with a hard boundary at zero. A random offset was also added to the data, which was selected from a uniform distribution (see Table \ref{table[flares]}). Examples of each simulated flare shape can be found in the top panels of Figure \ref{figure[flare_shape]}.

\begin{table}[htp]\caption{Details of Simulated Flare Parameters and Noise.}\label{table[flares]}
\centering
\begin{tabular}{ccc}
  \hline
 Parameters & Exponential & Gaussian \\
 \hline
 $L_{\textrm{\scriptsize{flare}}}$ & $U(100,200)$ & $U(100,200)$ \\
 $t_\textrm{\scriptsize{peak}}$ & $U(30, 300-L_{\textrm{\scriptsize{flare}}})$ & $U(0.4L_{\textrm{\scriptsize{flare}}}, 300-L_{\textrm{\scriptsize{flare}}})$ \\
 $A_{\textrm{\scriptsize{flare}}}$ & $10+N(0,4)$ & $10+N(0,4)$ \\
 $\sigma_{\textrm{\scriptsize{rise}}}$ & n/a & $U(0.1,3)$ \\
 $\sigma_{\textrm{\scriptsize{decay}}}$ & n/a & $U(5,20)$ \\
 Offset & $U(0,100)$ & $U(0,100)$ \\
 \hline
 White S/N & $i\in\mathbb{Z}:i\in[1,5]$ & $i\in\mathbb{Z}:i\in[1,5]$ \\
 $r$ & $U(0.81,0.99)$ & $U(0.81,0.99)$ \\
 Red S/N & $17+N(0,1)$ & $17+N(0,1)$ \\
  \hline
\end{tabular}\\
\tablecomments{$U$ Indicates Values Were Taken from a Uniform Distribution, $N$ Indicates Values Were Taken from a Normal Distribution and n/a Indicates ``Not Applicable''.}
\end{table}

\subsection{Synthetic QPPs}
\label{ssec:sim_qpp}
While some of the flares were left in their basic forms, as described above, various QPP-like signals were added to others, and we now give details of these modifications.
\subsubsection{Single-exponential-decaying sinusoidal QPPs}
\label{sec:simple_qpp}
The simplest form of QPP signal was based on an exponentially decaying periodic function. Such a signal has been used to model QPPs observed in both solar and stellar flares \citep[e.g.][]{2013ApJ...773..156A, 2015ApJ...813L...5P, 2016MNRAS.459.3659P, 2016ApJ...830..110C}. Here the QPP signal as a function of time, $I(t)$, (as measured in, e.g., flux or intensity), is given by
\begin{equation}
I(t) = A_{\textrm{\scriptsize{qpp}}}\exp\left(-\frac{t}{t_{e}}\right)\cos\left(\frac{2\pi t}{P}+\phi\right),
\label{eqn[qpp_simple]}
\end{equation}
where $A_{\textrm{\scriptsize{qpp}}}$ is the amplitude of the QPP signal, $t_e$ is the decay time of the QPP, $P$ is the QPP period, and $\phi$ is the phase. $A_{\textrm{\scriptsize{qpp}}}$ was varied systematically with respect to the amplitude of the simulated flare, $P$ was varied systematically with respect to the length of the flare, $L_{\textrm{flare}}$, and $t_e$ was varied systematically with respect to $P$. Details can be found in Table \ref{table[simple_qpp]}. For each simulated flare, $\phi$ was chosen randomly from a uniform distribution in the range $[0, 2\pi]$. Examples of the QPP signals added to two simulated flares can be seen in the middle panels of Figure \ref{figure[flare_shape]}. 

\begin{table*}[htp]\caption{Details of QPP Signals of Simulated Flares.}\label{table[simple_qpp]}
\centering
\begin{tabular}{lcccccc}
  \hline
 Type & \multicolumn{2}{c}{Number in HH1} & \multicolumn{2}{c}{Number in HH2} & Parameters & Variation \\
  & Exponential & Gaussian & Exponential & Gaussian & & \\
 \hline
 \multirow{4}{*}{Single QPP} & \multirow{4}{*}{25} & \multirow{4}{*}{25} & \multirow{4}{*}{16} & \multirow{4}{*}{16} & $L_{\textrm{\scriptsize{flare}}}/P$ & $[10, 20, 30]$  \\
 & & & & & $A_{\textrm{\scriptsize{qpp}}}/A_{\textrm{\scriptsize{flare}}}$ & $[0.1, 0.2, 0.3]$\\
  & & & & & $t_e/L_{\textrm{\scriptsize{flare}}}$ & $[\frac{1}{30}, \frac{1}{20}, \frac{1}{15}, \frac{1}{10}, \frac{2}{15}, \frac{1}{5}, \frac{2}{5}]$\\
  & & & & & $\phi$ & $U[0,2\pi]$ \\
 \hline
 \multirow{4}{*}{Two QPPs} & \multirow{4}{*}{2} & \multirow{4}{*}{2} & \multirow{4}{*}{0} & \multirow{4}{*}{0} & $P_2/P$ & $U(0.45, 0.55)$\\
 & & & & & $A_{\textrm{\scriptsize{qpp2}}}/A_{\textrm{\scriptsize{qpp}}}$ & $U(0.5, 0.8)$\\
  & & & & & $t_{e2}/t_e$ & $U(0.45, 0.55)$ \\
  & & & & & $\phi_2$ & $U[0,2\pi]$ \\
 \hline
 \multirow{2}{*}{Nonstationary QPPs} &  \multirow{2}{*}{2} &  \multirow{2}{*}{2} &  \multirow{2}{*}{0} &  \multirow{2}{*}{0} & $\nu_1$ & $0.001\nu_0$\\
    & & & & & $t_1$ & $100$ \\
 \hline
 Linear background & 1 & 2 & 0 & 0 & $C_1$ & $A_{\textrm{\scriptsize{flare}}}U(-1,1)$\\
 \hline
 \multirow{2}{*}{Quadratic background} & \multirow{2}{*}{2} & \multirow{2}{*}{1} & \multirow{2}{*}{0} & \multirow{2}{*}{0} & $C_1$ & $A_{\textrm{\scriptsize{flare}}}U(-1,1)$ \\
    & & & & & $C_2$ & $U(0,300)$ \\
 \hline
\end{tabular}
\tablecomments{We note that in the flares containing two QPPs, nonstationary QPPs and linear and quadratic background trends, the parameters $P$, $A_{\textrm{qpp}}$, $t_e$ and $\phi$ were defined in the same manner as for the ``Single QPP'' flares, i.e. randomly or systematically varied as described in this table.}
\end{table*}

\subsubsection{Two Exponentially Decaying sinusoidal QPPs}
\label{ssec:2_qpp}
A second QPP signal was added to a number of the simulated flares. This took the same form as the first QPP and so can also be described by equation \ref{eqn[qpp_simple]}. The amplitude of the second QPP, $A_{\textrm{qpp2}}$, was scaled systematically with respect to the amplitude of the first QPP, $A_{\textrm{qpp}}$, such that $A_{\textrm{qpp2}}<A_{\textrm{qpp}}$ (see Table \ref{table[simple_qpp]}). Similarly, the period and decay time of the second QPP were scaled systematically relative to the period of the first QPP. Recall that the decay time of the original QPP, $t_e$, was scaled relative to the period of the original QPP, $P$, so the decay time of the second QPP, $t_{e2}$, was also varied systematically relative to $t_e$. The phase was again selected from a uniform distribution in the range $[0, 2\pi]$. 

\subsubsection{Nonstationary sinusoidal QPPs}\label{ssec:non_stat}
In real flares the physical conditions in the flaring region evolve and change substantially during the event, and so nonstationary QPP signals are observed regularly \citep[e.g.][]{2019PPCF...61a4024N}. To take this into account, some of the input synthetic QPP signals were nonstationary, and specifically had nonstationary periods. Here we concentrate on varying the period with time, but a future study could, for example, examine the impact of a varying phase or amplitude on the ability of the hounds' methods to detect QPPs. The nonstationary signal was based on equation \ref{eqn[qpp_simple]}; however, the frequency of the sinusoid was varied as a function of time such that
\begin{equation}\label{eqn:non_stat}
f=f_0\left(\frac{f_1}{f_0}\right)^{t/t_1},
\end{equation}
where $f_0$ is the frequency at time $t=0$ and $f_1$ is the frequency at time $t=t_1$. Here $f_0=1/P$ and, as in Section \ref{sec:simple_qpp}, $P$ was varied systematically with respect to $L_{\textrm{flare}}$. For all simulated flares with nonstationary QPPs, $t_1=100$ and $f_1=1/(100P)$, meaning that the period increased with time, as was the case for the real QPPs observed by, for example, \citet{2018ApJ...858L...3K} and \citet{2019ApJ...875...33H}. All other parameters were varied in the manner described in Section \ref{sec:simple_qpp}.

\subsubsection{Multiple flares}\label{ssec:multiple}
In addition to the sinusoidal QPPs, simulations were produced where the QPPs consisted of multiple flares. In these simulations either one or two additional flares were added to the initial flare profile. The shapes of these flares were the same as the original flare. 

When one additional flare was incorporated, the timing of the secondary flare was selected randomly from a uniform distribution such that the peak of the secondary flare occurred during the decay phase of the original flare. The amplitudes of the secondary flares were scaled relative to the amplitude of the initial flare, where the ratio of the flare amplitudes was selected using a uniform random number generator in the range [0.3, 0.5] and the amplitude of the second flare was always smaller than the original (see Table \ref{table[simulations]}). For the remainder of this article, simulated flares containing two flares will be referred to as ``double flares.''

When two additional flares were incorporated, the amplitude of the tertiary flare was selected to be 60\% of the amplitude of the secondary flare. For these flares, the timing of the secondary flare was restricted to the first half of the flare decay phase. Two regimes were used to determine the timing of the tertiary flare: In the first regime, the timing was selected using a uniform random number generator and was allowed to occur anywhere in the second half of the decay phase (see Table \ref{table[simulations]}). The second regime was designed to produce a periodic signal so that the separation in time between the secondary and tertiary peaks was fixed at the time separation of the primary and secondary peaks. For the rest of this article, the first regime will be referred to as ``nonperiodic multiple flares,'' while the second regime will be referred to as ``periodic multiple flares.''

\subsection{Noise}
\label{ssec:noise}
Two types of noise were added onto each simulated flare. Firstly, white noise was added, which was taken from a Gaussian distribution, where the standard deviation of the Gaussian distribution was systematically varied relative to the amplitude of the flare. In flares that included a synthetic QPP signal, the amplitude of that signal was also  systematically varied with respect to the amplitude of the flare. This ensured that the amplitude of the white noise was, therefore, also systematically varied with respect to the QPP amplitude. 

In addition to the white noise, red noise was also added onto the simulated flares. Red noise is a common feature of flare time series, and if its presence is not properly accounted for by detection methods, it can lead to false detections \citep[e.g.][]{2016ApJ...825..110A}. The added red noise, $N_i$, can be described by the following equation:
\begin{equation}
N_i =rN_{i-1}+\sqrt{(1-r^2)}w_i, 
\label{eqn[red_noise]}
\end{equation}
where $i$ denotes the index of the data point in the time series, $r$ determines the correlation coefficient between successive data points, and $w_i$ denotes a white-noise component. Here $r$ was selected using a uniform random number generator in the range [0.81, 0.99]. $w_i$ was taken from a Gaussian distribution, centered on zero and with a standard deviation that was scaled systematically relative to the amplitude of the flare.

In this study, the noise was added to the simulated flare in an additive manner. In reality this is likely to be somewhat simplistic, and some multiplicative component is expected. Further studies are required to determine the impact of the multiplicative component on the detection of QPPs.

\subsection{Background trends}\label{ssec:background}
In real flare data, a background trend is often observed in addition to the underlying flare shape itself (which can also be considered as a background trend when searching for QPPs). This is particularly true in stellar white light observations, where the light curve can be modulated by, for example, the presence of starspots \citep{2015ApJ...813L...5P, 2016MNRAS.459.3659P} but can also be observed if the flare containing the QPPs occurs during the decay phase of a previous flare. To determine the impact of this on the ability of the detection methods to identify robustly QPPs, background trends were incorporated into some of the simulated flares. These backgrounds were either linear or quadratic, and the coefficients of the background trend were all varied with respect to the amplitude of the original flare. For the linear background trend, a variation of
\begin{equation}
L(t)=C_1t, 
\label{eqn[linear]}
\end{equation}
was added to the simulated flare time series, where $C_1$ was a constant chosen randomly from a uniform distribution to be some positive or negative fraction of the flare amplitude ($A_{\textrm{flare}}U(-1,1)$). As a constant offset was added to all simulated time series as standard, there was no need to include an additional constant offset in equation \ref{eqn[linear]}. Similarly, the quadratic background trends were given by 
\begin{equation}
Q(t)=C_1t+C_2t^2, 
\label{eqn[quad]}
\end{equation}
where $C_1$ was defined as above in the linear background trend and $C_2$ was chosen randomly from a uniform distribution in the range $0<C_2<300$. 

\begin{table*}[htp]\caption{Details of simulated Single, Double, and Multiple Flares.}\label{table[simulations]}
\centering
\resizebox{\textwidth}{\height}{\hskip-2.2cm\begin{tabular}{lcccccccc}
  \hline
 Type & \multicolumn{4}{c}{Number}& \multicolumn{2}{c}{Exponential} & \multicolumn{2}{c}{Gaussian}\\ 
 \cline{2-5} \cline{6-7}\cline{8-9}
 & \multicolumn{2}{c}{HH1} & \multicolumn{2}{c}{HH2}& \multicolumn{2}{c}{} & \multicolumn{2}{c}{}\\ 
 \cline{2-3}\cline{4-5}
  & E & G & E & G & Parameters & Variation & Parameters & Variation \\
 \hline
 Single & 1 & 0 & 19 & 22 & & & & \\
 \hline
 \multirow{3}{*}{Double} & \multirow{3}{*}{1} & \multirow{3}{*}{0} & \multirow{3}{*}{5} & \multirow{3}{*}{7} & $t_{\textrm{\scriptsize{peak2}}}$ & $t_{\textrm{\scriptsize{peak}}}+U(0,0.375L_{\textrm{\scriptsize{flare}}})$ &  $t_{\textrm{\scriptsize{peak2}}}$ & $t_{\textrm{\scriptsize{peak}}}+U(0,0.375L_{\textrm{\scriptsize{flare}}})$\\
 & & & & & $A_{\textrm{\scriptsize{flare2}}}$ & $U(0.1A_{\textrm{\scriptsize{flare}}},0.3A_{\textrm{\scriptsize{flare}}})$ & $A_{\textrm{\scriptsize{flare2}}}$ & $U(0.1A_{\textrm{\scriptsize{flare}}},0.3A_{\textrm{\scriptsize{flare}}})$\\
 & & & & & $L_{\textrm{\scriptsize{flare2}}}$ & $U(0.4L_{\textrm{\scriptsize{flare}}},0.6L_{\textrm{\scriptsize{flare}}})$ & $\sigma_{\textrm{\scriptsize{rise2}}}, \sigma_{\textrm{\scriptsize{decay2}}}$ & $0.1\sigma_{\textrm{\scriptsize{rise}}}, 0.1\sigma_{\textrm{\scriptsize{decay}}}$\\
 \hline
 \multirow{3}{*}{Nonperiodic multiple} & \multirow{3}{*}{3} & \multirow{3}{*}{3} &\multirow{3}{*}{4} & \multirow{3}{*}{3} & $t_{\textrm{\scriptsize{peak3}}}$ & $t_{\textrm{\scriptsize{peak}}}+U(0.375L_{\textrm{\scriptsize{flare}}}, 0.75L_{\textrm{\scriptsize{flare}}})$ &   $t_{\textrm{\scriptsize{peak3}}}$ & $t_{\textrm{\scriptsize{peak}}}+U(0.375L_{\textrm{\scriptsize{flare}}}, 0.75L_{\textrm{\scriptsize{flare}}})$ \\
  & & & & & $A_{\textrm{\scriptsize{flare3}}}$ & $0.6A_{\textrm{\scriptsize{flare2}}}$ & $A_{\textrm{\scriptsize{flare3}}}$ & $0.6A_{\textrm{\scriptsize{flare2}}}$\\
 & & & & & $L_{\textrm{\scriptsize{flare3}}}$ & $L_{\textrm{\scriptsize{flare3}}}$ & $\sigma_{\textrm{\scriptsize{rise3}}}, \sigma_{\textrm{\scriptsize{decay2}}}$ & $0.1\sigma_{\textrm{\scriptsize{rise}}}, 0.1\sigma_{\textrm{\scriptsize{decay}}}$\\
 \hline
 \multirow{3}{*}{Periodic multiple} & \multirow{3}{*}{4} & \multirow{3}{*}{4} & \multirow{3}{*}{1} & \multirow{3}{*}{7} & $t_{\textrm{\scriptsize{peak3}}}$ & $t_{\textrm{\scriptsize{peak}}}+2(t_{\textrm{\scriptsize{peak2}}}-t_{\textrm{\scriptsize{peak}}})$&  $t_{\textrm{\scriptsize{peak3}}}$ & $t_{\textrm{\scriptsize{peak}}}+2(t_{\textrm{\scriptsize{peak2}}}-t_{\textrm{\scriptsize{peak}}})$ \\
   & & & & & $A_{\textrm{\scriptsize{flare3}}}$ & $0.6A_{\textrm{\scriptsize{flare2}}}$ & $A_{\textrm{\scriptsize{flare3}}}$ & $0.6A_{\textrm{\scriptsize{flare2}}}$\\
 & & & & & $L_{\textrm{\scriptsize{flare3}}}$ & $L_{\textrm{\scriptsize{flare3}}}$ & $\sigma_{\textrm{\scriptsize{rise3}}}, \sigma_{\textrm{\scriptsize{decay2}}}$ & $0.1\sigma_{\textrm{\scriptsize{rise}}}, 0.1\sigma_{\textrm{\scriptsize{decay}}}$\\
 \hline
 \end{tabular}}\\
 \tablecomments{``E'' denotes flare shapes with exponential decays based on the flare shape of \citet{2014ApJ...797..122D}. ``G'' denotes flares shapes based upon two half-Gaussians.}
\end{table*}

\subsection{Real flares}\label{ssec:real}
In addition to the simulated flares, the hare-and-hounds exercises also contained a number of disguised real solar and stellar flares. The real flares were chosen predominantly from previously published results where QPP detections had been claimed. In addition, one flare where no QPPs had previously been detected was included in the sample. They were also chosen based on the number of data points within the flare, such that they would fit the model of the simulated flares, with each containing 300 data points. For each real flare, the time stamps were removed and an offset, chosen randomly from a uniform distribution, was added (in the same manner as with the simulated flares; see Section \ref{sec:sims}). Each flare was then saved in the same kind of file as the simulated flares and given a random ID number; thus, these flares were indistinguishable from the simulated ones. To test the impact of signal-to-noise (S/N) on the ability to detect the QPPs, additional red and white noise was added to each real flare, and these data were saved in a separate file and given a different randomly selected ID number.

\section{Hare-and-hounds Exercises}\label{sec:hh_setup}
The first hare-and-hounds exercise (HH1) concentrated on the quality of the detections. HH1 consisted of 101 simulated flares, and numbers of each type of simulated flare can be found in Tables \ref{table[simple_qpp]} and \ref{table[simulations]}. This sample contained simulated flares of all types and of various different S/N levels. The hounds were given no information about what was in the sample prior to analysis, and so the test was completely blind.

As there were only eight flares that did not contain QPPs in the HH1 sample (one single flare, one double flare, and six nonperiodic multiple flares), HH1 is not suited to testing the false-alarm rate of the hounds' methods. We therefore set up a second hare-and-hounds exercise, HH2, which contained 100 simulated flares, 60 of which contained no QPP signal, of which 41 were single flares. The remaining 40 simulated flares contained a single sinusoidal QPP, i.e. a single QPP signal described by equation \ref{eqn[qpp_simple]}. The numbers of each simulated flare type included in HH2 can be found in Tables \ref{table[simple_qpp]} and \ref{table[simulations]}. We note that HH2 was set up after the simulations had been described to the hounds and the results of HH1 discussed. However, the majority of hounds did not modify their methodologies between HH1 and HH2. The exceptions to this are JAM, who took measures to improve his methodology based on the results of HH1, and TVD, who automated the detection code between the HH1 and HH2 exercises. A discussion of the impact of these modifications is given in Sections \ref{ssec:trim} (for JAM) and  \ref{ssec:smooth} (for TVD). 

To investigate further the impact of detrending on the detection of QPPs, a third hare-and-hounds exercise was performed, HH3. Only TVD participated in this exercise, and the aim of HH3 was to  test specifically the smoothing method used by TVD to detrend the flares. HH3 contained 18 flares, with 11 based on an exponential shape and 7 based on the Gaussian shape. Each flare contained a single, exponentially decaying QPP, with $4< P< 20$, $1\le t_e/P \le4$, and S/N of either 2 or 5. 

The simulated flares included in HH1, HH2, and HH3 can be found at \url{https://github.com/ambroomhall}. 

\section{Methods of detection}\label{sec:method}
\begin{figure}[htp]
  \centering   \includegraphics[width=0.45\textwidth]{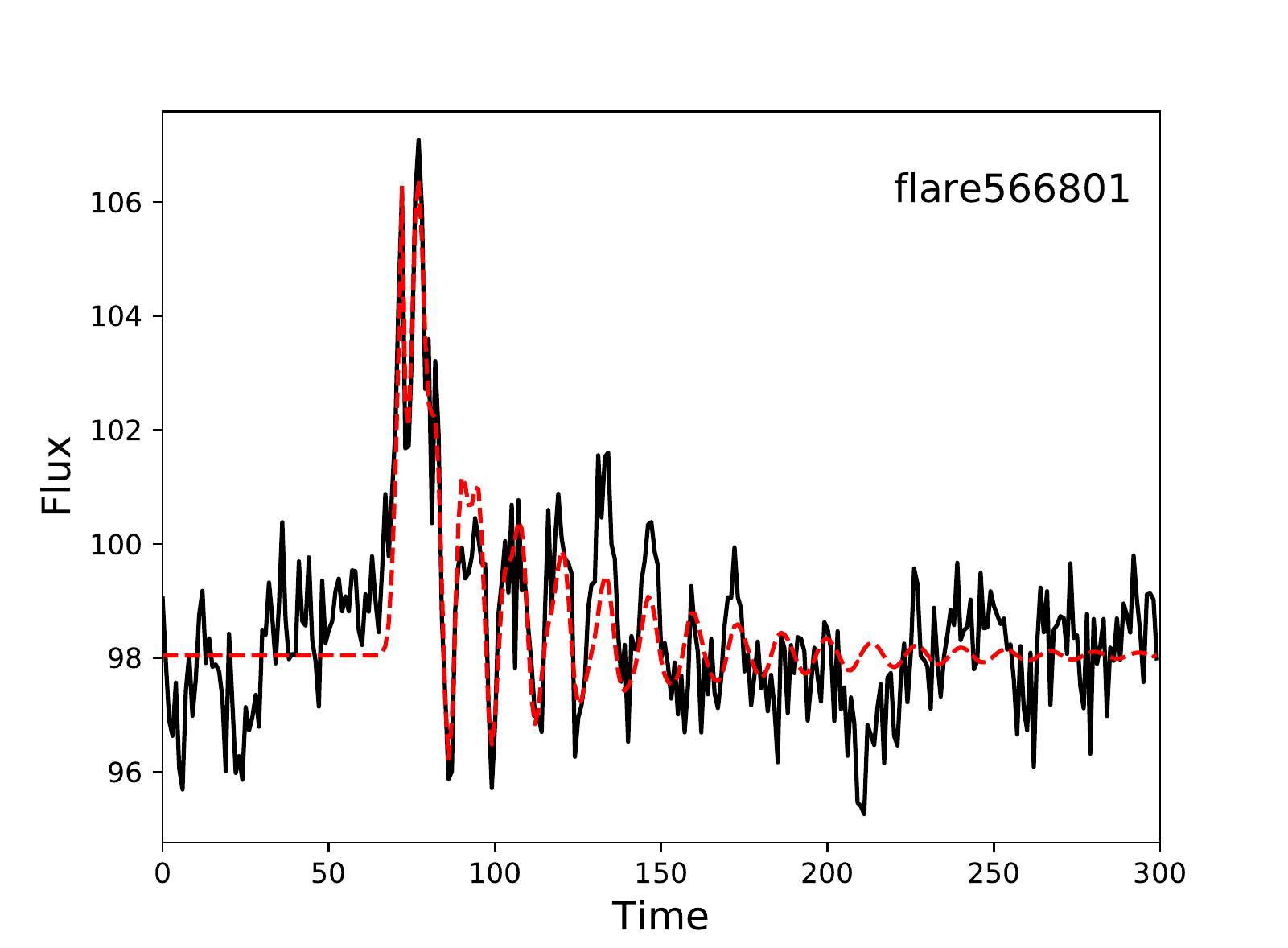}\\
   \caption{Flare 566801, which was based on the flare template of \citet{2014ApJ...797..122D} with $L_\textrm{flare}=134.1$, $t_\textrm{peak}=73.3$, and $A_{\textrm{flare}}=10.3$. The flare contained two QPPs with $P=13.4$, $t_e=53.6$, $A_\textrm{qpp}=3.1$, $P_2=8.4$, $t_{e2}=33.4$, and $A_\textrm{qpp2}=2.1$. The S/N of the flare was 5.0. The black solid line depicts the data given to the hounds, while the red dashed line shows the model.} \label{figure[566801_model]}%
\end{figure}

Eight methods were used to analyze the simulated flares, and we now detail those methods. In each method we will show an example analysis of Flare 566801, which was based on the \citet{2014ApJ...797..122D} template. The flare, which is shown in Figure \ref{figure[566801_model]}, had a S/N of 5.0 and contained two QPPs of periods 13.4 and 8.4. This flare was chosen because all hounds were successfully able to recover the primary period (of 13.4), although we note that this was only true for JAM after modifying his methodology for HH2.

\subsection{Gaussian Process Regression--JRAD}
Gaussian processes (GPs) have become a popular method for generating flexible models of astronomical light curves. Unlike analytic models that describe the entire time series by a fixed number of parameters (e.g. polynomials or sines), GPs are non-parametric and instead use ``hyperparameters'' to define a kernel (or autocorrelation) function that describes the relationship between data points. Splines and damped random walk models are two special cases of GP modeling that have been used extensively in astronomy. For full details on using GPs to model astronomical time series see \cite{foreman-mackey2017} and references therein.

We utilize the {\tt Celerite} GP package developed for Python \citep{foreman-mackey2017} qwing to its flexibility in generating kernel functions and speed for modeling potentially large numbers of data points.
In our QPP hare-and-hounds experiment, we are interested in describing a quasi-periodic modulation that decays in amplitude (e.g. equation 1). {\tt Celerite} comes with an ideal kernel for modeling such data: a stochastically driven damped harmonic oscillator, defined by \citet{foreman-mackey2017} as

\begin{equation}
    S(f) = \sqrt{\frac{2}{\pi}} \frac{S_0\,f_0^4}
    {(f^2-{f_0}^2)^2 + {f_0}^2\,f^2/Q^2}\,,
\end{equation}
\noindent
where $Q$ is the QF or damping rate of the oscillator, $f_0$ is the characteristic oscillation frequency of the QPPs, and $S_0$ governs the peak amplitude of oscillation. 

\begin{figure}[htp]
  \centering   \includegraphics[width=0.45\textwidth]{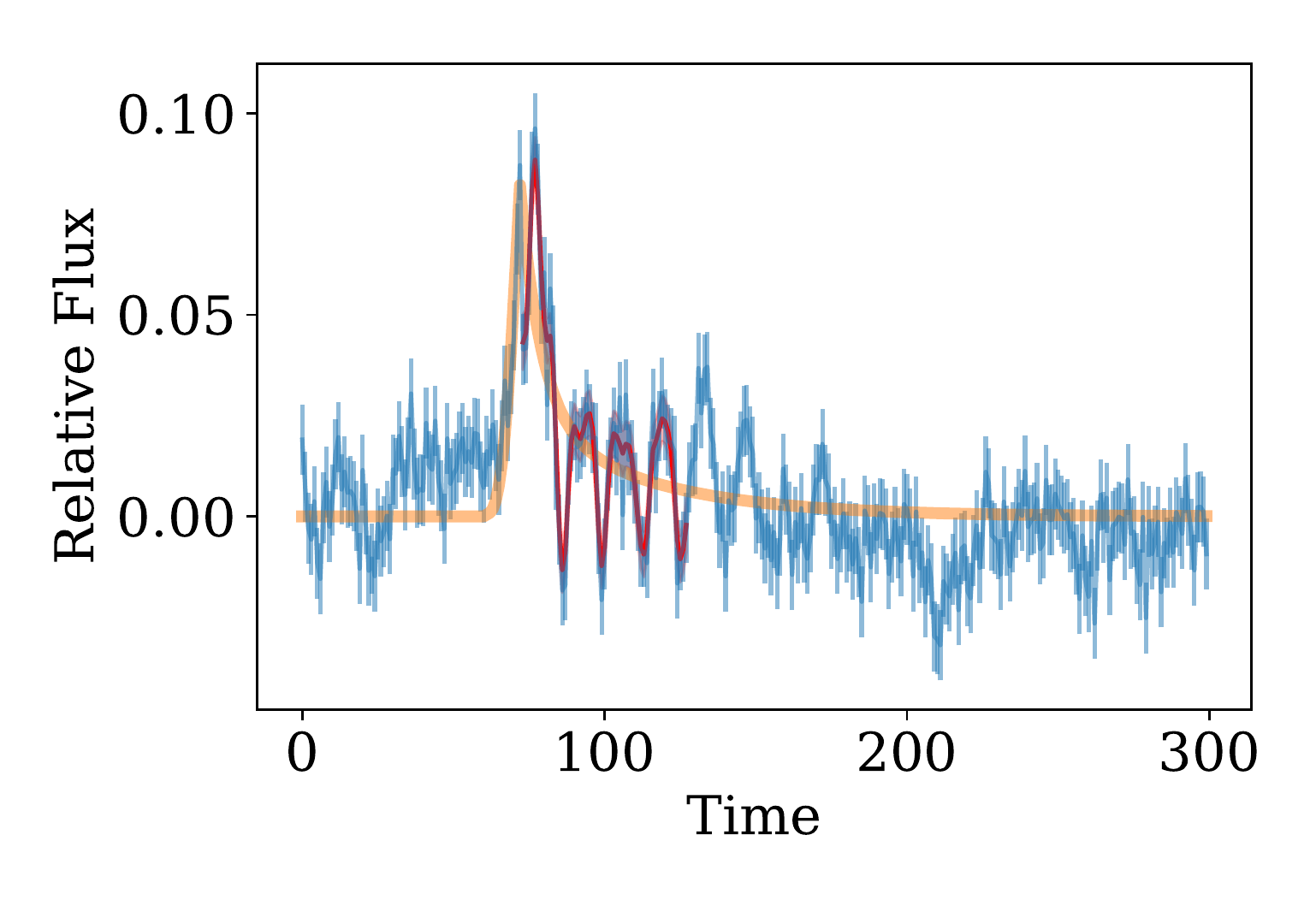}\\
   \caption{GP analysis performed for Flare 566801. Blue is the original simulated light curve. Orange is the \citet{2014ApJ...797..122D} flare model that was subtracted from the data. Red is the GP fit to the QPP. } \label{figure[JRAD]}%
\end{figure}

Since we were only interested here in identifying the QPP component, we first detrended any nonflare stellar variability and subtracted off a smooth flare profile from each event. This was accomplished by first subtracting a linear fit from each candidate event. The \citet{2014ApJ...797..122D} flare polynomial model was then fit to each event using least-squares regression, and this smooth flare was then subtracted from the data. An example of the \citet{2014ApJ...797..122D} flare polynomial model that was fitted to Flare 566801 can be seen in Figure \ref{figure[JRAD]}. Ideally this should leave only the QPPs (if present) in the data to be modeled by our GP. While this approach was fast and easy to interpret, we note that a {\it better} approach to detrending the flare event would be to fit the underlying flare {\it and} the GP simultaneously, e.g. using a Markov Chain Monte Carlo (MCMC) sampler. 

For simplicity, we fit our GP to the residual data that was left after the peak of the polynomial flare (i.e. in the decay phase), and only within 5 times the full-width-at-half-maximum (FWHM) of the flare (i.e. $5\times t_{1/2}$). This was done to avoid overfitting any remaining stellar variability or complex flare shapes that were not removed from our simple detrending procedure. We then followed the worked tutorial included with {\tt Celerite} to fit a damped harmonic oscillator (SHOTerm) GP kernel to our residual data, using the {\tt L-BFGS-B} sampler. This provided us estimates of the flare QPP timescale (period), decay time, and amplitude, as well as generating a model of each flare residual light curve. The QPP period was determined plausible for each simulated event if it was longer than three data points (well enough resolved to measure) and shorter than 200 time units (well constrained by the 300 time units simulated for each event).

\subsection{Wavelet Analysis--LAH}\label{ssec:lah}
Wavelet analysis is a popular tool used in many studies to analyze variations and periodic signals in solar and stellar flaring time series. A detailed description of wavelet analysis is given in \cite{torrence1998practical}, but the main points are mentioned here. The idea of wavelet analysis is to choose a wavelet function, $\Psi(\eta)$, that depends on a time parameter, $\eta$, and convolve this chosen function with a time series of interest. The wavelet function must have a mean of zero and be localized in both time and frequency space. The Morlet wavelet function is most often used when studying oscillatory signals, as it is defined as a plane wave modulated with a Gaussian,
\begin{equation}
\Psi({\eta}) = \pi^{-1/4}  e^{i \omega_0 \eta} e^{-\eta^2/2}
\end{equation} 
Here $\omega_0$ is the nondimensional associated frequency. The wavelet transform of an equally spaced time series, $x_n$, can then be defined as the convolution of $x_n$ with the scaled and translated wavelet function $\Psi$, given by
\begin{equation}
W_n(s) = \sum_{n'=0}^{N-1} x_{n'} \Psi^{*} \left[ \frac{(n' - n)\delta t }{s}  \right]
\end{equation}
Here $\Psi^{*}$ represents the complex conjugate of the wavelet function and $s$ is the wavelet scale.  By varying the scale $s$ and translating it along the localized time index $n$, an array of the complex wavelet transform can be determined. The wavelet power spectrum is defined as $|W_n(s)|^2$ and informs us about the amount of power that is present at a certain scale $s$ (or period) and can be used to determine dominant periods that are present in the time series $x_n$. A 1D global wavelet spectrum can also be calculated, defined as
\begin{equation}
\overline{W^2}(s) = \frac{1}{N} \sum_{n=0}^{N-1} |W_n(s)|^2
\end{equation}

In this exercise, the significance of enhanced power in the wavelet spectra was tested using a red-noise background spectrum. Following \citet{gilman1963power} and \citet{torrence1998practical},  this was estimated by a lag-1 autoregressive AR(1) process given by 
\begin{equation}
x_n = \alpha x_{n-1} + z_n
\end{equation}
where $\alpha$ is the lag-1 autocorrelation, $x_0 = 0$, and $z_n$ represents white noise.

\begin{figure*}[htp]
  \centering   \includegraphics[width=0.9\textwidth]{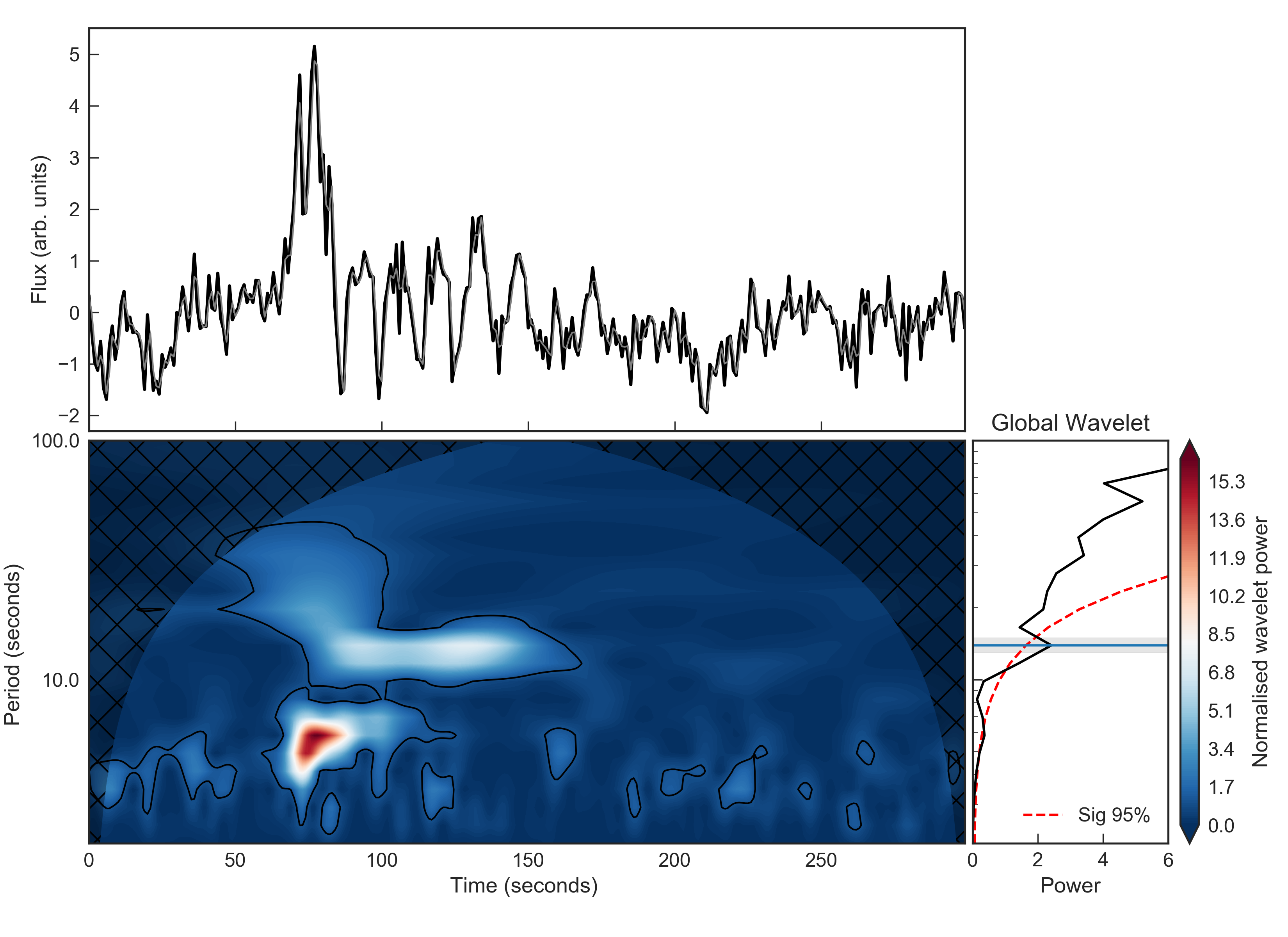}\\
   \caption{Wavelet analysis on the simulated Flare 566801. The flare time series is shown in the top panel, and the associated wavelet power spectrum and global wavelet spectrum are shown in the bottom panels. The normalized wavelet spectrum indicates regions of enhanced power at certain periods with regions above the 95\% confidence level marked by the thin solid lines. The shaded and hatched area is the cone of influence. The global wavelet spectrum is shown in the bottom right hand panel. The black line indicates the global wavelet power from the associated wavelet power spectrum and the red dashed line indicates the 95\% confidence level above the red-noise background model. For the hare-and-hound exercise, a detected period was defined as having global wavelet power above this confidence level. In this example, a horizontal line is drawn at the peak of the global spectrum at $\sim$13~s.} \label{figure[LAH]}%
\end{figure*}

For the hare-and-hounds test samples, the flare signals were \textit{not} detrended before employing the use of wavelet analysis. In this way, the red-noise component can be taken into account when searching for a significant period and avoids the introduction of a bias or error in choosing a detrending window size. In some cases the input flare series was smoothed by two data points to reduce noise. To be robust in the analysis of all the flares in this exercise, a detected period was defined as having a peak in the global power spectrum that lies above the 95\% confidence level. An example of this wavelet analysis performed on the simulated Flare 566801 is shown in Figure~\ref{figure[LAH]}, where a significant peak in the global spectrum is identified at $\sim13\,$ time units in agreement with the input period. A short-lived signal is also seen at around 6 time units that is just above the significance level. This period is slightly lower than, but not inconsistent with, secondary signal included in Flare 566801, which had an input periodicity of 8.4 time units.

\subsection{Automated Flare Inference of Oscillations (AFINO)--ARI}\label{ssec:ari}

The AFINO was designed to search for global QPP signatures in flare time series. The main feature of the method is that it examines the Fourier power spectrum of the flare signal and performs a model fitting and comparison approach to find the best representation of the data. AFINO is described in detail in \citet{2015ApJ...798..108I, 2016ApJ...833..284I}; here we summarize the key steps in the method. The first step in AFINO is to apodize the input time series data by normalizing by the mean and applying a Hanning window to the original time series. The results are not very sensitive to the exact choice of window function, but windowing is necessary in order to address the effects of the finite-duration time series on the Fourier power spectrum. The normalization, meanwhile, is for convenience only.

\begin{figure*}[htp]
  \centering   \includegraphics[width=0.9\textwidth]{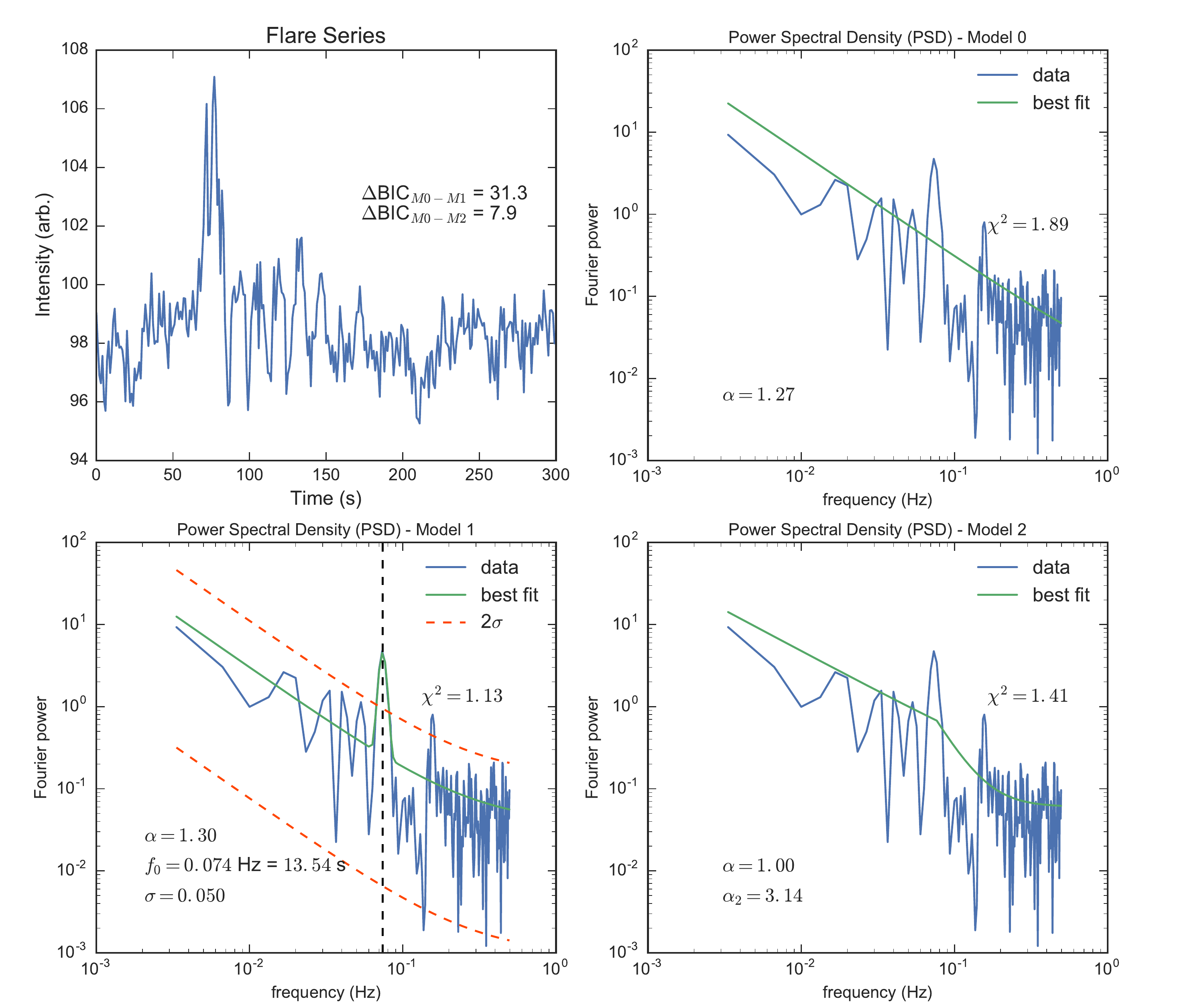}\\
   \caption{AFINO applied to the synthetic Flare 566801. The input flare time series is shown in the top left panel. The remaining panels show the best fits of three models to the Fourier power spectrum of the flare: a single power-law plus a constant (top right), a power law with a bump representing a QPP-like signature (bottom left), and a broken power-law plus  constant (bottom right). The BIC shows that the QPP-like model is strongly preferred over both the single power law and broken power law models. The best-fit frequency is ~0.074 Hz, corresponding to a period of ~13.5s, and is shown by the vertical dashed line in the bottom left panel. The $\Delta BIC$ values are indicated in the top left panel, where $M0$ is the single power-law model, $M1$ is the QPP model, and $M2$ is the broken power-law model.}
   \label{figure[afino]}%
\end{figure*}

The next stage, and the key element of the AFINO procedure, is to perform a model comparison on the Fourier power spectrum of the time series. AFINO is flexible regarding both the choice of models describing the relation between frequency and power, and the range of data being included in the fitting procedure. In this work, as in \citet{2016ApJ...833..284I}, AFINO is implemented testing three functional forms for the Fourier power spectra: including a single power law, a broken power law, and a power law plus Gaussian enhancement. The last model is designed to represent a power spectrum containing a quasi-periodic signature, or QPP, while the other models represent alternative hypotheses. These power-law models are based on the observation that power-law Fourier power spectra are a common property of many astrophysical and solar phenomena such as active galactic nuclei, gamma-ray bursts, stellar flares, and magnetars \citep{2010AJ....140..224C, 2011A&A...533A..61G, 2013ApJ...768...87H, 2015ApJ...798..108I}, and that such power laws can lead naturally to the appearance of bursty features in time series. This power law must therefore be accounted for in Fourier spectral models to avoid a drastic overestimation of the significance of localized peaks in the power spectrum \citep{2005A&A...431..391V, 2011A&A...533A..61G}. Figure \ref{figure[afino]} shows examples of the three models fitted to the power spectrum produced for Flare 566801.

In order to fit each model to the Fourier power spectrum, we determine the maximum likelihood $L$ for each model with respect to the data. For Fourier power spectra, the uncertainty in the data points is exponentially distributed \citep[e.g.][]{2005A&A...431..391V, 2010MNRAS.402..307V}. Hence, the likelihood function may be written as

\begin{equation}
L = \prod^{N/2}_{j=1} \frac{1}{s_j} \exp \left(-\frac{i_j}{s_j} \right),
\end{equation}
where $\mathcal{I}$ = ($i_1$,...,$i_{N/2}$)  represents the observed Fourier power at frequency $f_j$ for a time series of length $N$, and $S$ = ($s_i$,...,$s_{N/2}$) represents the model of the Fourier power spectrum. In AFINO, the maximum likelihood (or equivalently the minimum negative log-likelihood) is determined using fitting tools provided by SciPy \citep{scipy_ref}. Once the fitting of each model is completed, AFINO performs a model comparison test using the Bayesian information criterion (BIC) to determine which model is most appropriate given the data. The BIC is closely related to the maximum likelihood $L$, and the BIC comparison test functions similarly to a likelihood ratio test \citep[see][for a recent review]{2018AdSpR..61..655A}. The BIC (for large $N$) is given by

\begin{equation}
BIC = - 2 \ln(L) + k \ln(n)
\label{bic_eqn}
\end{equation}
where $L$ is the maximum likelihood described above, $k$ is the number of free parameters, and $n = N/2$ is the number of data points in the power spectrum. The key concept of BIC is that there is a built-in penalty for adding complexity to the model. Using the BIC value to compare models therefore tests whether the added complexity offered by the QPP-like model is sufficiently justified. This approach is intentionally conservative, with one of the primary goals of AFINO being to have a low false-positive--or Type I error--rate. The $k \ln (n)$ term is particularly significant for short data series where $n$ is not very large, such as in stellar flare light curves. 

To compare models, we calculate $dBIC$ = $BIC_j$ - $BIC_{QPP}$, for all non-QPP models $j$. The BIC for each model will be negative, and as the fitting code tries to minimize the BIC, the best-fitting model will be the one with the largest negative BIC value. Therefore, when the BIC value for the QPP-like model is lower than that of the other models - i.e., when $dBIC$ is positive for all alternative models $j$ - there is evidence for a QPP detection. For the purposes of this work, we divide the strength of evidence into different categories. When $dBIC < 0$ compared to all other models, there is no evidence of a QPP detection. If $0 < dBIC < 5$ compared to all other models, we identify weak evidence for a QPP signature. For $5 < dBIC < 10$, we identify moderate QPP evidence. Finally, events where $dBIC > 10$ compared to all other models indicate strong evidence for a QPP-like signature. For context and to more easily compare with other methodologies, the dBIC value can be expressed in more concrete probabilistic terms, or approximately translated to a $t$-statistic value \citep{10.2307/2291091, 10.2307/271063}. For example, a dBIC in the 6-10 range indicates approximately $>95$\% preference (or 2-$\sigma$) for one model over another, while a dBIC $>10$ corresponds to a $>99$\% preference for the minimized model.

For Flare 566802, when comparing a single power-law model to the QPP model, $dBIC=31.3$, indicating strong evidence for a QPP signature. Similarly, when comparing a broken power-law model to the QPP model, $dBIC=23.4$, again indicating strong evidence for a QPP signature. When comparing a broken power-law model to the single power-law model, $dBIC=7.9$, implying that the broken power law is a better representation than the single power law, but still not as good as the QPP model. Since the QPP-like model is strongly preferred over both alternatives, this event is recorded as a `strong' QPP flare. The QPP model correctly identifies the period of the QPP to within 0.1 units. 

\subsubsection{Relaxed AFINO--LAH in HH1}
The AFINO methodology described above in Section \ref{ssec:ari} was also employed independently by LAH. However, a somewhat ``relaxed'' version was implemented. Instead of testing three functional forms of the Fourier power spectrum, only two were considered, namely, a single power law and a power law with a Gaussian bump. These models were both fit to the data, a model comparison between them was performed, and a $dBIC$ was calculated. A flare from the HH1 sample with a $dBIC > 10$ was taken to have a significant QPP signature.

\subsection{Smoothing and Periodogram, [HH1 Untrimmed] versus [HH2 Trimmed + Confidence Level] -- JAM}\label{ssec:jam}

\begin{figure*}[htp]
  \centering   \includegraphics[width=0.45\textwidth]{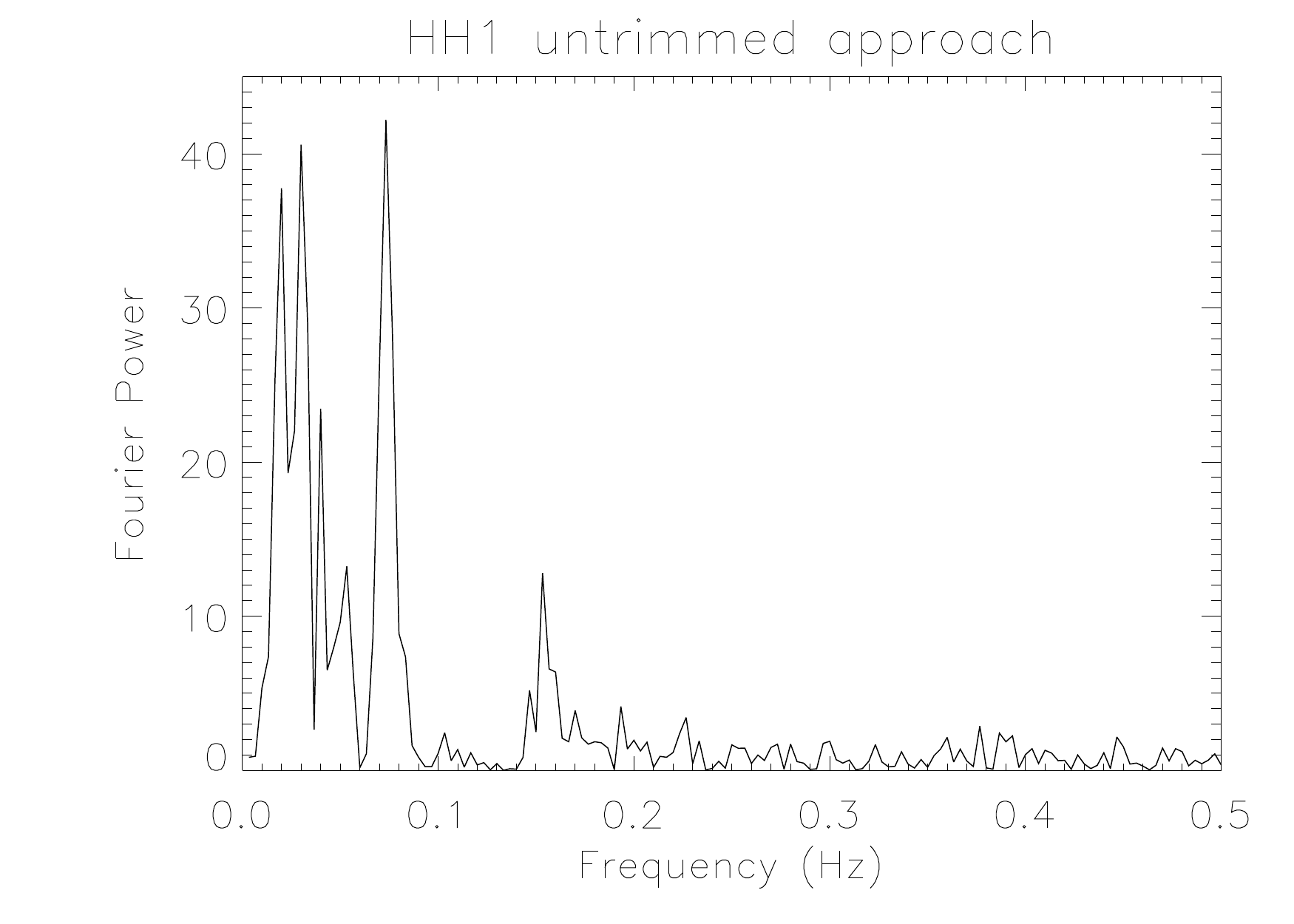}
  \includegraphics[width=0.45\textwidth]{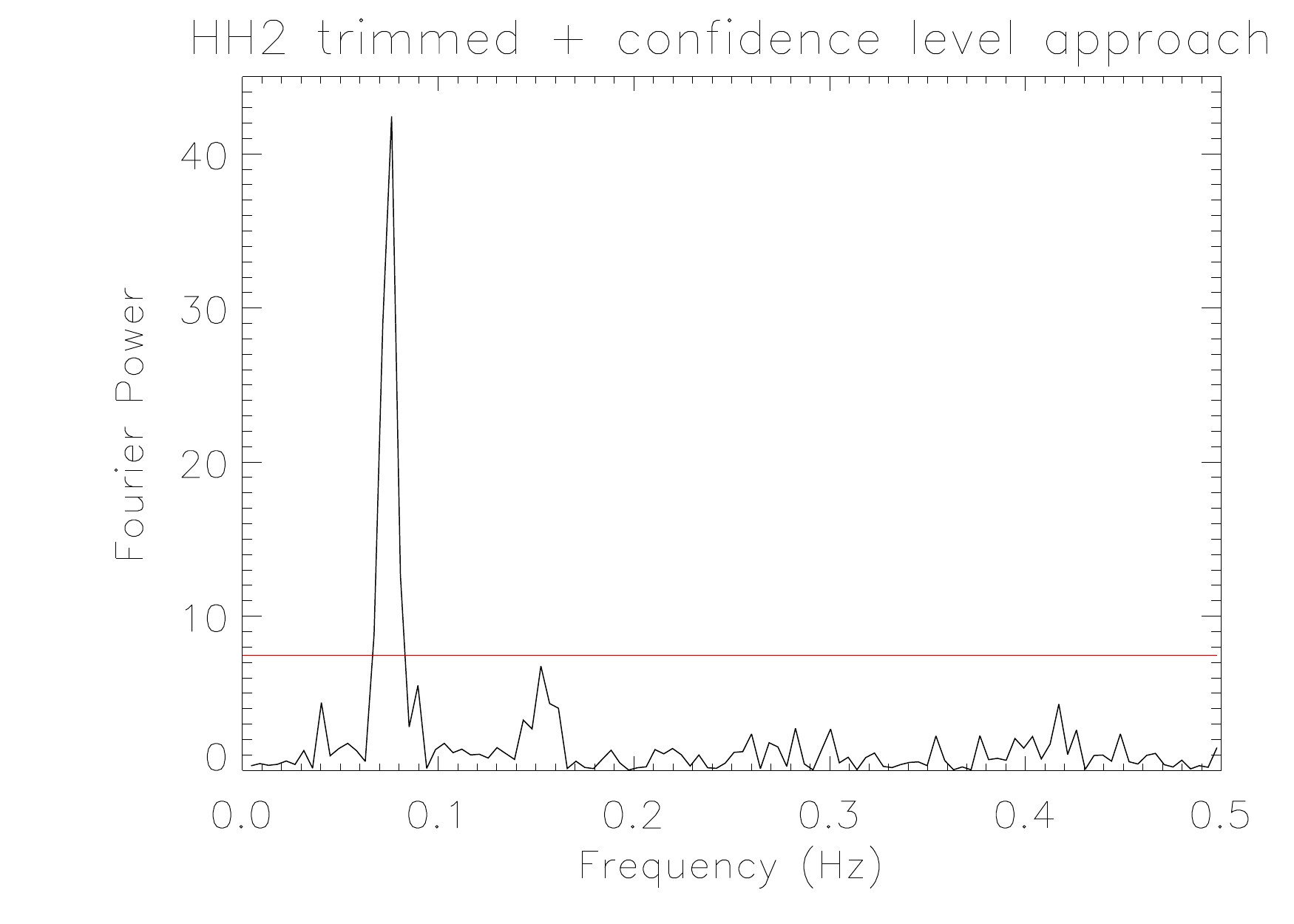}\\
   \caption{Frequency-power spectra produced by JAM for Flare 566801. Left panel: original method used in HH1, where the full time series was used to generate a smoothed light curve that was then subtracted from the original time series before the power spectrum was computed. Right panel: modified approach used for HH2, where the data were trimmed to start at the location of the local maximum before generating the smoothed light curve. In this improved method, a false-alarm probability was used to determine the significance of any peaks and the red horizontal line shows the 95\% confidence level. We note that Flare 566801 was in HH1, not HH2, but is used here to demonstrate the HH2 method employed by JAM for consistency.} \label{figure[JAM]}%
\end{figure*}

Under this methodology, we investigated the robustness of a simple and straightforward approach to oscillation detection. For each of the simulated flares of HH1, an overall trend for the data was generated by smoothing the flare light curve over a window of 50 data points. The smoothed flare light curve was then subtracted from the original signal to generate a residual, and then a Lomb-Scargle periodogram was generated from the residual. The Lomb-Scargle periodogram \citep{1976Ap&SS..39..447L, 1982ApJ...263..835S} is an algorithm for detecting periodicities in data by performing a Fourier-like transform to create a period-power spectrum. Although not relevant for the simulated data considered here, it is particularly useful if the data are unevenly sampled, as is often the case in astronomy. Further details can be found in \citet{2018ApJS..236...16V}. The frequency with the most power from the Fourier power spectrum was identified, and this single frequency was recorded for all HH1 flares. Under this methodology, it was straightforward to construct detrended data and obtain a dominant period from the periodogram. In some cases, no dominant peak was apparent in the periodogram, in which case no periodicity was recorded. In HH1 (only), the decision over whether to record a periodicity was made following a by-eye inspection of the periodogram and so was a subjective choice of the user. Figure \ref{figure[JAM]} shows an example of the periodogram produced for Flare 566801. A number of large peaks are visible at low frequencies, and so none were identified as detections following the by-eye inspection. The approach was not labor intensive. However, this simplistic approach suffered from an overall trend skewed by data from both before and after the flare peak and did not implement an objective method of assessing the significance of the detections. The approach was similar to the method in Section~\ref{ssec:tvd}, but the smoothing parameter, $N_\mathrm{smooth}$, was kept fixed at 50.

The approach was improved for HH2, in which the time series, $F(t)$, was trimmed to begin at the location of the local maximum $(dF/dt=0)$. In this way, the trimmed time series only considered the decay phase of the simulated HH2 flares. The trimmed time series was smoothed over a window of 12 data points to generate an overall trend. This trend was subtracted from the trimmed time series to generate a residual, and a Lomb-Scargle periodogram was constructed from the residual. The frequency with the most power from the Fourier power spectrum was identified, and the significance of this peak was assessed by comparing with a 95\% confidence level based on white noise. In this way, a single frequency was recorded only for HH2 flares where the detection was assessed to be significant. The right panel of Figure \ref{figure[JAM]} shows an example of a periodogram, for Flare 566801, produced using this method. A single peak is visible above the 95\% confidence limit, at a period of 13.1, which is close to the input period of 13.4. 

\subsection{Empirical Mode Decomposition (EMD)--TM and DK}
\label{ssec:EMD}
 It has been established that QPPs are not exclusively stationary signals, as the periods of QPPs can be seen to drift with time \citep[e.g.][]{2019PPCF...61a4024N}. Many traditional methods, such as the fast Fourier transform, are poorly equipped to handle nonstationary signals \citep[see, e.g., Table 1 in][]{2008RvGeo..46.2006H} as they attempt to fit the signal with spurious harmonics. The technique of EMD, however. makes use of the power of instantaneous frequencies in a meaningful way and, as the method is entirely empirical and relies only on its own local characteristic time scales, is well adapted to nonstationary datasets.

\begin{figure*}[htp]
	\begin{center}
	\includegraphics[width=0.9\textwidth]{{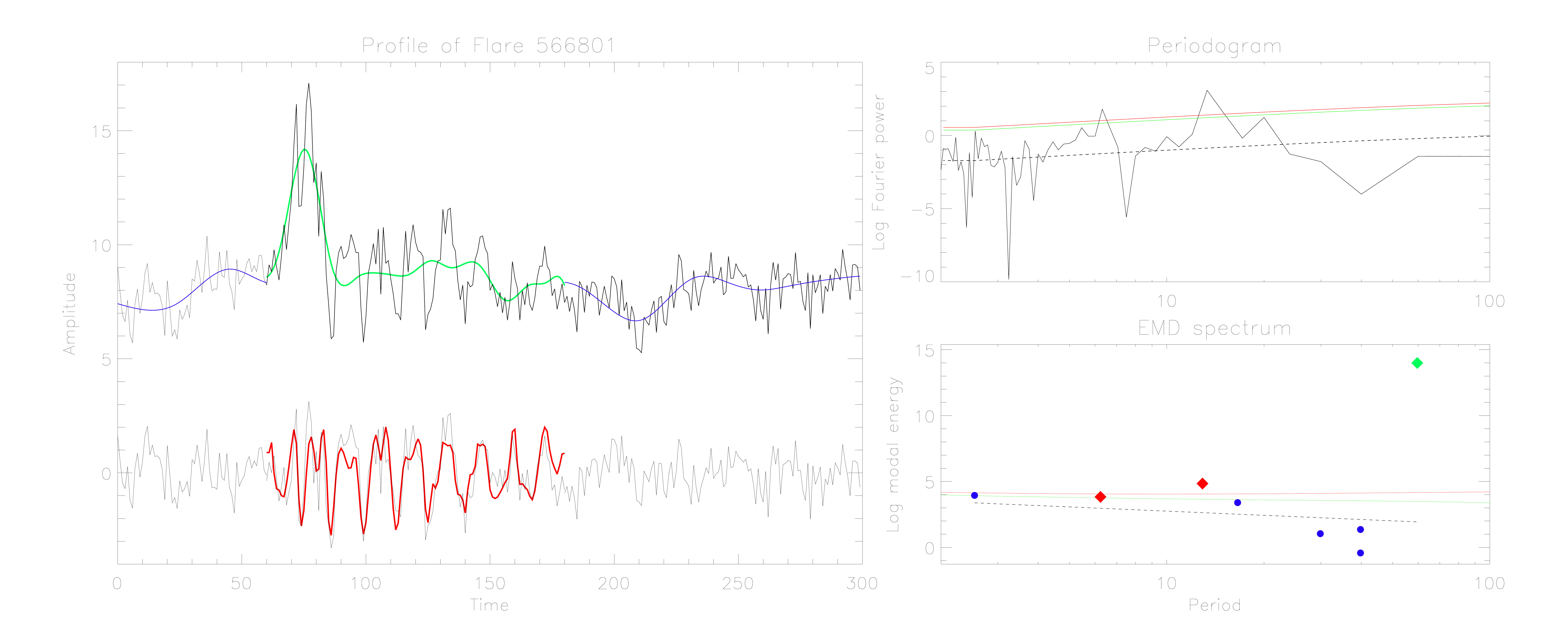}}
    \end{center}
\caption{\label{fig:EMDfigTREND}
EMD analysis of Flare 566801 with an appropriate choice of shift factor.
Left panel: The upper light curve gives the entire duration of the input signal, with EMD extracted trends overlaid in blue and green, separated into pre-flare, flaring and post-flare regions. Below is the detrended light curve overlaid in red by the combination of two statistically significant IMFs.  
Top right panel: Periodogram of the detrended signal with confidence levels of 95$\%$ (green) and 99$\%$ (red). Two significant peaks are observed at $\sim$~6.4 and 14.4.
Bottom right panel: EMD spectrum  of the original input signal with two significant modes, with periods 6.2 (at a confidence level of 95 $\%$) and 12.9 (99$\%$), shown as red diamonds. The trend is given as a green diamond. Blue circles correspond to noisy components with $\alpha \approx 0.89$. The 95$\%$ and 99$\%$ confidence levels are given by the green and red lines, respectively, with the expected mean value shown by the dotted line. }
\end{figure*}

EMD \citep[developed in][]{1998RSPSA.454..903H}, decomposes a signal into a number of intrinsic mode functions (IMFs). These IMFs are functions defined such that they satisfy two conditions: first, that the number of extrema and zero crossings must differ by no more than one, and second, the value of the mean envelope across the IMFs entire duration is zero. IMFs can therefore exhibit frequency and amplitude modulation and can be nonstationary, and they may be recombined to recover the input in a similar way to Fourier harmonics/ The IMF(s) with the largest instantaneous periods may be deducted from the signal as a form of detrending. In particular, the trends found for Flare 566801 can be seen in the upper light curve in the left panel in Figure \ref{fig:EMDfigTREND} and were subsequently subtracted from the signal. The detrended light curves can then be reanalyzed using EMD to give a new set of IMFs that are tested for statistical significance based on confidence levels of 95$\%$ and 99$\%$.
The process of decomposing a signal into IMFs is known as ``sifting," wherein an iterative procedure is applied. At each step, an upper and lower envelope is constructed via cubic spline interpolation of the local maxima and minima. A mean envelope can be obtained by averaging out these two envelopes, which is then subtracted from the input data to produce a new ``proto-IMF''--completing the process of one sift. The new ``proto-IMF'' is then taken to be the new input signal and this method is repeated until a stopping criterion is met. In this case, the stopping criterion is defined by the \lq\lq shift factor,\rq\rq which is given as the standard deviation between two consecutive sifts. Once the standard deviation drops below this value, the computation ceases and the ``proto-IMF'' is taken as an IMF. Then, this IMF is deducted from the raw signal, and the process restarts so that new IMFs can be sifted out. The \lq\lq shift factor\rq\rq influences the number of IMFs extracted and their associated periods. In general, if the value of the shift factor is too high, the IMFs remain obscured by noise and conversely if the value is too low, the IMFs decompose into harmonics \citep[a more detailed discussion can be found in][]{2010AADA..02..277W}.

A superposition of colored and white noise was assumed to be present in the original signal, where the relationship between Fourier spectral power $S$ and frequency $f$ can be described by $S\propto f^{-\alpha}$, where $\alpha$ is a power-law index usually described by a \lq\lq color\rq\rq. White noise is naturally denoted by $\alpha = 0$ as spectral energy is independent of frequency, and can be seen to dominate at high frequencies, whilst colored noise, given by $\alpha > 0$, has a greater significance over lower frequencies. By fitting a broken power law to the periodogram of the detrended signal, the value of $\alpha$ corresponding to colored noise can be found, as outlined in Section~\ref{ssec:CEP}, and this value is used when calculating the confidence levels.

Here the modal energy of an IMF is defined as sum of squares of the instantaneous amplitudes of the mode, and its period is given as the value generating the most significant peak given by the IMF's corresponding global wavelet spectrum. The total energy $E$ and period $P$ of IMFs extracted with EMD from colored noise are related via $E\propto P^{\alpha-1}$. These two properties may be represented graphically in an EMD spectrum \citep[e.g.][]{2018ApJ...858L...3K}, shown in the bottom right panel of Figure~\ref{fig:EMDfigTREND} for Flare 566801. 
Each IMF is represented by a single point corresponding to its dominant period and total energy.
The probability density functions for the energies of IMFs, obtained from pure colored noise, follow chi-squared distributions \citep[see][]{2016A&A...592A.153K}, which use the value of $\alpha$ estimated in the periodogram-based analysis to give confidence levels. It must be noted that the chi-squared energy distribution is not a valid model for the first IMF (corresponding to the extracted function with the shortest period), and so this IMF cannot be measured against the confidence level and hence must be excluded from analysis.
It is expected that the IMF(s) corresponding to the trend of the light curve will be significantly energetic and correspond to a large period, seen in the EMD spectrum in Figure~\ref{fig:EMDfigTREND} as a green diamond, substantially above the 95$\%$ and 99$\%$ confidence levels, given in green and red, respectively. 

In HH1, the time series were manually trimmed into three distinct phases; the pre-flare, flaring, and post-flare regions, and each region was individually investigated for a QPP signature. The time at which the gradient of the light curve rapidly increased was defined as the start time of the flaring region, which continued until the amplitude of the signal returned to its pre-flare level at which point the post-flare region began. For Flare 566801, the flaring section showed evidence of QPP-like behavior and the resulting periodogram (top right panel of Figure~\ref{fig:EMDfigTREND}) of the detrended light curve produced two statistically significant peaks above the 99$\%$ confidence level at $\sim$~6.4 and 14.4, agreeing with the input periods of 8.4 and 13.4. The detrended light curve was decomposed further into seven IMFs, of which two modes were detected to be statistically significant.
The significant IMFs give periods of $\sim$~6.2 and 12.9, with confidences of 95$\%$ and 99$\%$ respectively, which agrees well with both the periodogram-based results and input values. Their superposition is shown in red overlay in the left panel of Figure~\ref{fig:EMDfigTREND} and gives a reasonable visual fit to the input signal. 

The technique of detrending the light curve using EMD, producing a periodogram from the detrended signal, and performing EMD one further time was carried out for 26 datasets given in HH1 (a total of 78 trimmed light curves were processed with this methodology, corresponding to three subsets in each of 26 events). The 26 flares analyzed with EMD were chosen following a by-eye examination of all the datasets in the sample and were selected as the flares most likely to produce a positive detection. EMD was only performed on a limited number of the flares in HH1 owing to the time intensive nature of the technique, which requires a manual input of an appropriate choice of \lq \lq shift factor \rq \rq for an appropriate set of periodicities for each signal.

\begin{figure}[htp]
	\begin{center}
	\includegraphics[width=0.45\textwidth]{{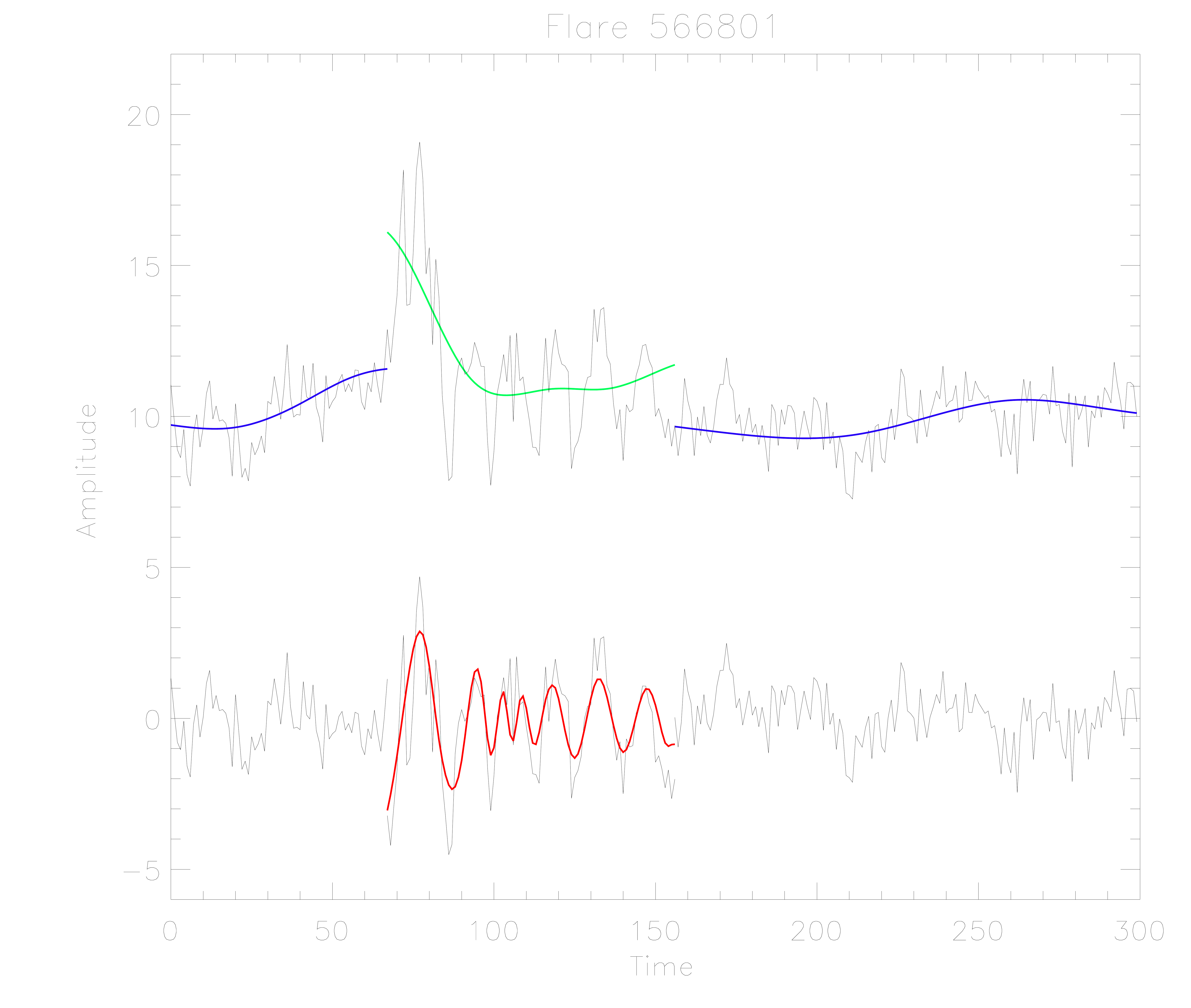}}
    \end{center}
\caption{\label{fig:EMDfig_badtrend}
Analysis of Flare 566801 with an inappropriate choice of shift factor. The upper light curve (black) is the raw signal, with trends extracted from EMD overlaid (blue and green). Below is the detrended light curve (black), with the statistically significant IMF overlaid in red. }
\end{figure}

Initially in HH1, due to user inexperience, insufficient care was taken over the choice of this value, leading to poorly selected trends and IMFs suffering from the effects of mode mixing, decreasing the accuracy of recovered periodicities.  This is partially reflected in the relatively poorer agreement between input and output periods in Section~\ref{sssec:EMD_HH1}.
An example of this is shown in Figure~\ref{fig:EMDfig_badtrend} where a too large shift factor has been chosen to appropriately determine the trend of the flare region. Note how the characteristic rise and exponential decrease are not seen in the trend and how the trends of the three regions do not join smoothly. A better-fitted shift factor gives a trend that bisects the input signal approximately through the midpoints of its apparent oscillations (seen in Figure~\ref{fig:EMDfigTREND}), allowing for a better representation of the QPPs once detrended. This rough choice of shift factor gave an output of a single IMF, with a period of 17.7, which has just a poor agreement with the input value. Moreover, a clear evidence of another common issue in the EMD analysis, a so-called mode mixing problem, can be observed at $\sim$~110 in this example, where the time scale of the oscillation dramatically changes. Such intrinsic mode leakages appeared due to a poor choice of shift factor, which could adversely affect the estimation of the QPP timescales, and hence should be avoided.

When using EMD to detrend a flare signal, a lower shift factor should be selected, as this increases the sensitivity of the technique. In particular, special care must be taken in the choice of the shift factor in cases where the time scale of the flare (e.g. the flare peak width measured at the half-maximum level) is comparable to that of apparent QPPs, such as in Flare 566801, providing the method with enough sensitivity to decompose the intrinsic oscillations from the flare trend.
The value must also be selected carefully such that the extracted trend may retain a classical flare-like shape. Such a profile may introduce artifacts from rapid changes in gradient, which may be fitted with spurious harmonics, and so an appropriate choice of shift factor acts to minimize this effect through manual inspection. 

\subsection{Forward Modeling of QPP Signals--DJP}\label{ssec:djp}

This method is adapted from the Bayesian inference and MCMC sampling techniques recently applied to perform coronal seismology using standing kink oscillations of coronal loops.
Coronal loops are frequently observed to oscillate in response to perturbations from solar flares or CMEs.
Such oscillations have been studied intensively both observationally and theoretically, and so detailed models have been developed.
The strong damping of kink oscillations is attributed to resonant absorption, which may have either an exponential or a Gaussian damping profile depending on the loop density contrast ratio \citep{2013A&A...551A..40P,10.3389/fspas.2019.00022}.
In studies of standing kink oscillations, it is therefore natural to consider several different models, such as the shape of the damping profile.
\citet{2017A&A...600A..78P} also considered the presence of additional longitudinal harmonics and the change in their period ratios due to effects of density stratification or loop expansion, a time-dependent period of oscillation, and a possible low-amplitude decayless component.

The method is based on forward-modeling the expected observational signature for given model parameters, while MCMC sampling allows large parameter spaces to be investigated efficiently.
The benefit of this approach over more general signal analysis methods is that it potentially allows greater details to be extracted in the data.
For example, \citet{2017A&A...600A..78P} demonstrated that the presence of weak higher harmonic oscillations in kink oscillations would be recovered by a model that takes their strong damping into account, whereas they would have negligible signatures in periodogram and wavelet analysis.
The interpretation of the different components of the model (e.g. background trend and different oscillatory components) is done when defining the forward-modeling function compared with, for example, EMD, which produces several IMFs that must be interpreted afterwards.
The method also does not require the signal to be detrended (if the trend is also described by the model), which avoids the choice of trend affecting the results.

On the other hand, the usefulness of the method is based on the particular model being the correct one (or one of them if several models are considered).
In the case of QPPs there are several possible mechanisms that have been proposed.
Ideally each competing model could be applied to the data for an event and then compared, for example, using Bayes factors.
However, models relating the observational light curve to the physical parameters currently do not exist for some of the proposed mechanisms.
For example, the mechanism of generating QPPs by the dispersive evolution of fast wave trains has a characteristic wavelet signature, but the detailed form of it is only revealed by computationally expensive numerical simulations.

\begin{figure*}[htp]
\centering
\includegraphics[width=0.9\textwidth]{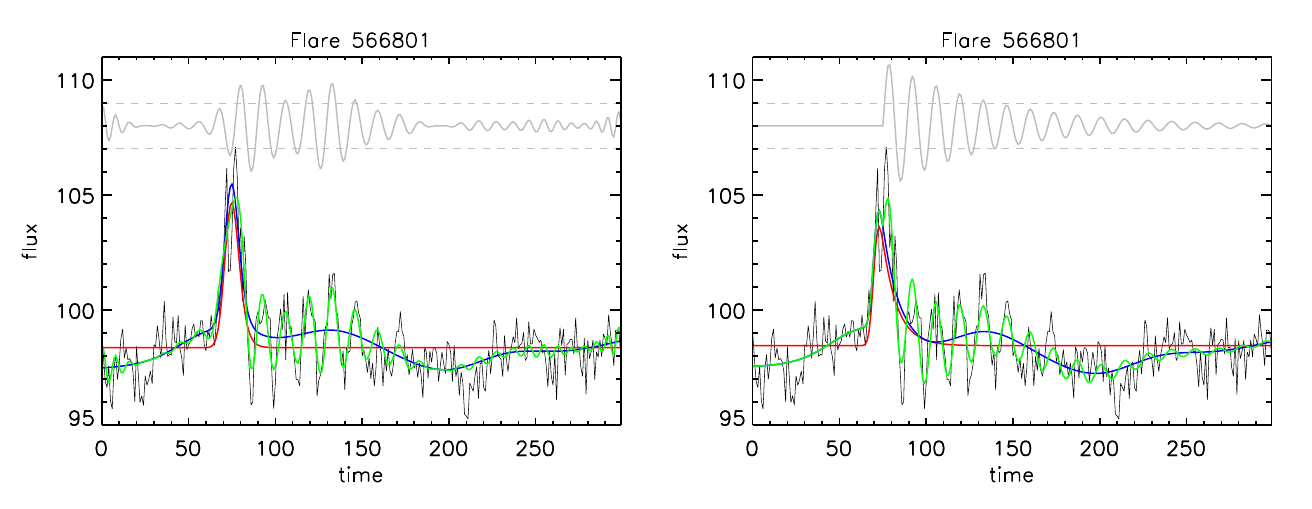}\\
\caption{Method of forward-modeling QPP signals based on the Bayesian inference and MCMC sampling used in \citet{2017A&A...600A..78P}. The left panel shows a combination of a spline-interpolated background, Gaussian noise, a flare described by equation \ref{eq:emg}, and a QPP signal with a continuous amplitude modulation rather than defined start and decay times. The right panel shows a model fit that contains a single flare, an exponentially decaying sinusoidal QPP (with the potential for a nonstationary period), a spline-based background, and Gaussian noise. In each panel the black line shows the simulated data for Flare 566801, the red line shows the flare component of the fit, based on equation \ref{eq:emg}, the blue line shows a combination of the flare fit and the background, while the green shows the overall fit. We note that all the components were fitted simultaneously and are only separated here for clarity. Gray lines correspond to the detrended signal (shifted for visibility). The gray dashed horizontal lines denote the estimated level of (white) noise in the signal. }
\label{figure[djp1]}
\end{figure*}

\citet{2016A&A...585L...6P,2016A&A...589A.136P,2017A&A...600A..78P} use  smooth background trends based on spline interpolation.
The background varying on a timescale longer than the period of oscillation is necessary for the definition of a quasi-equilibrium on top of which an oscillation occurs.
However, a smooth background does not allow impulsive events with rapid, large-amplitude changes, such as flares, to be well described.
\citet{2017A&A...607A...8P} considered the case of kink oscillations, which have a large shift in the equilibrium position associated with the impulsive event that triggered the oscillation.
This was done by including an additional term describing a single rapid shift in the equilibrium position of the coronal loop.
In that work the shifts only took place in one direction, and so a hyperbolic tangent function was suitable to describe it.
In this paper, the large changes in light curves due to flares instead have both a rising and decaying phase, and so an exponentially modified Gaussian (EMG) function is more suitable, which has the form

\begin{eqnarray}
\mathrm{EMG} \left( x \right) =&A& \frac{\lambda}{2} \exp \left( \frac{\lambda}{2} \left(2 \mu + \lambda \sigma^2 - 2 x \right) \right) \nonumber \\
&\times&\mathrm{erfc} \left( \frac{\mu + \lambda \sigma^2 - x}{\sqrt{2} \sigma} \right)
\label{eq:emg}
\end{eqnarray}
where $\mathrm{erfc}\left(x\right) = 1 - \mathrm{erf}\left(x\right)$ is the complementary error function,
$A$ is a constant determining the amplitude,
$\mu$ and $\sigma$ are the mean and standard deviation of the Gaussian component, respectively, and
$\lambda$ is the rate of the exponential component.
The EMG function has a positive skew due to the exponential component, which allows it to describe a wide range of flares, having a decay phase greater than or equal to the rise phase. An example of the EMG function fitted to Flare 566801 can be seen in Figure \ref{figure[djp1]}.

Figure~\ref{figure[djp1]} shows the results for models based on a QPP signal with a continuous amplitude modulation, with defined start and decay times, and an exponentially damped sinusoidal oscillation. (A Gaussian damping profile was also tested, but the Bayesian evidence supported the use of an exponential damping profile.) The green lines represent the model fit based on the maximum a posteriori probability (MAP) values for the model parameters. The blue lines correspond to the background trend component of the model, and the gray lines are the detrended signals. The MCMC sampling technique used in \citet{2017A&A...600A..78P,2017A&A...607A...8P,2018ApJ...860...31P} estimates the level of noise (here assumed to be white) in the data by comparing with the forward-modeled signal. This level is indicated in the figures by the gray dashed horizontal lines. A simple criterion for QPP detection is to therefore require several oscillation extrema to exceed this level. In addition to Flare 566801, shown in Figure \ref{figure[djp1]}, this technique was used to analyze the nonstationary QPP flares and so will be discussed further in Section \ref{ssec:non_stat_out}. 

\subsection{Periodogram-based significance testing -- CEP}\label{ssec:CEP}

\begin{figure*}
\includegraphics[width=\linewidth]{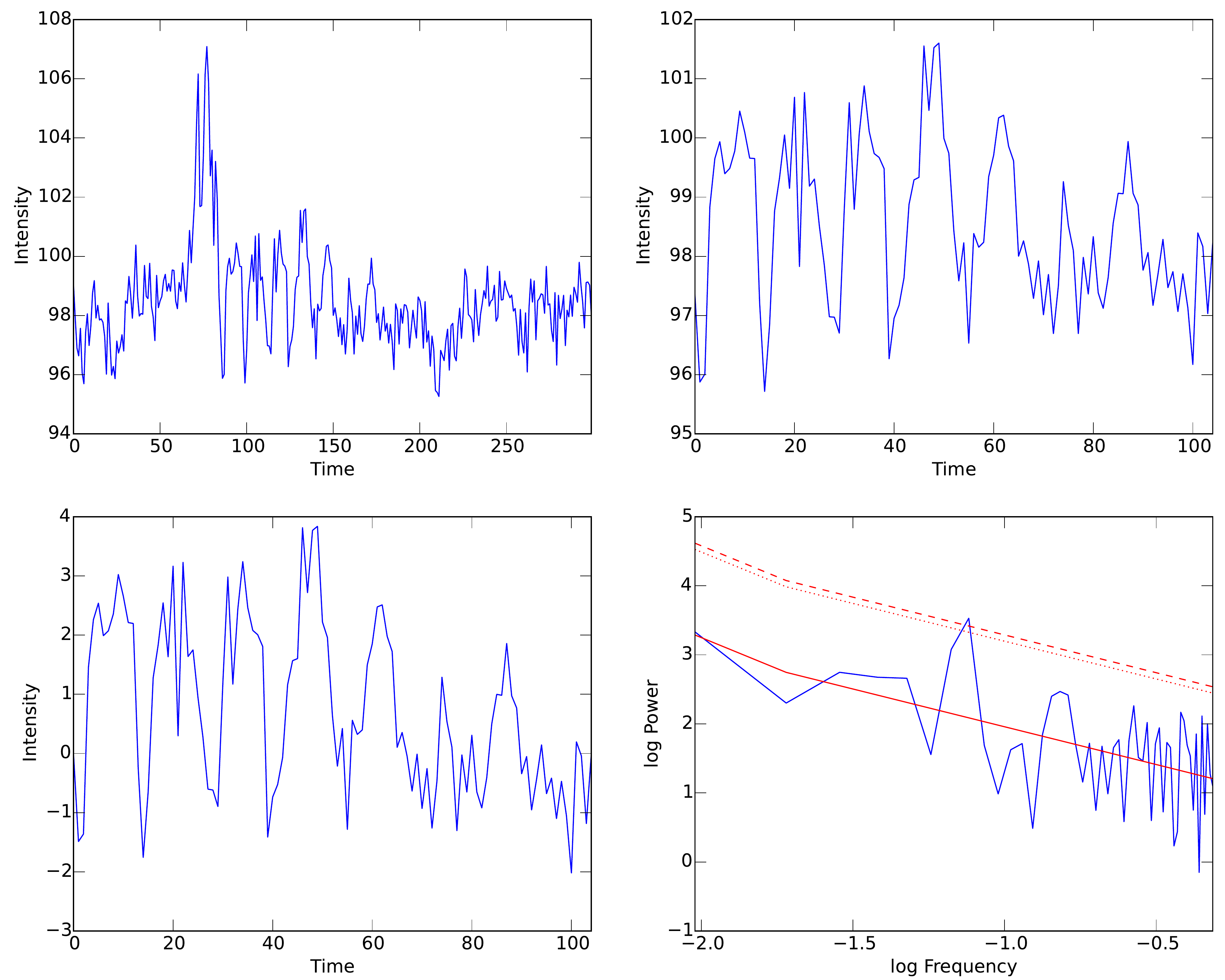}
\caption{Illustration of the steps involved for the analysis method described in Section~\ref{ssec:CEP} (CEP). Top left: The original simulated light curve for Flare 566801. Top right: The section of light curve that showed the best evidence of a QPP signal in the periodogram after manual trimming. Bottom left: The trimmed section of light curve after a linear interpolation between the first and last data points had been subtracted, to remove some of the background trend. Bottom right: The periodogram corresponding to the data shown in the bottom left panel. The solid red line shows the fitted broken power-law model, while the dotted and dashed red lines show the 95\% and 99\% confidence levels, respectively.}
\label{figure[cep]}
\end{figure*}

This significance testing method (CEP) is based on that described in detail in \citet{2017A&A...602A..47P}, with the main difference being that it does not account for data uncertainties since none exist for the synthetic data. To begin with, the simulated light curves were manually trimmed so that only the flare time profile was included. A linear interpolation between the start and end values was subtracted as a very basic form of detrending. The detrending performed for Flare 566801 can be seen by comparing the top right and bottom left panels of Figure \ref{figure[cep]}. Since the calculation of the periodogram assumes that the data are cyclic, subtracting this straight line removes the apparent discontinuity between the start and end values. This step will not alter the probability distribution of the noise in the periodogram, while it will act to suppress any steep trends in the time series data, which have been shown to reduce the S/N of a real periodic signal in the periodogram \citep{2017A&A...602A..47P}. Lomb-Scargle periodograms were then calculated for each of these flare time series with a linear trend subtracted. 

The presence of trends and colored noise in time series data results in a power-law dependence between the powers and the frequencies in the periodogram. Therefore, to account for this, a broken power-law model with the following form was fitted to the periodogram:
\begin{eqnarray}
	\log\left[\hat{\mathcal{P}}(f)\right] =
		\begin{cases}
			-\alpha\log\left[f\right] + c  \qquad \qquad \text{if } f < f_{break} \\
			-\left(\alpha - \beta\right)\log\left[f_{break}\right] - \beta\log\left[f\right] + c  \\
			 \qquad\qquad\qquad\qquad\qquad\text{if } f > f_{break}\,,
		\end{cases}
\end{eqnarray}
where $\hat{\mathcal{P}}(f)$ is the model power as a function of frequency, $f$; $f_{break}$ is the frequency at which the power-law break occurs; $\alpha$ and $\beta$ are power-law indices; and $c$ is a constant. The break in the power law accounts for the fact that there may be a combination of white and red noise in the data, and in some cases the amplitude of the red noise may fall below that of the white noise at high frequencies. An example of the power law model fitted to Flare 566801 can be seen in Figure \ref{figure[cep]}. The noise follows a chi-squared, 2 degrees of freedom (dof) distribution in the periodogram, and the noise is distributed around the broken power law \citep{2005A&A...431..391V}. For a pure chi-squared,2 dof distributed noise spectrum, the probability of having at least one value above a threshold, $x$, is given by
	\begin{equation}
		\text{Pr}\left\{X>x\right\} = \int_{x}^{\infty} \mathrm{e}^{-x'} \mathrm{d}x' = e^{-x}\,,
	\end{equation}
where $x'$ is a dummy variable representing power in the periodogram. For a given false-alarm probability, $\epsilon _{N}$, the above probability can be written as
	\begin{equation}
		\text{Pr}\left\{X>x\right\} \approx \epsilon _{N}/N\,,
	\end{equation}
where $N$ is the number of values in the spectrum \citep{2002MNRAS.336..979C}. Hence, a detection threshold can be defined by
	\begin{equation}
    	x = \ln\left(\frac{N}{\epsilon _N}\right)\,.
    \end{equation}
To account for the fact that the above expression is only valid when the power spectrum is correctly normalized (with a mean equal to one), and that the noise is distributed around the broken power law, the confidence level for the periodogram is found from $\log[\hat{\mathcal{P}}_j] + \log[x\langle \mathcal{I}_j / \hat{\mathcal{P}}_j\rangle]$, where $\mathcal{I}_j$ is the observed spectral power at frequency $f_j$. This confidence level gives an assessment of the likelihood that the periodogram could contain one or more peaks with a value above a particular threshold power purely by chance, if the original time series data were just noise with no periodic component. The confidence level used as the detection threshold for this study was the 95\% level, which corresponds to a false-alarm probability of 5\% (or, in other words, a 5\% chance that the periodogram could contain one or more peaks above that threshold as a result of the noise). In addition, only peaks corresponding to a period greater than four times the time cadence and less than half the duration of the trimmed time series were counted, as it is not clear that periodic signals with periods outside of this range can be detected reliably. Although the 95\% confidence level was used as the detection threshold for this analysis, many of the detected periodic signals had powers well above the 95\% level in the periodogram.

This method is sensitive to the choice of time interval used for the analysis (this will be discussed further in Section~\ref{ssec:trim}); hence, the start and end times of the section of light curve used for the analysis were manually refined where there appeared to be a periodic signal in the data but the corresponding peak in the periodogram was not quite at the 95\% level. This process is described in more detail in \citet{2017A&A...608A.101P}. Figure~\ref{figure[cep]} shows the trimmed time series for Flare 566801 and the power spectrum. This method identified a statistically significant peak at $14.0 \pm 0.5$ time units, which is in good agreement with the input period.

\subsection{Smoothing and periodogram -- TVD}\label{ssec:tvd}
\begin{figure*}
\centering
\includegraphics[width=0.9\linewidth, trim=4cm 8.5cm 4cm 7cm]{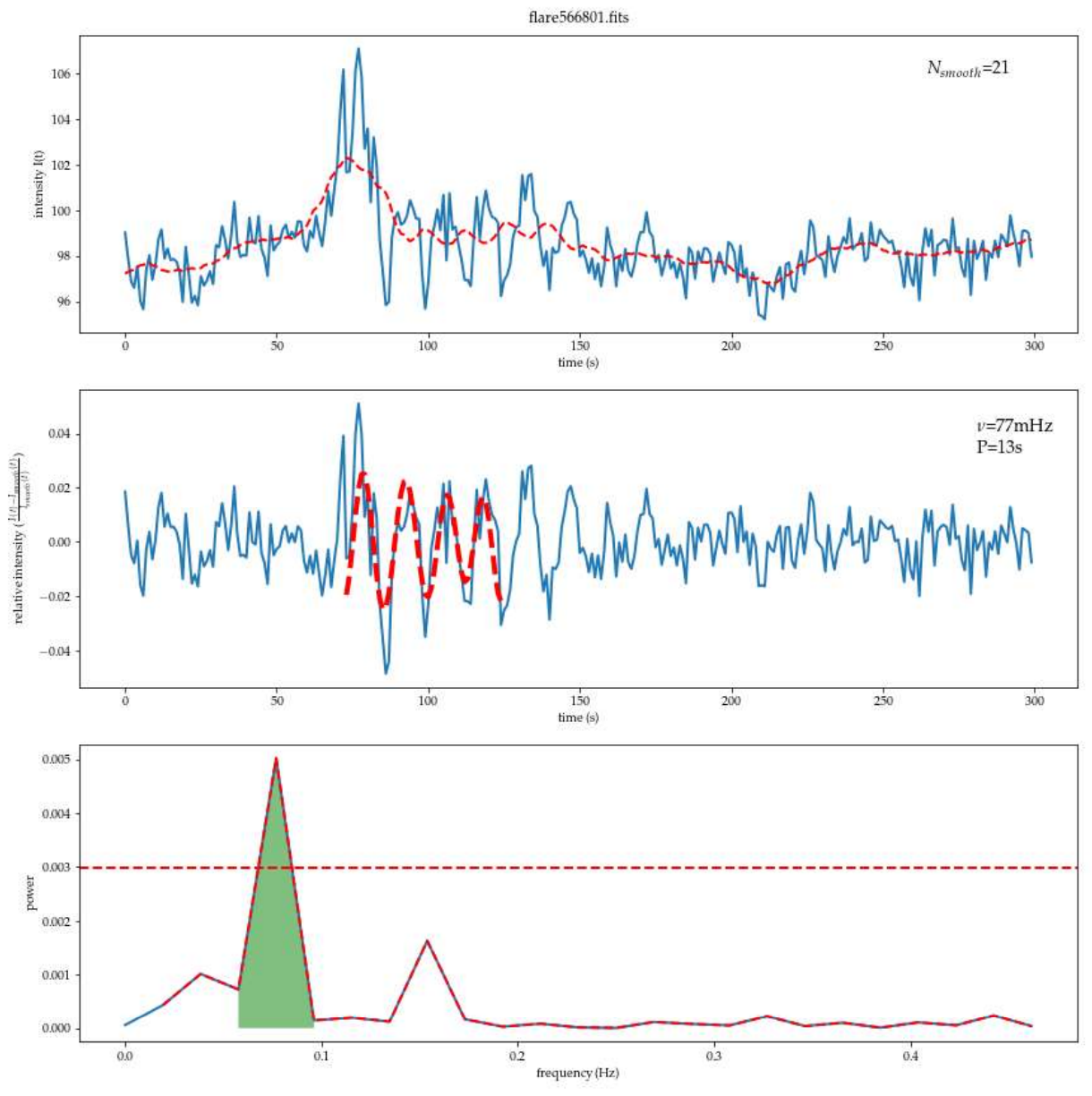}
\caption{Overview of the analysis method of TVD. The top panel shows the raw light curve $f(t)$ as a function of time with the blue solid line and overplotted is the smoothed light curve $f_\mathrm{smooth}(t)$ as a red, dashed line with $N_\mathrm{smooth}$ in the key. The middle panel shows the relative flux $\frac{I(t)-I_\mathrm{smooth}(t)}{I_\mathrm{smooth}(t)}$. The bottom panel shows the Lomb-Scargle periodogram of the signal in the middle panel, and the peak frequency and period are indicated in the key of the middle panel. The horizontal red dashed line in the bottom panel is the false-alarm level (95\% level). The area shaded in green in the bottom panel is used to reconstruct the QPP signal, which is then shown with the red dashed line in the middle panel, overplotted on the relative signal. The length of this reconstructed curve shows the time interval $[t_\mathrm{flare},t_\mathrm{flare}+2\tau]$. }
\label{figure[tvd]}
\end{figure*}

TVD largely followed the method described in \citet{2011ApJ...740...90V}. In the first instance, the flare light curve $f(t)$ was smoothed using a window of length $N_\textrm{smooth}$ (with the python function \code{uniform\_filter}, which is part of \code{SciPy}). An initial value for the smoothing parameter was chosen manually and later adjusted during the procedure. The smoothed light curve $I_\textrm{smooth}(t)$ was considered to be the flare light-curve variation without the QPPs and noise. The original signal and the smoothed signal are shown in the top panel of Figure~\ref{figure[tvd]}. The maximum of the smoothed light curve is reached at $t_\mathrm{flare}=\mathrm{argmax}_t(I_\mathrm{smooth}(t))$. We have fitted the smoothed light curve with an exponentially decaying function $a+b\exp{(-t/\tau)}$ in the interval $[t_\mathrm{flare},\,300]$. From this fit with the exponentially decaying function, we have selected the QPP detection interval to $[t_\mathrm{flare},t_\mathrm{flare}+2\tau]$. In that interval, we have computed the residual in the detection interval by subtracting and normalizing to the background and call this the QPP signal $I_\mathrm{QPP}(t)=I(t)-I_\textrm{smooth}(t)$, which is shown in the middle panel of Figure~\ref{figure[tvd]}. From this QPP light curve, we have constructed a Lomb-Scargle periodogram (see bottom panel of Figure~\ref{figure[tvd]}). In the periodogram, we have selected the frequency with the most power and have retained it as significant if its false-alarm probability was less than 5\%. In Figure \ref{figure[tvd]} it can be seen that a peak is visible above the 95\% false-alarm level at 13s, in good agreement with the input periodicity. The false-alarm probability was computed with the assumption that the QPP signal was compounded with white noise. After this procedure, the smoothing parameter $N_\textrm{smooth}$ was manually and iteratively adjusted. In the second iteration, the smoothing parameter was taken to be roughly corresponding to the detected period in the first iteration, and so on. This led to a rapid convergence, in which attention was paid to capture the impulse phase of the flare sufficiently well, in order not to introduce spurious oscillatory signal. 

Between HH1 and HH2 TVD automated his method. This involved systematically testing different smoothing windows, $N_\textrm{smooth}$, to remove the background trend: Smoothing windows of widths from 5 to 63 were tested where the smoothing width was increased by two in each iteration. For each detrended time series, a periodogram was found and the false-alarm probability and frequency of the largest peak recorded. The optimal smoothing window was deemed to be the one that produced a peak in the power spectrum with the lowest false-alarm probability. While automation makes the process less time-consuming for the user, there were some pitfalls, and these are discussed in Section \ref{ssec:smooth}. For some of the flares TVD flagged that the results looked untrustworthy. This was often where long smoothing windows were selected for detrending the flare, meaning that the underlying flare shape was not removed correctly, leading to spurious peaks in the resultant power spectrum that dominated over the real QPP signal. In other instances the obtained periodicity did not match the periodicity visible in the residual time series. Identifying these cases relied on TVD's data analysis experience. When discussing the results of HH2 (Section \ref{sec:false}), we consider both the raw results and those obtained when the results flagged as untrustworthy were removed. 
 
\section{Results of the Hare-and-hounds Exercises}\label{sec:results}

\subsection{HH2: False-Alarm rates}\label{sec:false}
The aim of the second hare-and-hounds exercise (HH2) was to allow the false-alarm rate of the various methods to be determined. Although analysis of the flares in HH2 was performed after the analysis of the HH1 flares, we present the results of HH2 first to establish how often various detection methods make false detections, before considering  how precise those detections are, using HH1. HH2, therefore, contained a roughly even split between flares containing no QPP signal (60), flares containing a single, sinusoidal QPP (32), and periodic multiple flares (8; see Tables \ref{table[simple_qpp]} and \ref{table[simulations]}). 

Table \ref{table[HH2]} gives the number of false detections returned by each method, which are defined as the number of detections claimed for simulated flares that did not contain a QPP. For HH2, LAH and ARI both used the AFINO method in exactly the same manner, and so the results are identical (this was not the case for HH1). The AFINO, wavelet (LAH), and periodogram method employed by CEP were all reliable, making low numbers of false detections. The periodogram method employed by TVD also produced a low number of false detections; however, this comes with a caveat: TVD detrended the data by removing a smoothed version of the time series before determining the periodogram, where the width of the smoothing window was determined on a flare-by-flare basis. In HH2, TVD automated the selection of the optimal width for the smoothing window. The raw results from this automated method are denoted TVD1 in Table \ref{table[HH2]}. However, for some of the flares this width was surprisingly long, leading TVD to question the results. These manually filtered results are denoted TVD2 in Table \ref{table[HH2]}, which indicates that the false-alarm rate was far higher before manual intervention was incorporated. The primary difference between the periodogram methods employed by JAM and TVD was in the detrending: both detrended by removing a smoothed component, but JAM used the same smoothing window for each flare, while TVD used a flare-specific smoothing window. The method employed by JAM produces a large number of false detections, which. combined with the previous discussion concerning the automation of TVD's code, suggests that detrending needs to be done with great care. The GP method employed here also produces a large number of false detections, suggesting that a better method for estimating the statistical significance of the results is required.

\begin{table*}[htp]\caption{Statistics of detections in HH2.}\label{table[HH2]}
\centering
\begin{tabular}{lcccccccccccc}
  \hline
 Hounds & \multicolumn{2}{c}{Claimed} & \multicolumn{2}{c}{Claimed} & \multicolumn{2}{c}{Total Number}  & \multicolumn{2}{c}{Precise} & \% of Precise & TSS & HSS & Precision\\
 & \multicolumn{2}{c}{Detections} & \multicolumn{2}{c}{Detections} & \multicolumn{2}{c}{of False}  & \multicolumn{2}{c}{Detections} & Claimed & & & \\
  & \multicolumn{2}{c}{(No QPP)} & \multicolumn{2}{c}{(QPP)} & \multicolumn{2}{c}{Detections}  & \multicolumn{2}{c}{} & Detections & & & \\
  & $N$ & \% & $N$ & \% & $N$ & \% & $N$ & \% & & & & \\
 \hline
 AFINO (LAH \& ARI) & 0 & 0 & 8 & 25 & 1 & 13 & 7 & 18 & 88 & 0.18 & 0.20 & 1.00\\
 Wavelet (LAH) & 1 & 2 & 13 & 33 & 2 & 14 & 12 & 30 & 94 & 0.28 & 0.32 & 0.92 \\
 Periodogram (CEP) & 2 & 3 & 12 & 30 & 2 & 14 & 12 & 30 & 100 & 0.27 & 0.30 & 0.86\\
 Periodogram (TVD1) & 18 & 30 & 28 & 70 & 33 & 73 & 13 & 33 & 46 & 0.03 & 0.03 & 0.42\\
 Periodogram (TVD2) & 3 & 5 & 13 & 33 & 5 & 31 & 11 & 28 & 85 & 0.23 & 0.25 & 0.79\\
 Periodogram (JAM) & 29 & 48 & 25 & 63 & 41 & 76 & 13 & 33 & 52 & -0.16 & -0.16 & 0.31\\
 GP (JRAD) & 23 & 38 & 29 & 73 & 43 & 83 & 9 & 23 & 31 & -0.16 & -0.16 & 0.28\\
 \hline
\end{tabular}
\tablecomments{$N$ denotes the number of flares detected for each category. The second column shows the number of detections claimed in flares where no QPP was present. The percentage is calculated using the total number of simulated flares not containing a QPP, i.e. 60. The third column shows the number of detections claimed for flares where a QPP was included, which includes flares that either contained a single sinusoidal QPP or a periodic multiple flare. The percentage is calculated using the total number of QPP flares in HH2, i.e. 40. Precise detections are defined as those claimed detections within 3 units of the input periodicity, with any claimed detection more than 3 units from the input periodicity being classified as ``imprecise''. The fourth column shows the total number of false detections i.e. the sum of the claimed detections where no QPP was present and the imprecise detections. The percentage is determined using the total number of claimed detections (i.e. the sum of columns two and three). The fifth column shows the total number of precise detections. For the precise detections the percentage is calculated using the total number of simulated QPP flares i.e. 40. The final column gives the percentage of claimed detections that are precise, calculated using columns five and two. TVD1 indicates the raw results from TVD's automated method. TVD2 indicates results when manual filters were employed. The final three columns show True Skill Statistic (TSS) and the Heidke Skill Score (HSS), and precision, as defined in Section \ref{sssec:skill}.}
\end{table*}

\begin{figure*}[htp]
  \centering   \includegraphics[width=0.9\textwidth]{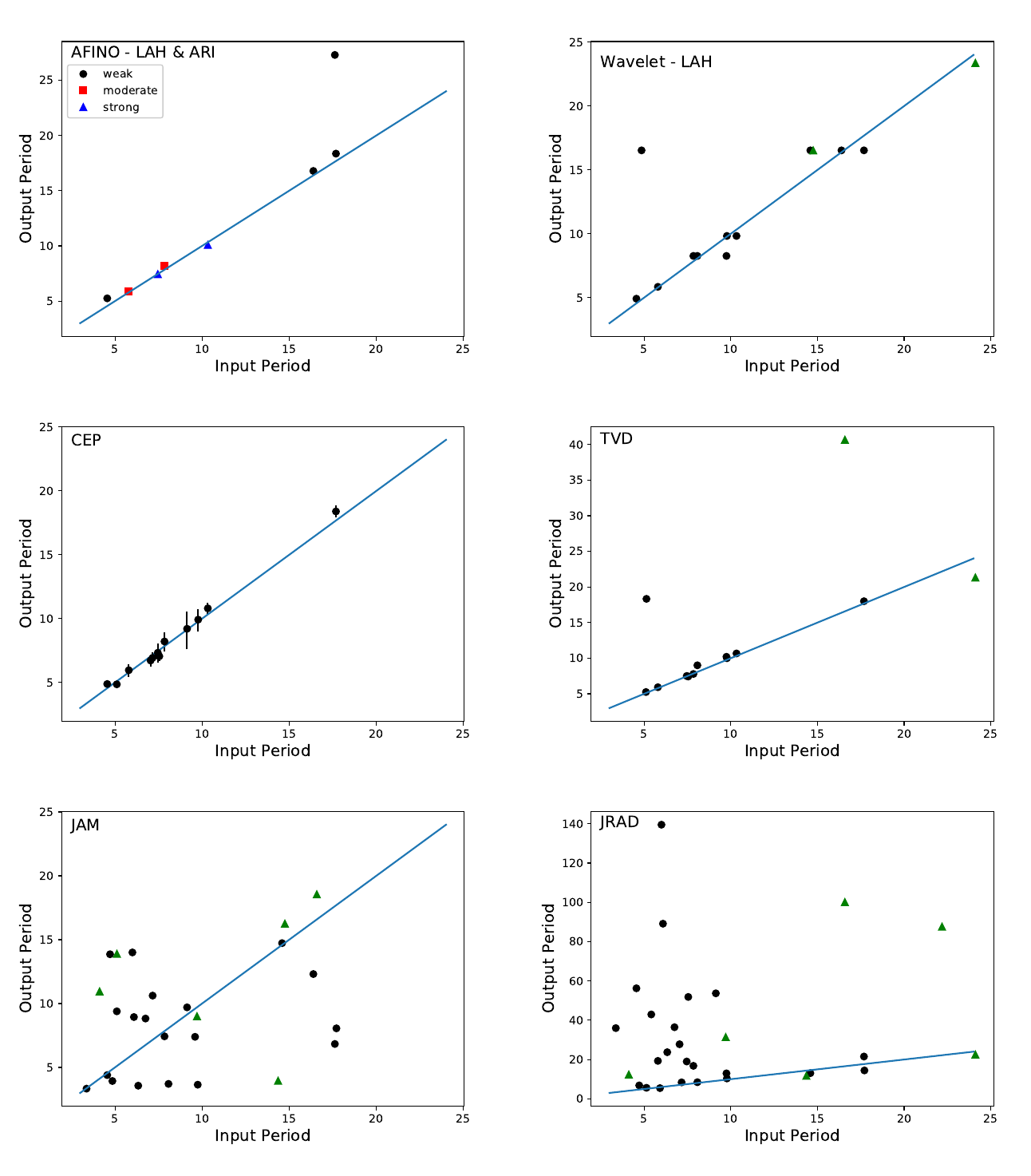}\\ 
   \caption{Results of HH2 analysis, where the output period from the various detection techniques are compared to the input period. In the top left panel a legend is included to describe the symbols, which refer to the strength of the AFINO detections (see Section \ref{ssec:ari}). In all other panels the black circles denote flares where a single sinusoidal QPP was included, and the green triangles indicate detections in simulations containing multiple periodic flares.}
   \label{figure[HH2_results]}
\end{figure*}

Table \ref{table[HH2]} shows that the four methods (AFINO, Wavelet, CEP, TVD2) that claimed low numbers of detections in flares where no QPPs were included all made relatively low numbers of detections $(<35\%)$; however, for all four methods those detections are precise, with at least 85\% of detections lying within three units of the input period. Table \ref{table[HH2]} also gives the total number of false detections (i.e. those in flares where no QPPs were present and imprecise detections). This sum constitutes a small percentage of the total number of claimed detections made by the AFINO, Wavelet, and CEP methods. In statistical hypothesis testing erroneous outcomes of statistical tests are often referred to as type I or type II errors. A type I error is said to occur if the null hypothesis, in this case that the data contain only noise, is wrongly rejected. In this article that would constitute claiming a detection of a QPP when no QPP was included in the simulated flare. Type II errors occur when the null hypothesis is wrongly accepted. Here that would mean failing to claim a detection when a QPP was present. Type I errors are generally regarded as far more serious than type II errors. In other words, it is far better to sacrifice a high detection rate (i.e. make type II errors) in favor of making false detections (type I errors), and so by adopting cautious approaches we can be confident in any detections these methodologies make. Conversely, the three methods that produced a higher number of false detections (TVD1, JAM, JRAD) also produced less precise detections: Although the methods claimed detections in over 60\% of flares containing QPPs, $\le52\%$ of those detections were within 3 units of the input period. In other words, approximately half of the detections claimed by these methods were imprecise and so can be considered as false alarms or type I errors. This is highlighted in Figure \ref{figure[HH2_results]}, which compares the periods obtained by the various methods with the input periods. 

The range of input periods for the single sinusoidal QPP simulated flares in HH2 was $3.3<P<17.8$. We can see from Figure \ref{figure[HH2_results]} that detections were made across the entire range of input periods. The apparent gap in detections between approximately $11<P<15$ occurs because there were few simulations included in that range. 

\begin{figure*}[htp]
  \centering   \includegraphics[width=0.45\textwidth]{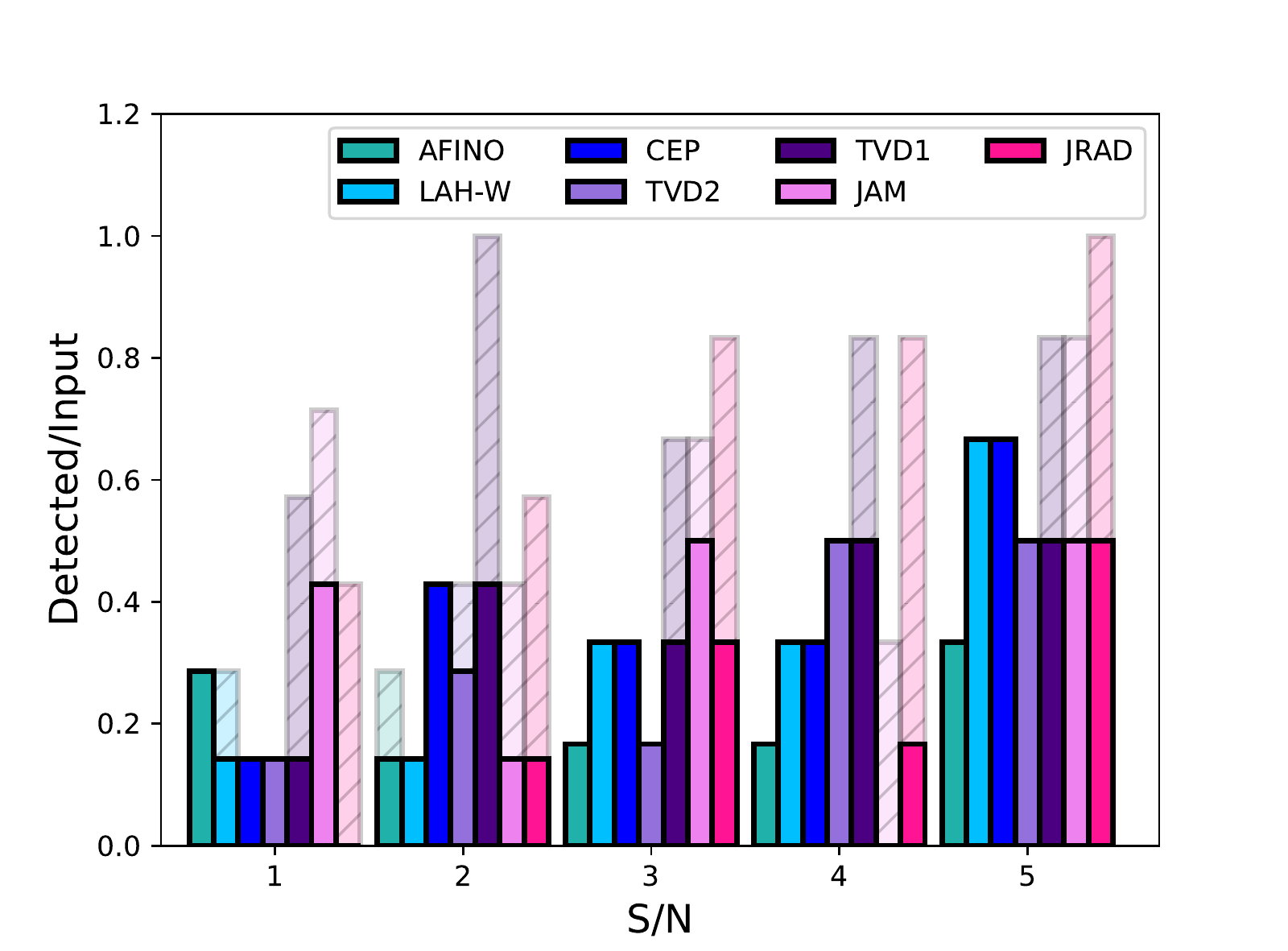}
  \centering   \includegraphics[width=0.45\textwidth]{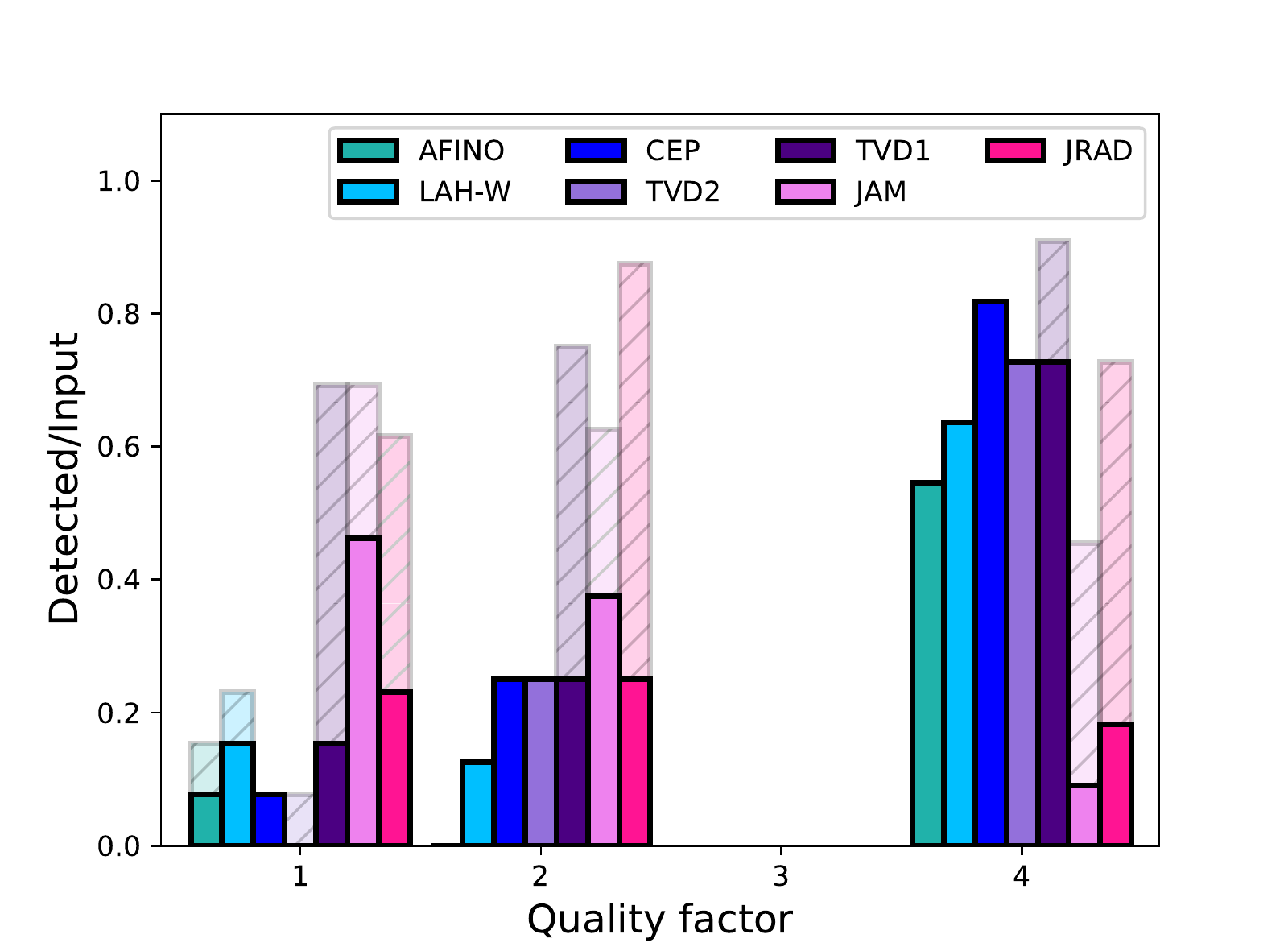}\\
   \caption{Left: histogram showing the distribution of S/N for the claimed detections of simple sinusoidal QPPs in HH2. Right: histogram showing the distribution of the QF for the claimed detections of simple sinusoidal QPPs in HH2. In both panels the number of claimed detections has been normalized by the total number of simulated QPPs with that S/N (or QF) included in HH2. The pale bars with hatching include all claims, whereas the darker bars with no hatching only include those claims within 3 units of the input QPP (i.e. the precise detections).}
   \label{figure[HH2_hist]}
\end{figure*}

The left panel of Figure \ref{figure[HH2_hist]} shows how the claimed detections were distributed in terms of QPP S/N. For the majority of methods, there is a weak dependence on QPP S/N; however, precise detections are made even for low-S/N QPPs. In particular, the AFINO method appears to work equally well at low and high S/N. On the other hand, the success of the wavelet technique employed by LAH appears to show a stronger dependence on S/N, with a systematic increase in the number of precise detections obtained with increasing S/N. 

The QF of a signal is defined as the ratio of the lifetime to period. The right panel of Figure \ref{figure[HH2_hist]} shows that the various techniques were far more successful at detecting QPPs with higher QFs than lower QFs. We note here that there were no QPPs with a quality factor of 3 in HH2. It is also interesting to note the large number of imprecise detections (as indicated by the pale, hashed bars) with low QFs made, in particular, by JAM and JRAD. However, low-QF QPPs also account for the individual imprecise detections made by AFINO, LAH's wavelet technique, and TVD's periodogram technique. However, we note, from the left panel of Figure \ref{figure[HH2_hist]} that these QPPs are also low S/N.  

\begin{figure}[htp]
  \centering   \includegraphics[width=0.45\textwidth]{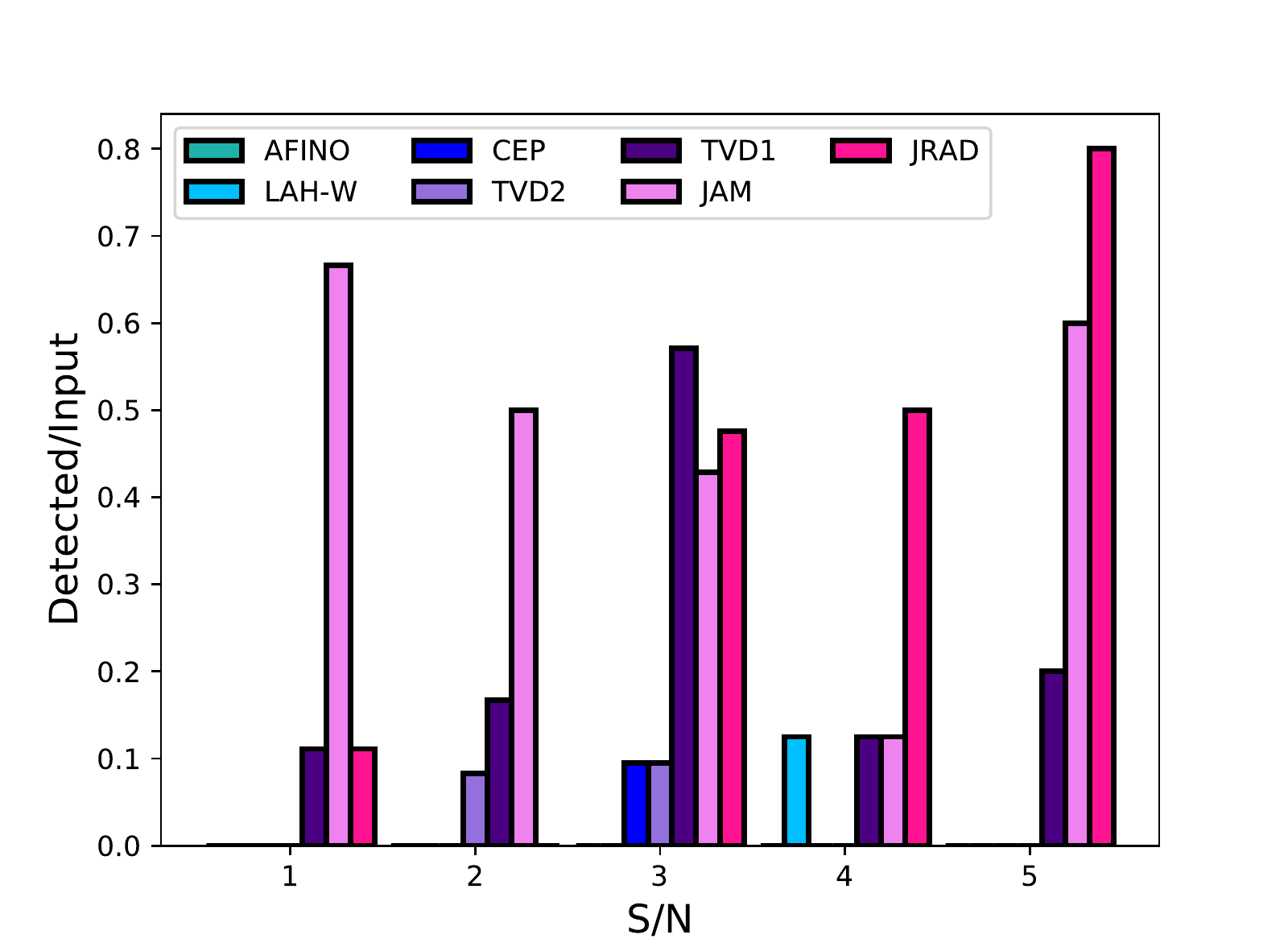}\\
   \caption{Histogram showing the distribution of S/N for the false detections. The number of claimed detections has been normalized by the total number of simulated flares with no QPPs and that S/N included in HH2.}
   \label{figure[HH2_hist_false]}
\end{figure}

Figure \ref{figure[HH2_hist_false]} shows how the false detections depend on S/N. Since these flares do not contain QPPs, the S/N refers to the flare itself. However, for those flares that do contain QPPs both the amplitude of the QPP and the noise are scaled relative to the amplitude of the flare itself, so the measurements are equivalent. As the numbers of false detections for AFINO, wavelet (LAH) and the periodogram methods of CEP and TVD2 are low, it is hard to make any conclusions from this. For TVD1 and JAM's methods there is no clear dependence on S/N, whereas the GP method of JRAD appears to produce more false detections at low S/N.

\subsubsection{Skill Scores}\label{sssec:skill}
As a final measure of the ability of the hounds to detect QPPs, we have also determined two skill scores and the ``precision.'' Skill scores \citep[see, e.g.][]{1976MWRv..104.1209W} provide a quantitative measure by which we can compare the performance of the hounds' methods. These statistics are commonly used in solar physics for assessing the effectiveness of flare forecasting methods \citep[e.g.][and references therein]{2008ApJ...688L.107B, 2012ApJ...747L..41B, 2015ApJ...798..135B, 2016ApJ...829...89B, 2019SoPh..294....6D}. In order to calculate the scores, the results first need to be sorted into four classes: true positive (TP), true negative (TN), false positive (FP) and false negative (FN). Here TP would include all precise detections of QPPs, TN would incorporate those flares correctly identified as not containing QPPs, FP would comprise of those flares that did not contain QPP but where detections were claimed, and FN would contain those flares that contained QPPs but where no detection was claimed. We would also contain imprecise detections in the FN category as although QPP detections were claimed, these did not correspond to the period of the input QPP. However, we note that in some cases the real QPP may have been detected but that the period of that QPP was not precisely estimated because of, for example, the limited resolution of the data or the impact of the red noise on the signal. However, this classification system means that in HH2 $\rm TP+FN=40$, the total number of flares in the sample containing QPP. Similarly, $\rm TN+FP=60$, i.e. the total number of flares that did not contain QPP. We combine these categories to give two skill scores, namely, the True Skill Statistic \citep[TSS;][]{Hanssen1965} and the Heidke Skill Score \citep[HSS;][]{Heidke1926}. The TSS is given by
\begin{equation}\label{eqn:TSS}
    	\textrm{TSS}=\frac{\textrm{TP}}{\textrm{TP+FN}}-\frac{\textrm{FP}}{\textrm{FP+TN}}.
\end{equation}
The TSS is sometimes favored over the HSS because it is not sensitive to variations in $\rm (TP+FN)/(TN+FP)$. However, since in HH2 each hound considered the same sample, that is not an issue here. The HSS compares the observed number of detections to those obtained by random. HSS is given by
\begin{equation}\label{eqn:HSS}
    	\textrm{HSS}=\frac{2(\textrm{TP}\times\textrm{TN}-\textrm{FN}\times\textrm{FP})}{\textrm{(TP+FN)(FN+TN)+(TP+FP)(FP+TN)}}.
\end{equation}
Values of both skill scores, which produce similar results, are given in Table \ref{table[HH2]} for each hound participating in HH2. The negative scores given to JAM and JRAD can be interpreted as showing that these methods perform worse than if the flares containing QPP were selected randomly. However, AFINO, LAH-wavelet and CEP all produce positive scores, while the improvement in the methodology between TVD1 and TVD2 is clearly highlighted. We note that while these values may be considered low, the skill scores do not differentiate between type I and type II errors, and, as already mentioned, the above methods prefer to take a cautious approach in an effort to minimize type I errors (FPs), even if that means making more type II errors (FNs). We therefore also quote the precision, which is given by
\begin{equation}\label{eqn:precision}
    	\textrm{Precision}=\frac{TP}{\textrm{(TP+FP)}}.
\end{equation}
As can be seen in Table \ref{table[HH2]}, AFINO and LAH-wavelet show very high precision, with CEP and TVD2 not far behind. The other methods show low precision.

\subsection{HH1: The quality of detections} \label{sec:quality}

\begin{figure*}[htp]
  \centering   \includegraphics[width=0.9\textwidth]{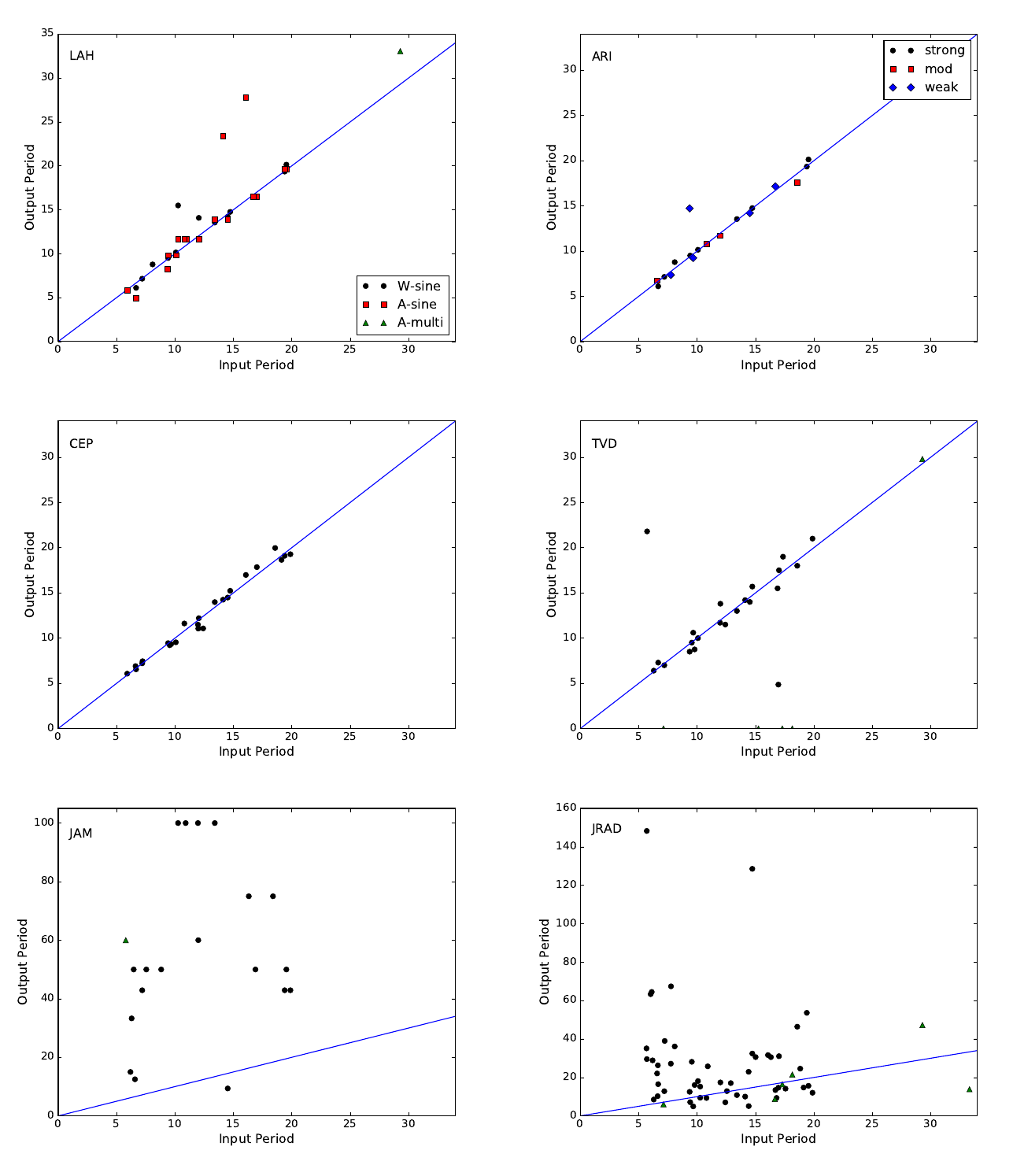}\\
   \caption{Results of HH1 analysis, where the output period from the various detection techniques are compared to the input period. In the top two panels legends are included to describe the symbols. For the top left panel ``W'' indicates that the wavelet technique was used, ``A'' indicates that the AFINO method was used, ``sine'' indicates that simulated flares where some form of sinusoidal QPP was included and ``multi'' indicates that a periodic multiple flare was detected. In the top, right panel the symbols indicate the strength of the confidence in the detection (see Section \ref{ssec:ari} for details). In all other panels the black circles denote flares where a sinusoidal QPP was included, and the green triangles indicate detections in simulations containing multiple periodic flares.}
   \label{figure[HH1_results]}
\end{figure*}

In HH1 72 (out of 101) of the input simulated flares contained some form of simulated QPP and over 21 (out of 101) were real flares, leaving only 7 flares with no form of QPP signal, making it difficult to assess the false-alarm rate in HH1. We therefore concentrate on the quality of those detections made. Table \ref{table[hh1_detections]} in Appendix \ref{sec:appendix} gives a breakdown of the types of QPPs that were detected by each method. Figure \ref{figure[HH1_results]} and Table \ref{table[HH1]} demonstrate that, for five detection methods (AFINO applied by LAH and ARI, wavelet approach employed by LAH, and the periodogram methods of CEP and TVD), when a detection is claimed, it tends to be robust, with over 80\% of claimed periodicities being within 3 units of the input periodicity. However, the other two methods (the combined detrending and periodogram method used by JAM and the Gaussian processing with a least-squares minimization utilized by JRAD) are far less reliable. 

\begin{table}\caption{Statistics concerning the quality of detections made in HH1.}\label{table[HH1]}
\centering\hskip-2cm
\resizebox{0.55\textwidth}{!}{
\begin{tabular}{lccccc}
  \hline
 & \multicolumn{2}{c}{Claimed} & \multicolumn{2}{c}{Precise } & \% of Precise  \\
 Hounds & \multicolumn{2}{c}{Detections} & \multicolumn{2}{c}{Detections} & Claimed  \\
 \cline{2-3}\cline{4-5}
  & Number & \% & Number & \% & Detections \\
 \hline
 AFINO (ARI) &  18 & 25 & 17 & 24 & 94 \\
 AFINO (LAH) &  18 & 25 & 15 & 21 & 83 \\
 Wavelet (LAH) &  12 & 17 & 11 & 15 & 92 \\
 Periodogram (CEP) &  24 & 33 & 24 & 33 & 100 \\
 Periodogram& 23 & 61 & 21 & 55 & 91 \\
  (TVD)\tablenotemark{*}  & & & & & \\
 Periodogram (JAM) & 20 & 28 & 0 & 0 & 0 \\
 GP (JRAD) & 56 & 78 & 9 & 13 & 16 \\
 \hline
\end{tabular}}
\tablecomments{\footnotesize{The claimed detections includes all detections reported for flares that contained some form of QPP signal, and the percentage is calculated using the total number of QPP flares in HH1, i.e. 72. Precise detections are defined as those claimed detections within three units of the input period. Here the percentage is again calculated using a total number of simulated QPP flares in HH1 (i.e. 72). The final column gives the percentage of claimed detections that are precise.}
\tablenotetext{*}{\footnotesize{TVD only analyzed 58 of the flares, and so the percentage of claimed and precise detections is calculated using the total number of QPP flares in this sample, which is 33.}}}
\end{table}

Table \ref{table[HH1]} also shows that the percentage of flares in which detections were claimed is fairly low for four of the five reliable methods (both AFINO methods, LAH's wavelet method, and CEP's periodogram method). This is an example of good practice: it is better to miss detections (type II errors or FNs) than to wrongly claim detections (type I errors or FPs). These methods all adopt this strategy: making a number of type II errors rather than risking type I errors. 

For the AFINO method all of the moderate and strong detections are precise, while all but one of the weak detections is precise. The same was true for HH2 (see Figure \ref{figure[HH2_results]} and Table \ref{table[HH2]}). In theory the moderate and strong detections correspond to those above a 95\% confidence level (see Section \ref{ssec:ari}). However, the high precision achieved at the expense of very few type I errors, even for the weak detections, suggests that this may, in fact, be an underestimate of the confidence level. It is possible that alternative measures of the quality of a model, such as the Akaike information criterion, which has a less stringent penalty for increasing the number of free parameters, may produce fewer type II errors, without increasing the risk of type I errors. However, determining this would require further testing beyond the scope of this paper.

In HH1 TVD's method was not automated, and so this method was only able to analyze 58 of the flares. However, this method did produce a high percentage of precise detections, with over 90\% of detected periodicities lying within 3 units of the input periodicity. We also note that the methodology claimed a far higher proportion of detections than the other four reliable methods, discussed in the above paragraph (see Table \ref{table[HH1]}). This, combined with the reliability of any detections made, is important, as TVD's method relies on detrending, and thus these results show that if detrending is performed in the nonautomated manner described in Section \ref{ssec:tvd} robust and reliable results can still be obtained.

\begin{figure*}[htp]
  \centering   \includegraphics[width=0.45\textwidth]{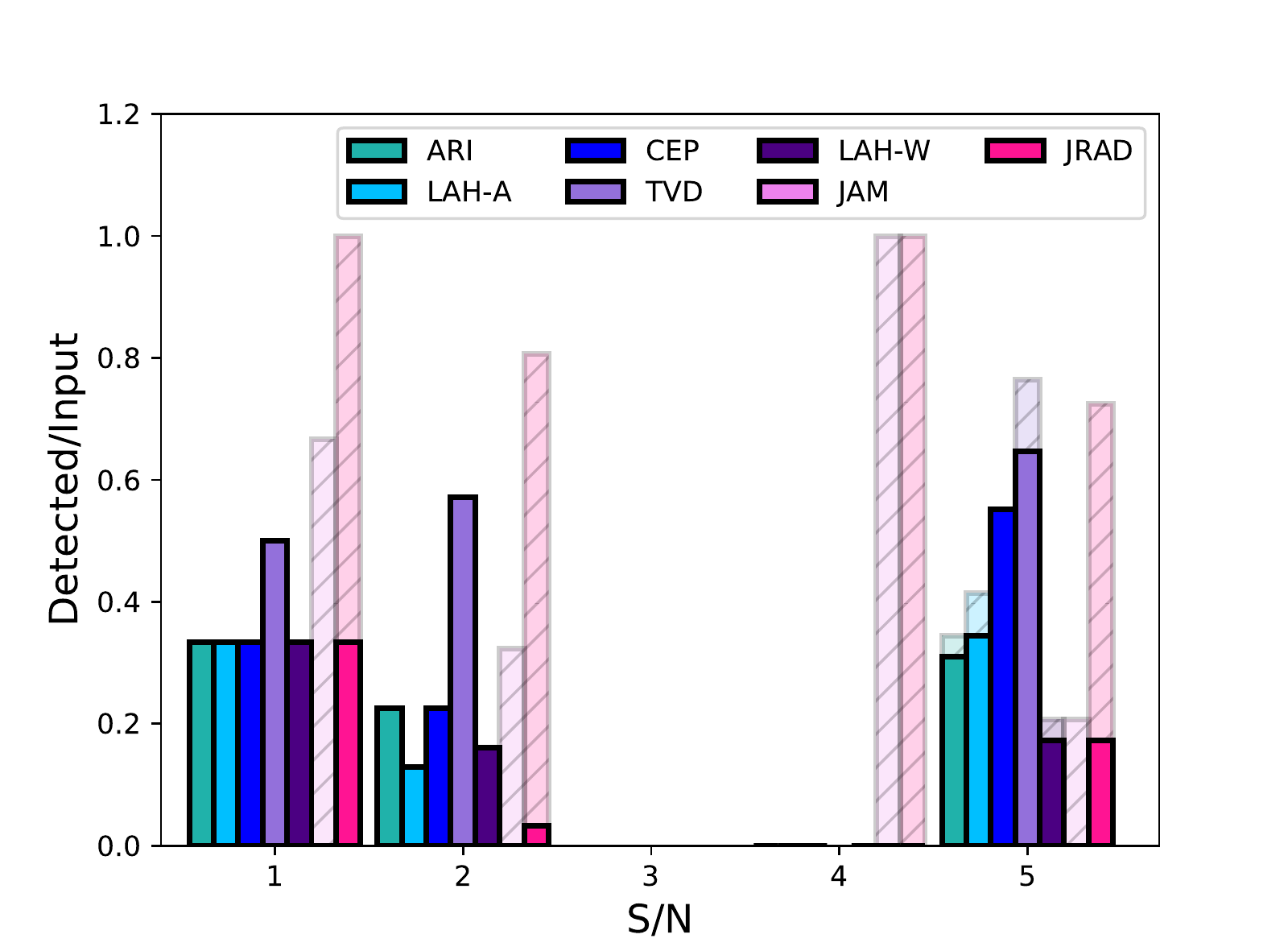}
  \centering   \includegraphics[width=0.45\textwidth]{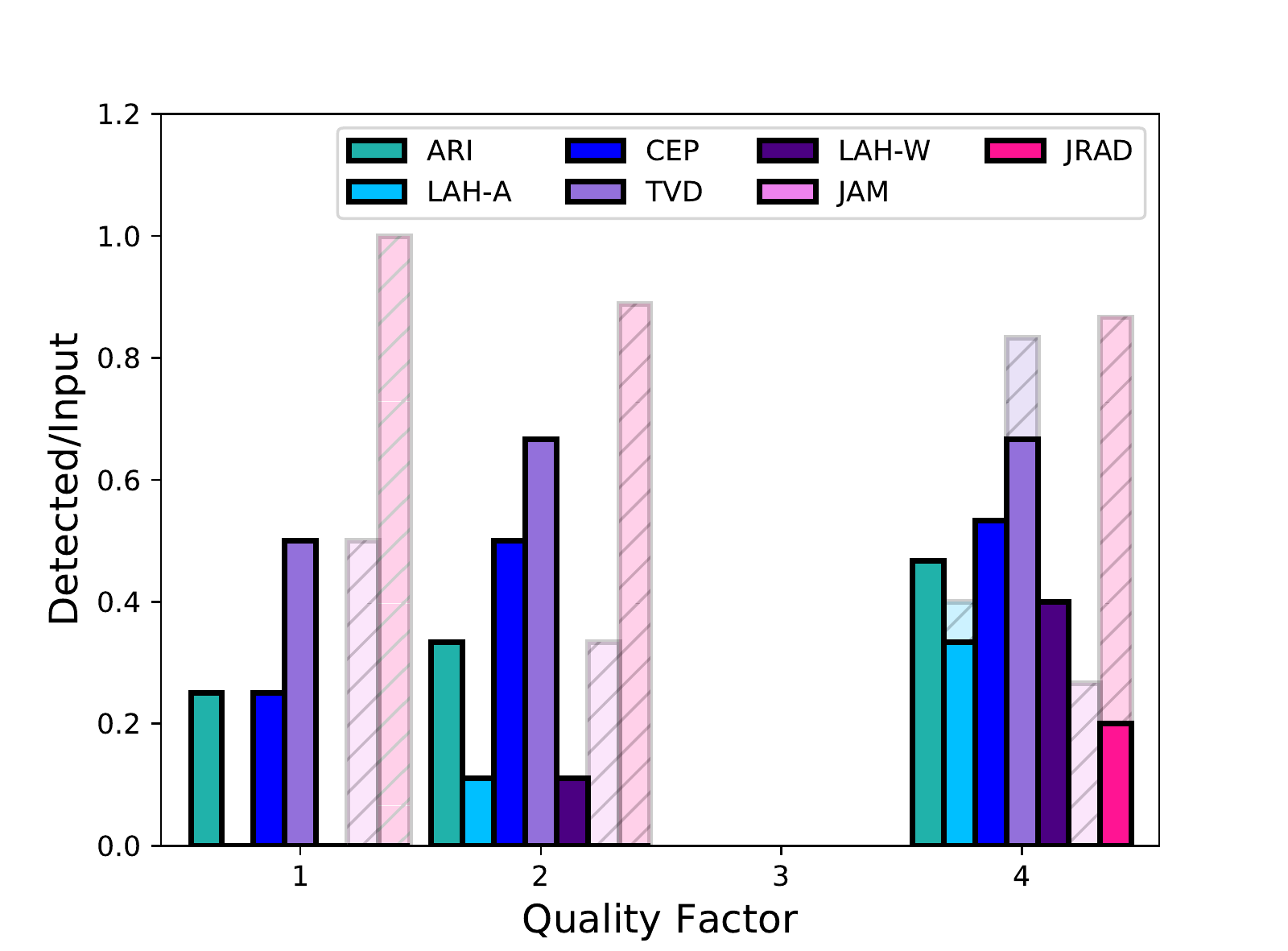}\\
   \caption{Left: histogram showing the distribution of S/N for the claimed detections of simple sinusoidal QPPs in HH1. Right: histogram showing the distribution of the QF for the claimed detections of sinusoidal QPPs in HH1. In both panels the number of claimed detections has been normalized by the total number of simulated QPPs with that S/N included in HH1. The pale bars with hatching are include all claims, whereas the darker bars with no hatching only include those claims within three units of the input QPPs (i.e. the precise detections).}
   \label{figure[HH1_hist]}
\end{figure*}

Figure \ref{figure[HH1_hist]} shows histograms of the S/N and QF for the detections made for the different methods in HH1. Here we only considered simulated flares in which some form of sinusoidal QPP was included but note that this covers all forms (including two sinusoidal QPPs, nonstationary QPPs, and those with varying backgrounds). As with HH1, there is little dependence on S/N, with precise detections being made at both low and high S/N. In contrast to HH1, the dependence on QF is less obvious. 

\subsubsection{Comparison of AFINO methods}
\begin{figure}[htp]
  \centering   \includegraphics[width=0.45\textwidth]{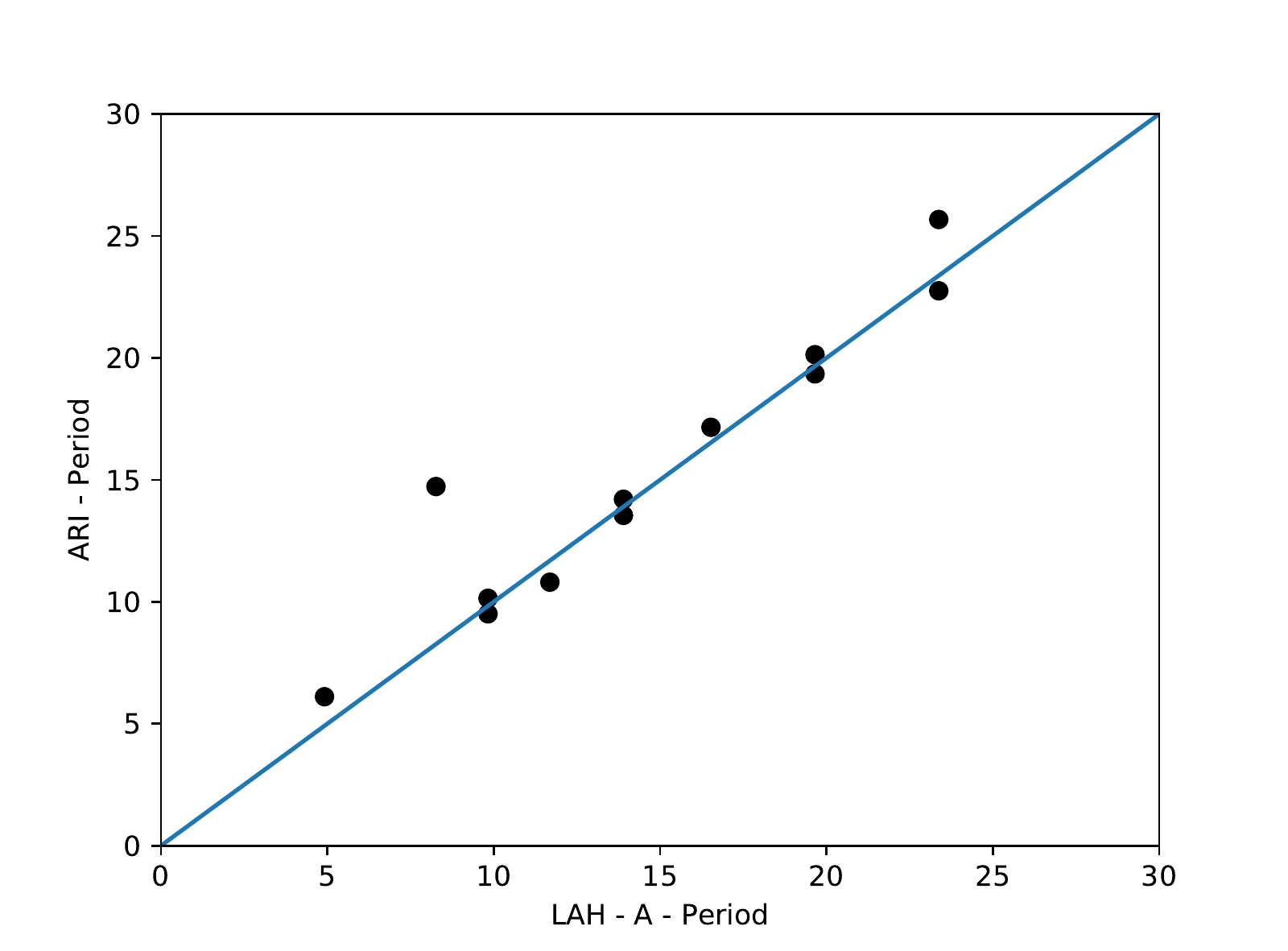}\\
   \caption{Comparison of periods claimed by the AFINO methods used by LAH and ARI.}
   \label{figure[AFINO_comp]}
\end{figure}

Both LAH and ARI used AFINO to detect QPPs in HH1, with LAH using a ``relaxed'' version. Figure \ref{figure[AFINO_comp]} shows that 12 detections were made by both methods and the periods claimed are in good agreement. In addition, 14 detections were claimed by LAH but not by ARI, including two false detections and two imprecise detections (see Figure \ref{figure[HH1_results]} and Table \ref{table[HH1]}), while nine detections were claimed by ARI but not by LAH (all flares containing simulated QPPs and all precise claims). Overall these results indicate that, as one would expect, the full AFINO method is more robust and reliable and hence should be used where possible.

\subsubsection{Empirical Mode Decomposition results}\label{sssec:EMD_HH1}
We consider the EMD results separately, as this method was only applied to 26 flares because of the time intensive nature of the methodology (see Section \ref{ssec:EMD} for details). The flares analyzed were selected from HH1 to be the most promising candidates following a by-eye examination.

DK and TM also took a different approach to many of the other hounds by splitting the simulated time series into three sections: pre-flare, flare, and post-flare. Unknown to the hounds, when simulating the flares the hare only included QPPs that occurred immediately after the peak. This is somewhat restrictive: while in real flares QPPs are predominantly detected during the impulsive phase of the flare, QPPs have previously been detected during the pre- \citep{2016ApJ...833..206T} and post-flare phases. Since the number of variables involved in simulating the flares and QPPs was already relatively large, the timing of the start of the QPPs was not varied with respect to the flare itself, but this could be the focus of a future study. In terms of this study, however, it means that any detections in the pre-flare phases can be considered false. ``EMD'' claimed QPP detections in the pre-flare phase of nine flares, and ``EMD-Fourier'' claimed detections in 10 flares (see Table \ref{table[EMD]}). These false detections are most likely observed as a result of the red noise that was added to the simulated data.

It is possible that, for high-QF simulated QPPs, the signal extends into the post-flare phase, meaning that any detections in this phase may be real. However, we note that only 4 out of 11 post-flare ``EMD'' detections and 3 out 11 post-flare ``Fourier'' detections would be considered as precise. This implies that both EMD-based techniques are making false detections in the post-flare phase as well.

Indeed, in all flare phases EMD found IMFs to be significant above a 95\% confidence level that transpired to be artifacts of colored noise. However, we note that a high proportion of the false pre- and post-flare periodicities were relatively long in comparison to the length of the data. Therefore, incorporating a caveat to ensure that, for example, at least one full cycle of a period is included in the data would substantially reduce the number of type I errors. 

In addition, many flares were analyzed with an inappropriate choice of shift factor, leading to poor trends and extracted IMFs where the expected periodicities were obscured. As discussed in Section~\ref{ssec:EMD}, the output is extremely sensitive to the choice of shift factor. However, given sufficient experience with the technique and a good grasp of the physical characteristics expected from your fittings, choosing a suitable shift factor becomes considerably more straightforward. It is the responsibility of the user to gain enough experience to be confident in their results, potentially through practice with simulated data, such as those utlized here. At the time of HH1, sufficient care was not taken over the choices of shift factor, which likely contributed to the poorer fit between the input and output periods. 

Another area where user experience is vital is in the selection of modes that are incorporated in the background trend. We remind the reader that, detrending was carried out through manual selection of the longest-period mode(s) and it is left to the user to incorporate as many modes as deemed reasonable as part of the trend. Whilst this was usually restricted to the highest one or two modes, this still remains a subjective process and raises the question of the reliability of detrending. 

\begin{table}\caption{Statistics of Detections by EMD Technique.}\label{table[EMD]}
\centering
\begin{tabular}{ccccc}
  \hline
  & \multicolumn{2}{c}{EMD} & \multicolumn{2}{c}{EMD-Fourier}  \\
  & Flares & Periodicities & Flares & Periodicities \\
 \hline
 Pre-flare & 8 & 9 & 10 & 10 \\
 Flare & 15 & 17 & 24 & 30 \\
 Precise Flare & 13 & & 18 & \\
 Post-flare & 11 & 11 & 9 & 12 \\
 \hline
\end{tabular}
\tablecomments{The ``Precise Flare'' detections are those detections made during the ``Flare'' phase that are within three units of the input QPP period.}
\end{table}

\begin{figure*}[htp]
  \centering   \includegraphics[width=0.45\textwidth]{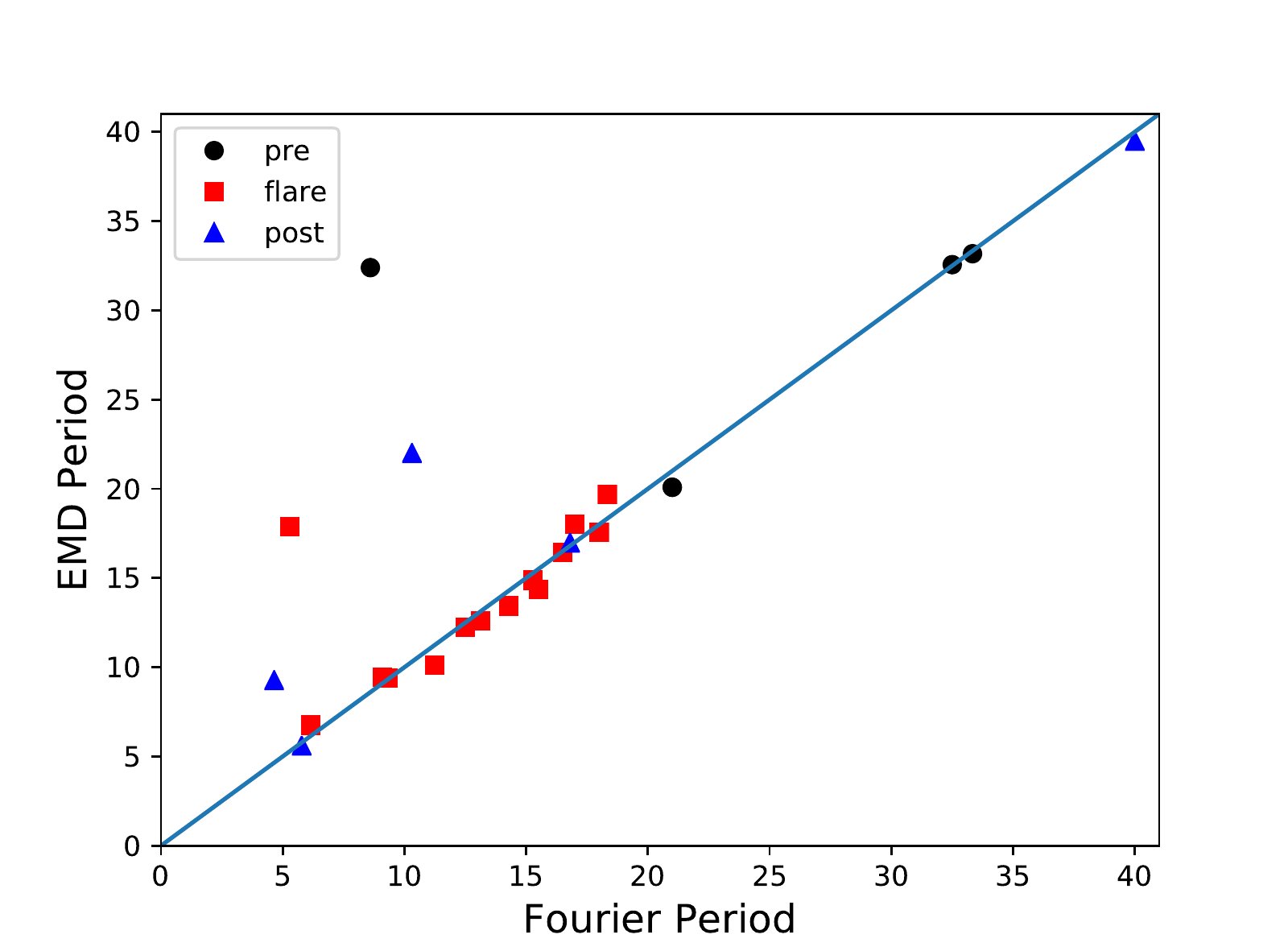}
  \centering   \includegraphics[width=0.45\textwidth]{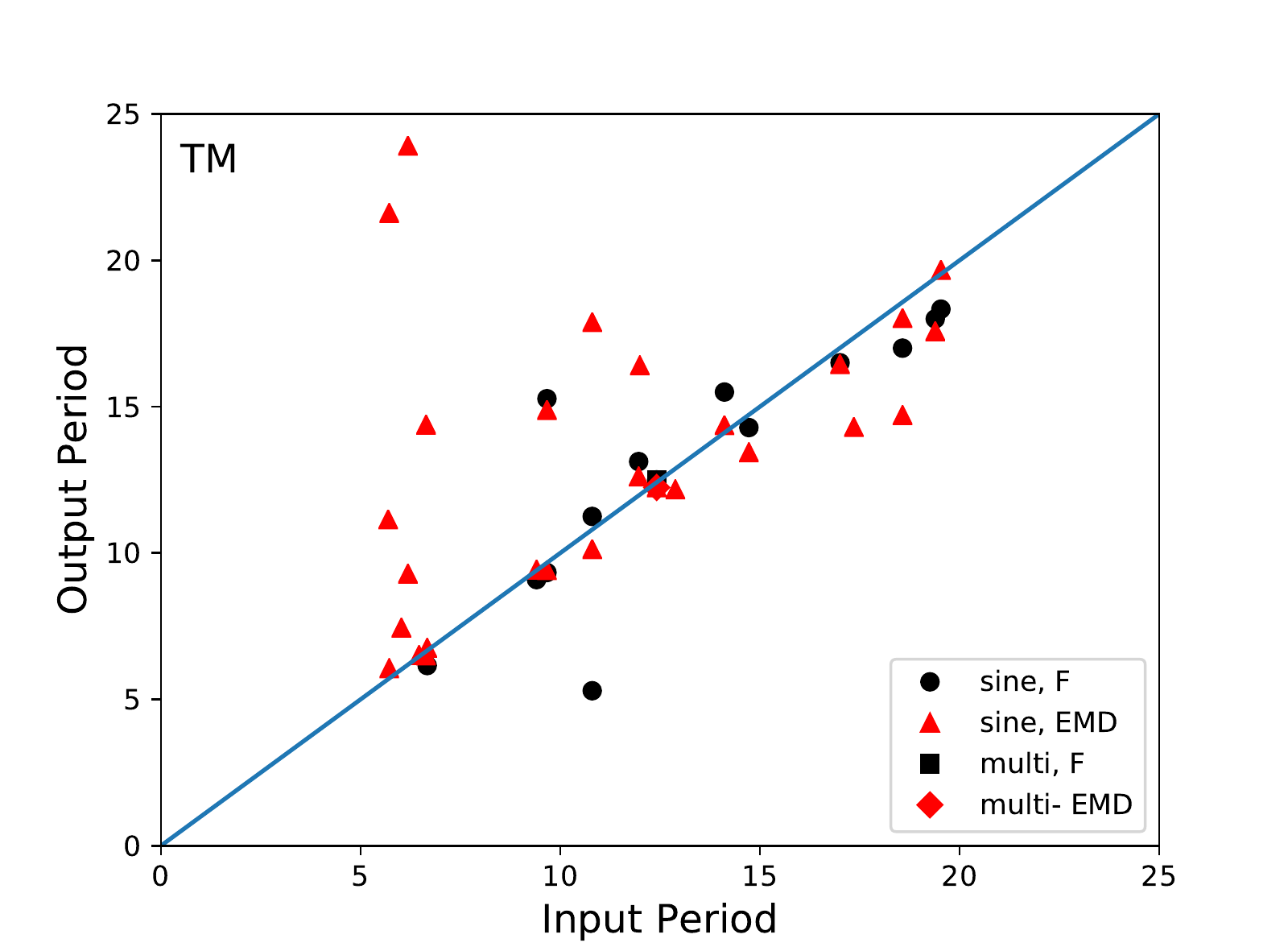}\\
   \caption{Left: comparison of the periodicities detected by the two methods incorporating EMDs. Right: comparison of claimed EMD detections made in the ``flare'' section with the input QPP periods.}
   \label{figure[HH1_EMD]}
\end{figure*}

As described in Section \ref{ssec:EMD} DK and TM used two methods for determining the significance of the detections. Table \ref{table[EMD]} shows that the two methods claimed different numbers of detections. While there was some overlap in the set of flares in which detections were claimed, in some cases detections were claimed by the Fourier method alone, and in other cases detections were claimed by the EMD method alone. The left panel of Figure \ref{figure[HH1_EMD]} compares those flares where detections were claimed in the same phase by both methods. For the majority of cases the two methods produce consistent periods, but not in all cases, including one from the ``flare'' section. Interestingly, both methods produce consistent false detections in the ``pre-'' and ``post''-flare phases, indicating that, when using EMD to detrend the data, insisting that detections are made by both methods is not a definitive way of ruling out type I errors.

The right panel of Figure \ref{figure[HH1_EMD]} compares the periods of claimed detections made in the flare section with the QPP periods input into the simulations. We note that although in some cases multiple detections were claimed, the method was not able to correctly pick out both periodicities in the two simulated flares examined that contained two sinusoidal QPPs. When the EMD threshold method was used to identify significant periodicities in the flare itself, 87\% of claimed detections were precise, which is slightly lower than, but still comparable to, the other robust detection methods (see Table \ref{table[hh1_detections]} in Appendix \ref{sec:appendix}). This suggests that if a periodicity is present in the data, the EMD technique is a good method of finding it. However, when the Fourier spectrum was used to identify significant periodicities, only 75\% of the claimed detections were precise, suggesting that this method is not as robust in the search for stationary oscillatory patterns in the signal. 

\subsubsection{Real Flares}\label{sssec:real}
In total, 21 of the simulated flares in HH1 were based on real data. As described in Section \ref{ssec:real}, in some cases the original data were included, but in others additional white noise was added. Although the majority of the claimed detections were in the original time series, there were some claimed detections in time series where additional noise was added. Detections were claimed for both solar and stellar flares, and there is no clear evidence to suggest that the QPPs were more likely to be detected in solar flares than stellar flares or vice versa (see Table \ref{table[hh1_detections]} in Appendix \ref{sec:appendix}).

\begin{figure}[htp]
  \centering   \includegraphics[width=0.45\textwidth]{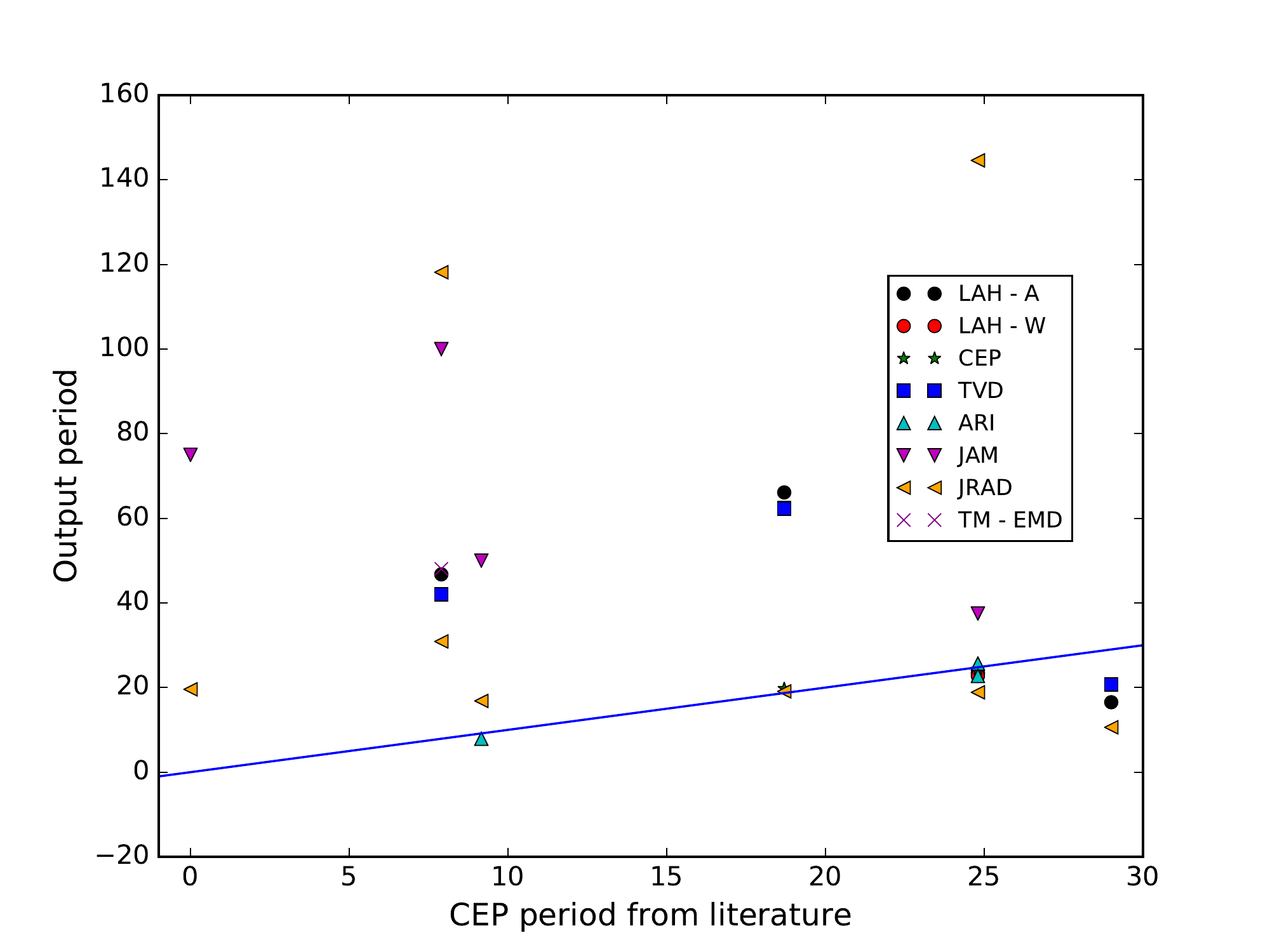}\\
   \caption{Comparison of the periods of claimed detections in real flare data with those found in the literature, nominally by \citet{2016MNRAS.459.3659P} for the stellar flares and \citet{2017A&A...608A.101P} for the solar flares.}
   \label{figure[HH1_real]}
\end{figure}

Figure \ref{figure[HH1_real]} compares the claimed periodicities obtained by the hounds (including CEP) with those found by CEP in \citet{2016MNRAS.459.3659P} for the stellar flares and in \citet{2017A&A...608A.101P} for the solar flares. One of the stellar flares included in HH1 was not found to have any periodicities by \citet{2016MNRAS.459.3659P} and so has been assigned a periodicity of zero in Figure \ref{figure[HH1_real]}. Since these flares are based on real data, there is no way to independently know whether a QPP signal is in fact present, or whether the results presented in this paper or the previously published literature are correct. However, it is notable that the majority of detections presented here lie far from the 1:1 line, indicating a mismatch with the prior literature for these events. In HH1, CEP claimed two detections of real flares, both of which were based on solar data, and both claimed that periodicities were consistent with the original detections. Interestingly, both of these cases had additional noise added to the flare. However, there were three HH1 flares containing solar data that did not have additional noise added to them, and CEP claimed no detections in these flares. This is likely to be because of differences in the choice of how to trim the flare, highlighting the important role trimming makes in the detection of QPPs by this method (see Section \ref{ssec:trim} for further discussion on this). CEP claimed no detections for the stellar flares in HH1; however, we note that in \citet{2016MNRAS.459.3659P} CEP employed a different methodology to detect the flares that involved detrending and wavelet techniques. The full AFINO method, employed by ARI in HH1, and LAH's wavelet technique produce results that are all consistent with those found in the above-mentioned literature. While some of the other techniques do produce some claims that are consistent with the literature results, they also claim some disparate periodicities. However, we note here again that this does not mean that the detections are incorrect. The majority of disparate detections in Figure \ref{figure[HH1_real]} lie well above the 1:1 line, indicating that the hounds are detecting longer periods than CEP. This could be a result of the methodology employed by CEP, which, by trimming. may focus on short-lived, small-period QPPs.

\subsection{Impact of trimming}\label{ssec:trim}
Since the exact shape of a periodogram is known to depend on the choice of interval for the time series data used to calculate the periodogram, in this section we show how this choice of time interval can affect the number of detections of periodic signals.

\begin{figure*}[htp]
  \centering   \includegraphics[width=0.9\textwidth]{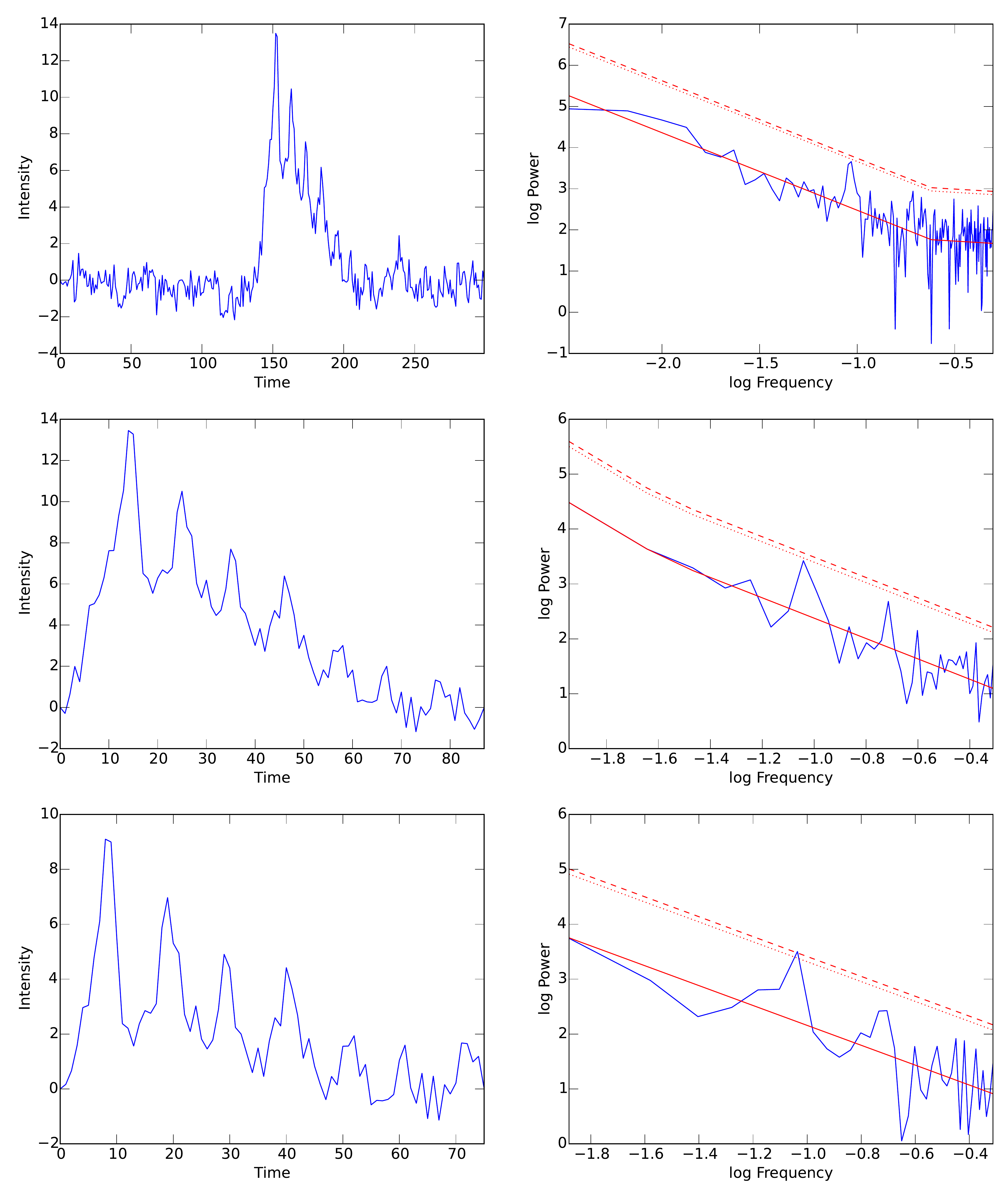}\\
   \caption{Demonstration of how the choice of time interval impacts the periodogram for Flare 629040. As a consequence, the significance level of the peak corresponding to the QPP signal, as determined by the method described in Section~\ref{ssec:CEP} (CEP), is changed. The light curves are shown on the left and the corresponding periodograms on the right. In each panel on the right the solid red line shows the fitted broken power-law model, while the dotted and dashed red lines show the 94\% and 99\% confidence levels, respectively. Top: using the whole simulated light curve provided (``whole''). Middle: trimming the light curve so that only the flare is included (``flare''). Bottom: trimming the light curve manually to the section of the flare that gives the highest significance level of the peak in the periodogram corresponding to the QPP signal (``manual''). Only in the manually trimmed light curve is the QPP signal assessed to be significant above the 99\% level.}
   \label{figure[trim_CEP_example]}
\end{figure*}

For the periodogram-based significance testing method employed by CEP, described in Section~\ref{ssec:CEP}, three different time intervals were tested for each synthetic flare. These were a manual trim to the section of light curve within the flare that gave the most significant peak in the periodogram (referred to as ``manual''), a trim to include the whole flare (referred to as ``flare''), and no trimming, where the entire provided light curve was used for the analysis (referred to as ``whole''). Figure \ref{figure[trim_CEP_example]} shows how trimming the data impacts the periodogram for Flare 629040. 

\begin{figure*}[htp]
  \centering   \includegraphics[width=0.45\textwidth]{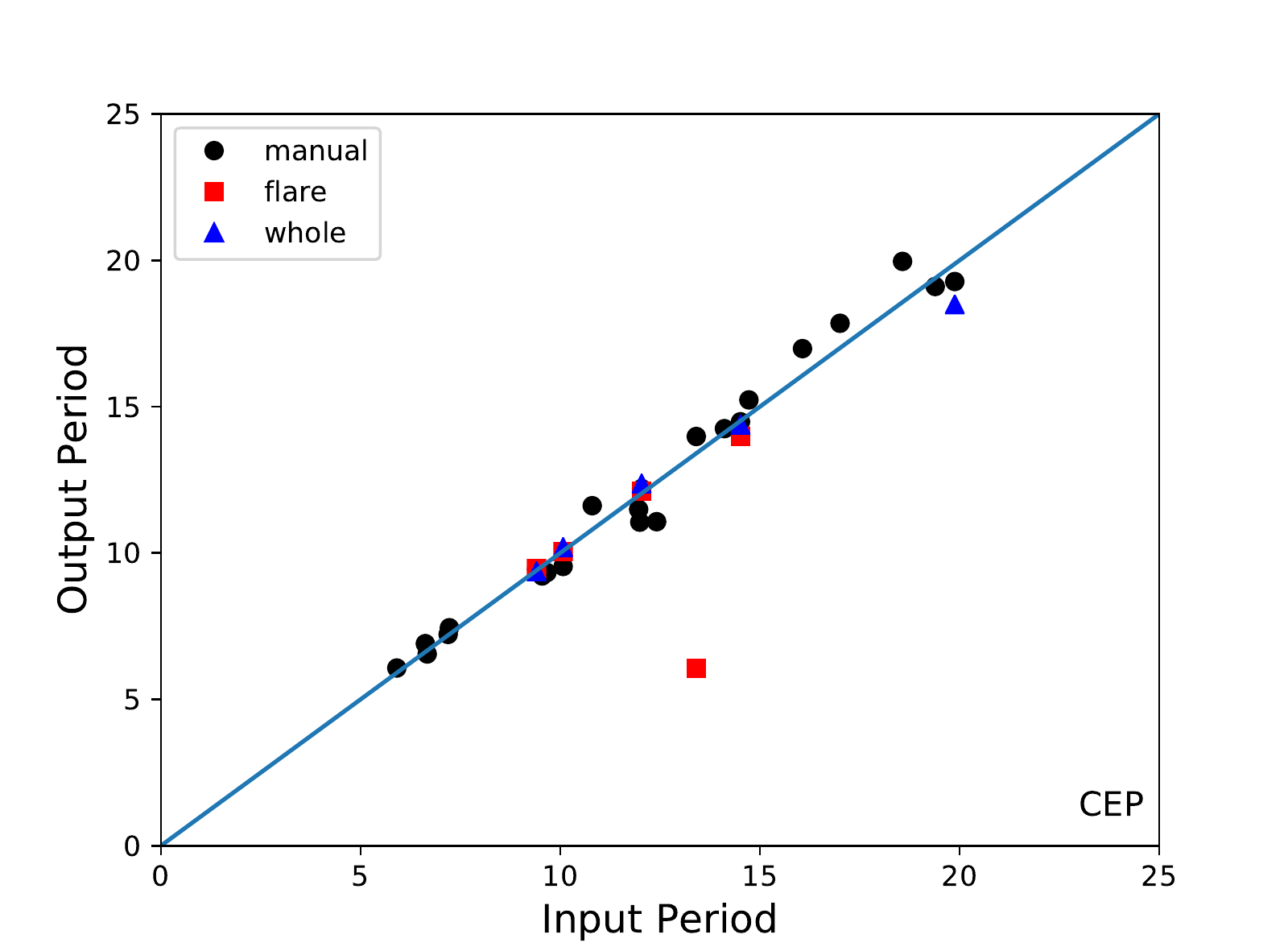}
  \centering   \includegraphics[width=0.45\textwidth]{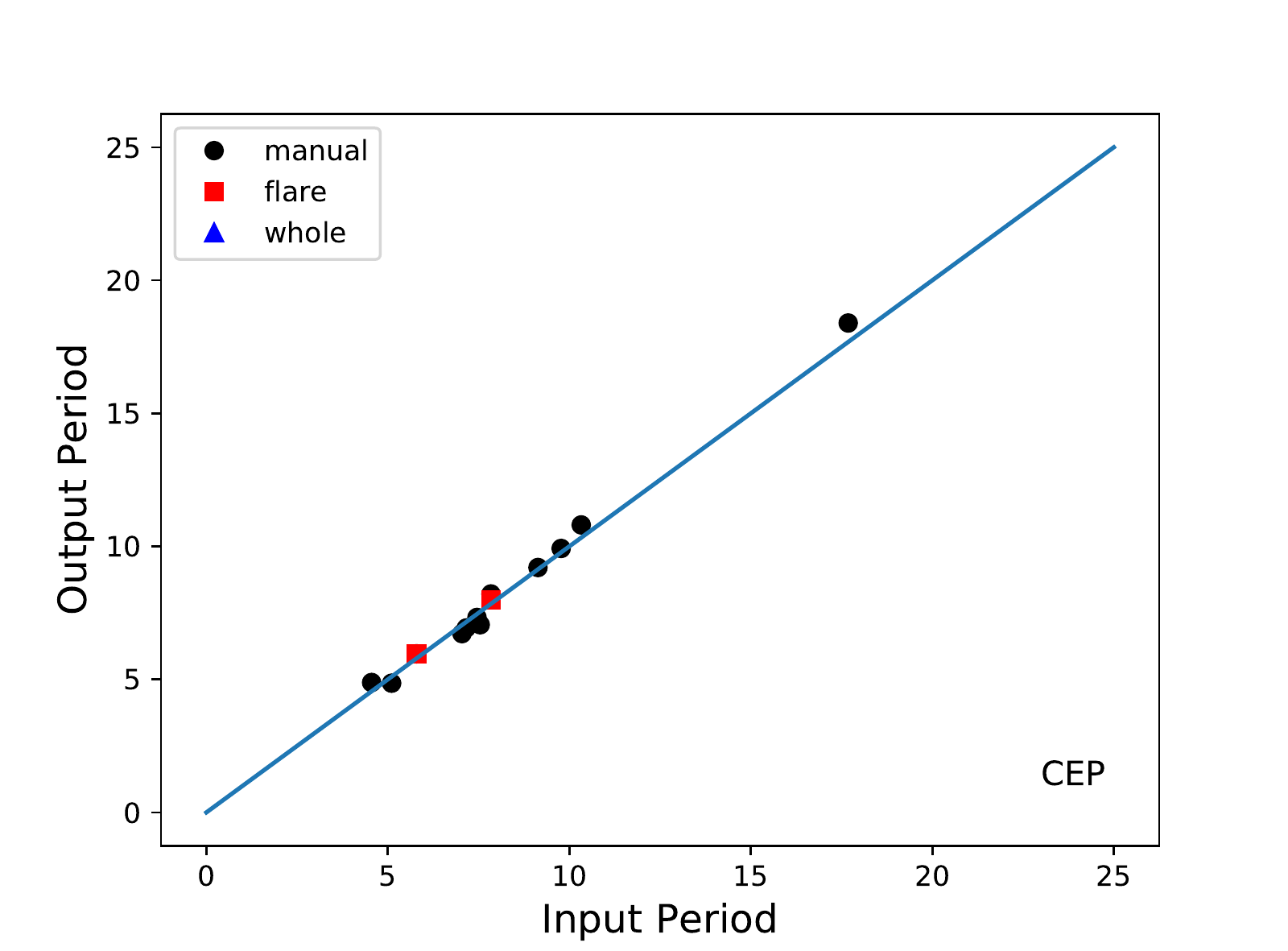}\\
   \caption{Left: Scatter plot showing detections of QPPs made in HH1 by using the method of CEP when the data were trimmed by different amounts. Right: Same as left panel but for HH2.}
   \label{figure[trim_CEP]}
\end{figure*}

For the first case (``manual'') the same time intervals as those used with this method to obtain the results in HH1 (Section \ref{sec:quality}) were chosen. As mentioned above, this approach resulted in 25 flares being identified as containing a periodic signal above the 95\% confidence level (23 sinusoidal QPP flares, two real flares, and no false detections). When the light curve was trimmed to include the whole flare but nothing more (``flare''), only five detections were made above the 95\% level (all sinusoidal QPP flares). Finally, when no trimming was performed and the whole light curve was used (``whole''), six detections were made above the 95\% confidence level, but one of these was a false detection (the other five were sinusoidal QPP flares). Figure \ref{figure[trim_CEP]} shows a comparison between the simulated (input) and detected (output) QPP periods for the different trimming methods. Only one detection lies more than 3 units from the input period.

This test was repeated for HH2. For the manually optimized time intervals used to obtain the results for this method in Section~\ref{sec:false}, 14 flares were found to contain a periodic signal above the 95\% level, but two of these were false detections. When the light curves were trimmed to contain the whole flares, the number of detections reduced to 2, although both were precise detections of single sinusoidal QPPs. Finally, when the whole light curves were used, no detections were made. Hence this shows that the choice of time interval is an important factor when applying this method, since the time interval can be chosen to avoid any steep changes in the light curve that might otherwise reduce the S/N ratio of a periodic signal in the periodogram \citep{2017A&A...602A..47P}.

These results imply that (a) when detections are claimed they tend to be robust regardless of trimming, (b) trimming to focus on the time span containing the QPPs substantially improves the likelihood of detecting QPPs, and (c) there is no benefit to trimming around the ``flare'' compared to taking the ``whole'' dataset. However, we note that none of the time series simulated here are substantially longer than the flare itself, which may not necessarily be the case in real data.

\begin{figure*}[htp]
  \centering   \includegraphics[width=0.45\textwidth]{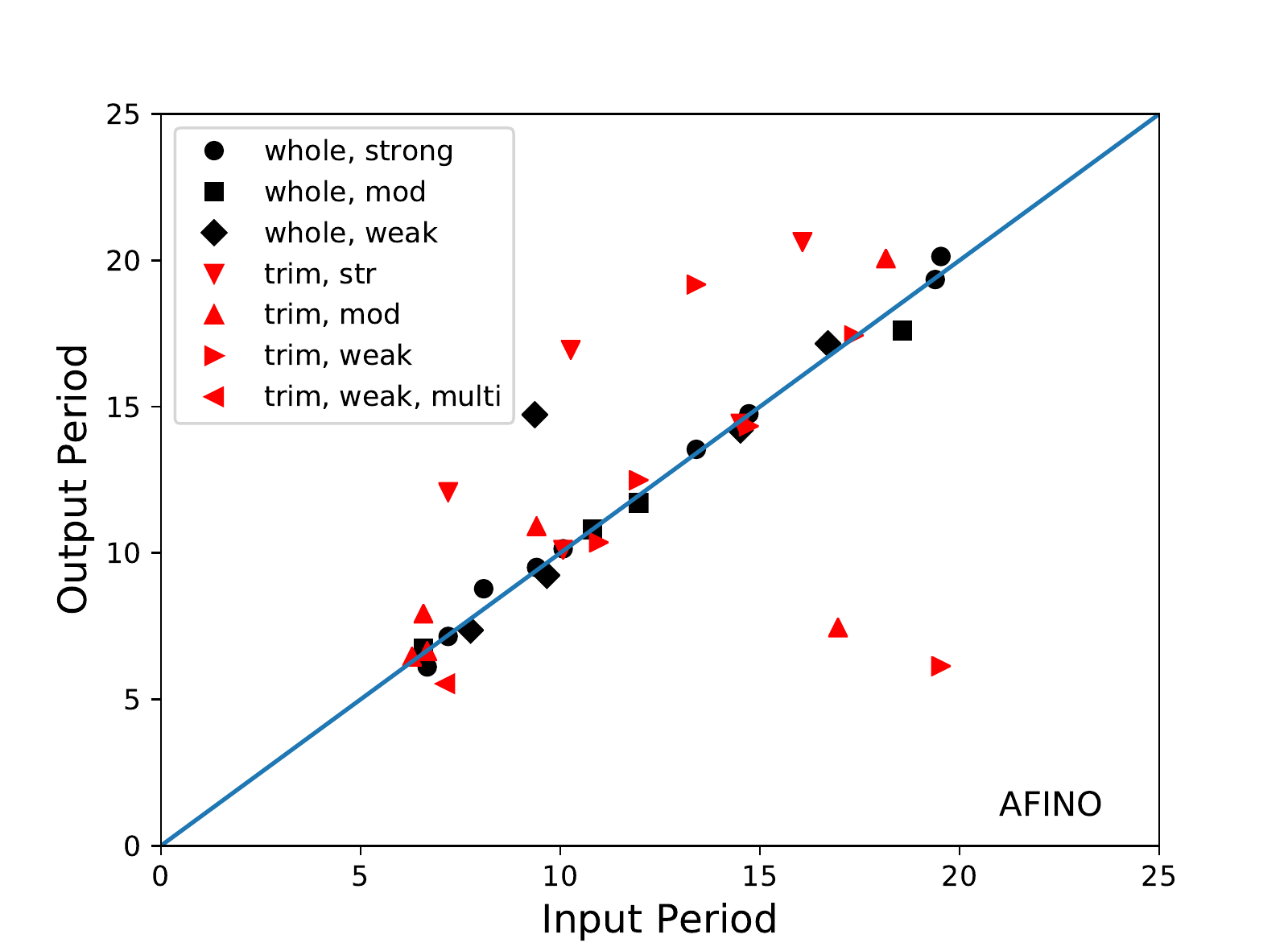}
  \centering   \includegraphics[width=0.45\textwidth]{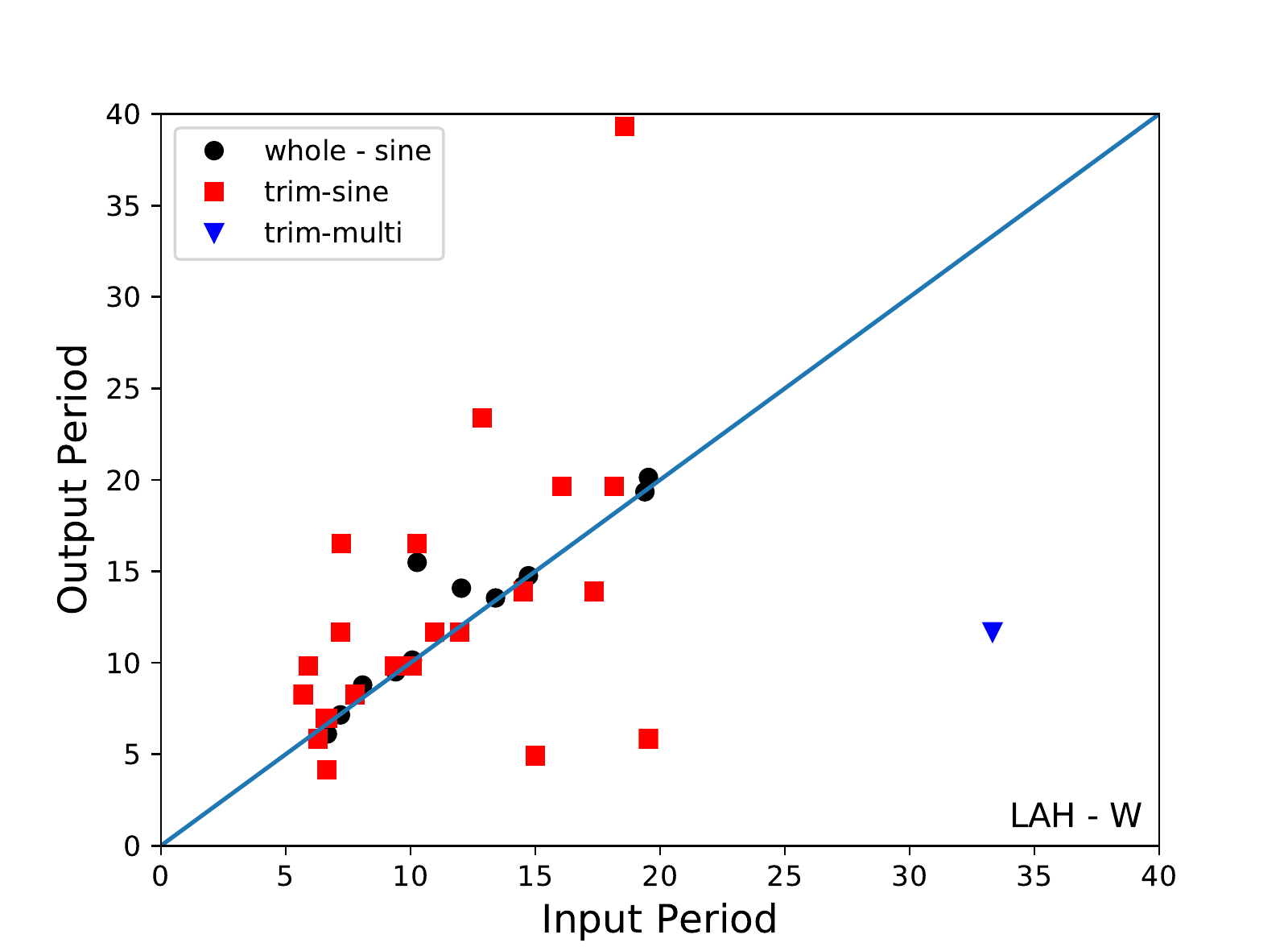}\\
   \caption{Left: Scatter plot showing detections of QPPs made in HH1 by using the AFINO method when the data were trimmed by different amounts. ``Whole'' refers to the whole time series and ``trim'' refers to time series trimmed using the ``manual'' trimming of CEP. The different symbol shapes indicate the strength of the detection as discussed in Section \ref{ssec:ari}. Right: Scatter plot showing detections of QPPs made in HH1 by using the LAH's wavelet method when the data were trimmed by different amounts.}
   \label{figure[trim_LAH]}
\end{figure*}

We recall here that neither the AFINO method nor LAH's wavelet method trimmed the data when looking for QPPs. To test the impact of trimming on these techniques, the methodologies were re-run on trimmed data  using the manual-trim timings of CEP. This was done for HH1 only, and the results can be seen in Figure \ref{figure[trim_LAH]}. 

For the full AFINO method, originally employed by ARI, 18 detections of sinusoidal QPPs and no periodic multiplets were claimed, with 17 of these detections considered to be precise. When the data were trimmed, LAH found that the full AFINO method produced 17 sinusoidal QPPs and one periodic multiplet detection, but only 12 of these were precise. No false detections were made in either case. However, we note that although there was some overlap, the set of simulated flares in which detections were made when the data were trimmed was not identical to the set of flares in which detections were made when the whole time series was used. 

When LAH's wavelet method was applied to the full time series of the simulated flares included in HH1, 12 detections of sinusoidal QPPs were claimed, of which 11 were considered to be precise (no periodic multiplet detections were claimed). As can be seen in Figure \ref{figure[trim_LAH]}, this increased to 26 claimed sinusoidal QPP detections and one periodic multiplet detection when the data were trimmed, with only 16 precise detections. 

This loss of precision may indicate that the AFINO and wavelet methods work best when considering the whole time series. It may also be an indication that the trimming applied for one method may not necessarily be the optimal trimming for another method. Another explanation for the loss of precision could be due to the reduction in resolution in the Fourier domain due to the reduced number of data points. For example, the lack of improvement in AFINO when examining the trimmed data can be explained in terms of the low number of data points in the trimmed time series: AFINO explicitly penalizes short data series (Equation \ref{bic_eqn}), so this is apparently enough to counteract any ``enhancement'' of the signal from trimming, at least in these cases.

Figure \ref{figure[JAM]} shows periodograms produced by the two methods used by JAM for HH1 and HH2, respectively. The primary difference between these methods was that for HH2 the time series were trimmed to start at the location of the local maximum, i.e. the peak of the flare. Comparison of the two panels shows that the additional trimming performed in the HH2 methodology removes the low-frequency noise from the spectrum, leaving just the peak from the QPPs. With hindsight it is possible to see that this peak is also present in the periodogram produced by the HH1 methodology; however, without prior knowledge it would not be possible for a user to distinguish between the QPP peak and the noise peaks. In both methods the background trend was removed before computing the periodogram by subtracting a smoothed version of the light curve. The difference between the two periodograms is likely to occur because sharp features, such as the impulsive rise phase of a flare, will not be sufficiently removed by subtracting a smoothed version of the light curve. Starting the time series after the sharp rise phase means that smoothing does a far better job of characterizing the background trend, thus reducing the low-frequency noise in the periodogram. A cautionary note here would be that in real flares there is no guarantee that the QPPs will start after the impulsive flares, and so limiting your search to the decay phase could lead to missed detections. However, as already discussed, type II errors are far less serious than type I errors, and so it is better to employ this strategy than risk FPs.

\subsection{Nonstationary QPPs}\label{ssec:non_stat_out}

\begin{figure}[htp]
  \centering   \includegraphics[width=0.45\textwidth]{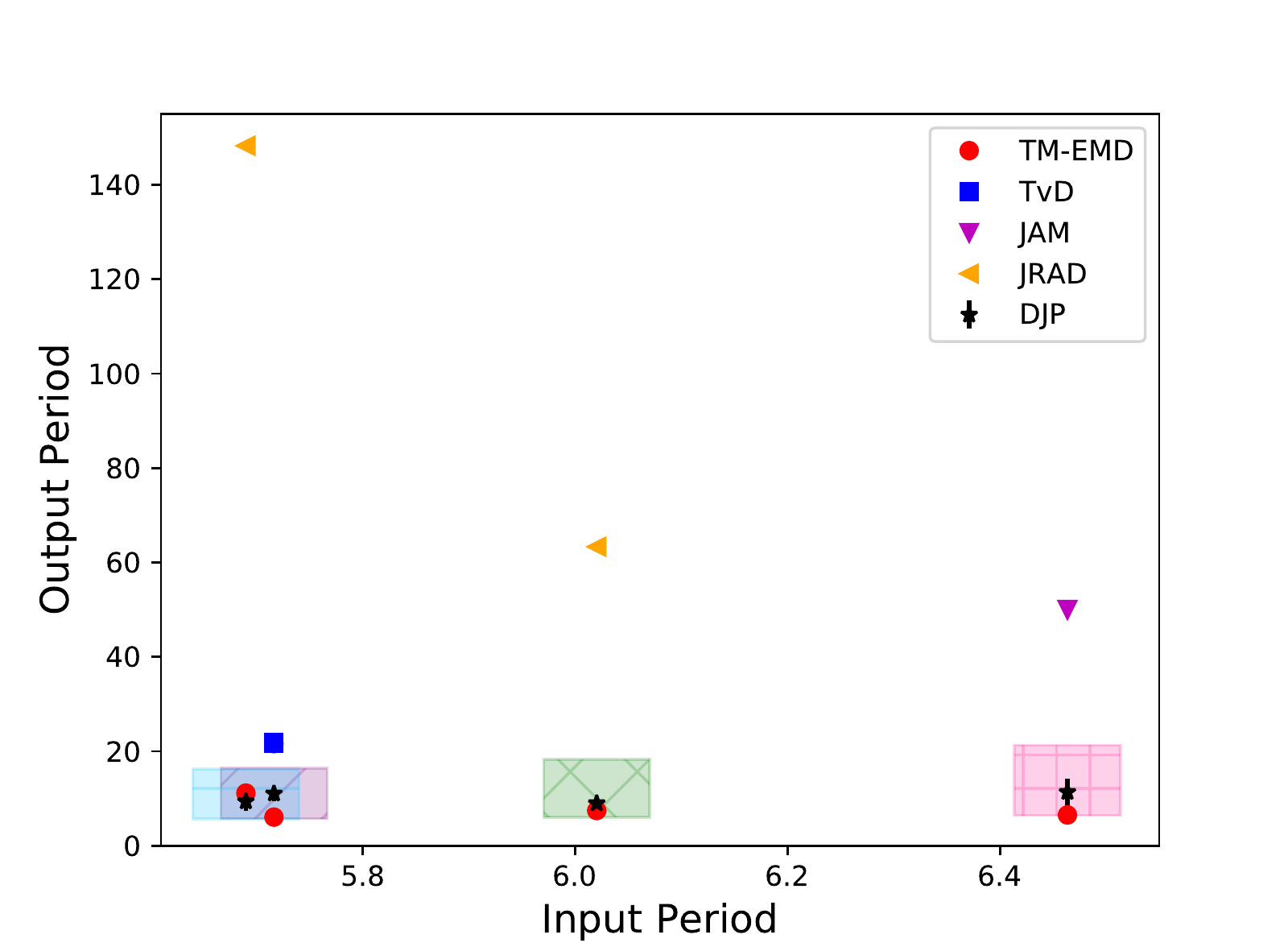}\\
   \caption{Scatter plot showing detections of nonstationary QPPs made in HH1. The input period is $1/\nu_0$ using equation \ref{eqn:non_stat}. The ordinate range indicated by the boxes shows the variation in period from $1/\nu(t=0)=1/\nu_0$ to $1/\nu(t=t_e)$ i.e. the period when the amplitude of the signal has decreased by a factor of e. The abscissa range is arbitrarily chosen to be centered on $1/\nu_0$ and of width 0.1. Difference colors/hatchings are used to differentiate between the different simulated flares.}
   \label{figure[non_stat]}
\end{figure}

Four nonstationary QPPs were included in HH1, but the majority of methods were unable to make robust detections of these QPPs (LAH--W, LAH--A, ARI, and CEP all failed to detect any of these QPPs; TVD, JAM, and JRAD claimed detections, but they were imprecise, as shown in Figure \ref{figure[non_stat]}). This is not completely surprising since periodogram-based methods, such as those employed by AFINO, CEP, TVD, and JAM, are better suited to detecting signals with stationary periods. EMD, on the other hand, makes no \textit{a priori} assumptions on the stationarity (or shape) of the periodicity. This is reflected by the fact that TM--EMD was able to precisely detect the periodicities of the included nonstationary QPPs (we note that TM only analyzed three of the four nonstationary QPPs blind, but once it became clear that EMD was capable of detecting nonstationary QPPs, TM analyzed the fourth nonstationary QPP but employed the same strategy as used in the blind tests). The EMD - Fourier method for assessing the significance of the detrended signal did not detect any of the nonstationary QPPs.

\begin{figure*}[htp]
	\begin{center}%
	\includegraphics[width=\textwidth]{{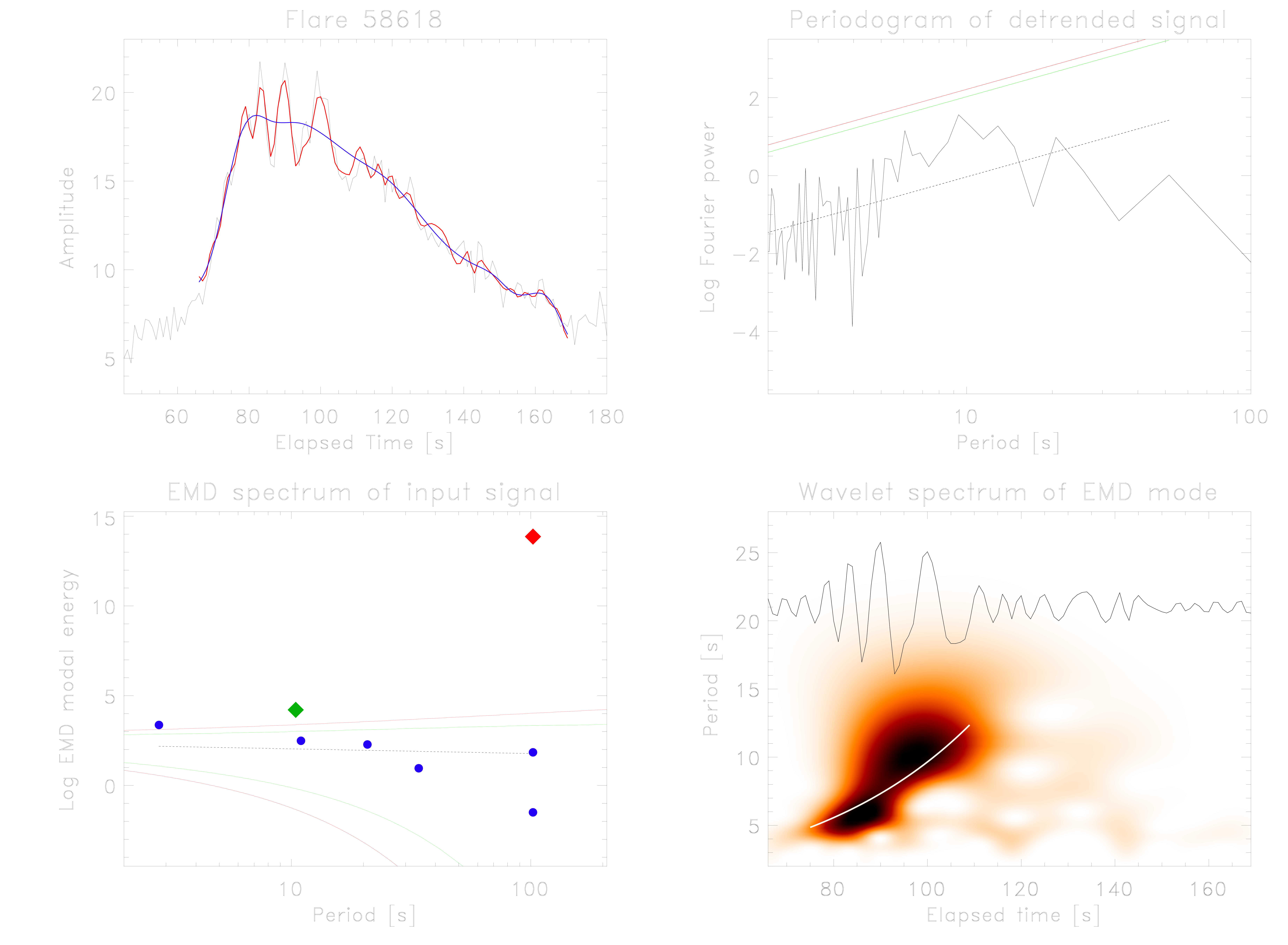}}
    \end{center}
\caption{\label{fig:EMDmethodcomp}
Top left: trimmed profile of Flare 58618 in black with the extracted EMD mode overlaid in red and trend in blue. Top right: periodogram of the detrended signal with confidence levels of 95$\%$ (green) and 99$\%$ (red). Bottom Right: Morlet wavelet spectrum of the statistically significant IMF (shown in black). The white line shows the approximation of the obtained period-time dependence by the chirp function (see Section \ref{ssec:non_stat}). Bottom left: EMD spectrum  of the original input signal with the significant mode shown as a green diamond. The trend is given as a red diamond. Blue circles correspond to noisy components with $\alpha \sim 0.89$. The 95$\%$ and 99$\%$ confidence levels are given by the green and red lines, respectively, with the expected mean value shown by the dotted line. }
\end{figure*}

\begin{figure*}[htp]
  \centering   \includegraphics[width=0.9\textwidth]{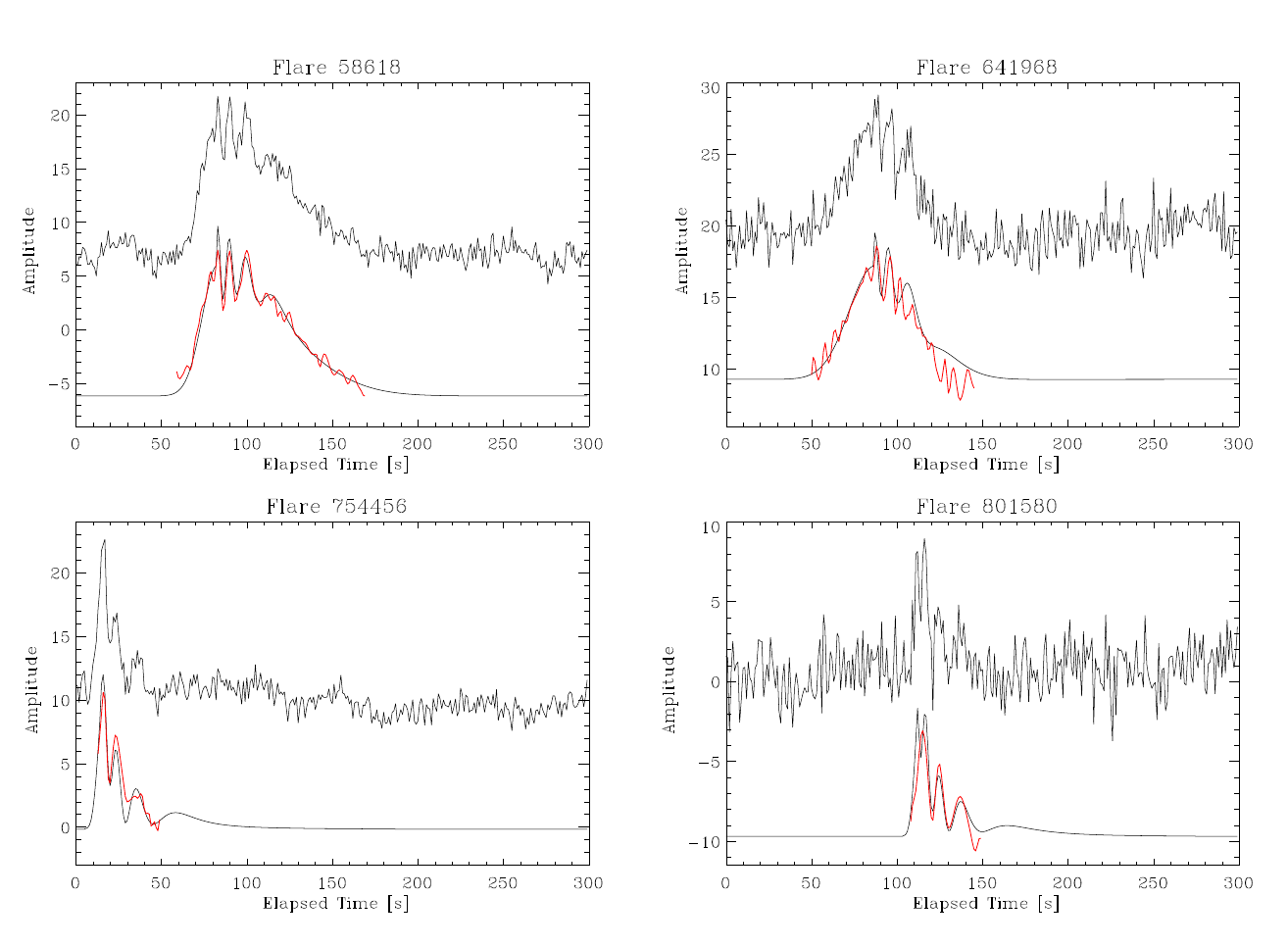}\\ 
   \caption{EMD analysis of four flares from HH1 containing nonstationary QPPs. In each panel the upper curve (solid black) is the raw input signal. Below the light curve is the input model with all noise removed, containing a trend and nonstationary QPP signal, which was given to TM by the hare only following the analysis for comparative purposes (black). Overlaid onto the input model is the statistically significant IMF of the (manually selected) flare phase, shown in red. For the cases of Flares 58618, 641968 and 754456, it was unknown to TM that the signals were of nonstationary origin and were analyzed under the same lack of assumptions of the other flares looked at in HH1. Flare 801580 was analyzed separately with the knowledge that it contained a nonstationary signal. }
   \label{figure[EMD_nonstat]}
\end{figure*}

Figure~\ref{fig:EMDmethodcomp} shows the results of the EMD methodology on one of the simulated flares, Flare 58618, which had a nonstationary QPP included. Figure~\ref{fig:EMDmethodcomp} also shows the Morlet wavelet spectrum of the EMD mode, which was found to be significant in the flaring section of the original signal. It clearly illustrates the increase of the oscillation period with time from about 75 to 110, which was approximated by the functional form $P(t)=P_0\left({P_1}/{P_0}\right)^{({t-t_0})/({t_1-t_0})}$, with the following parameters:
$P_0 \approx 4.9$, $P_1  \approx 12.7$, $t_0=75$, and $t_1=110$. The EMD-obtained mode gave a significant mean periodicity of 10.4\,s, which lies within a reasonable window of the fitting. Hence, this technique, although time intensive, has clear benefits when used in tandem with other traditional methods to extract nonstationary signals. Figure \ref{figure[EMD_nonstat]} shows the EMD analysis of the four nonstationary QPP flares included in HH1, including Flare 58618. It can be seen that the IMF obtained from the EMD analysis closely matches the input signal for all flares and thus demonstrates the ability of EMD to extract nonstationary QPP signals from the data. 

\begin{figure*}[htp]
\centering
\includegraphics[width=0.9\textwidth]{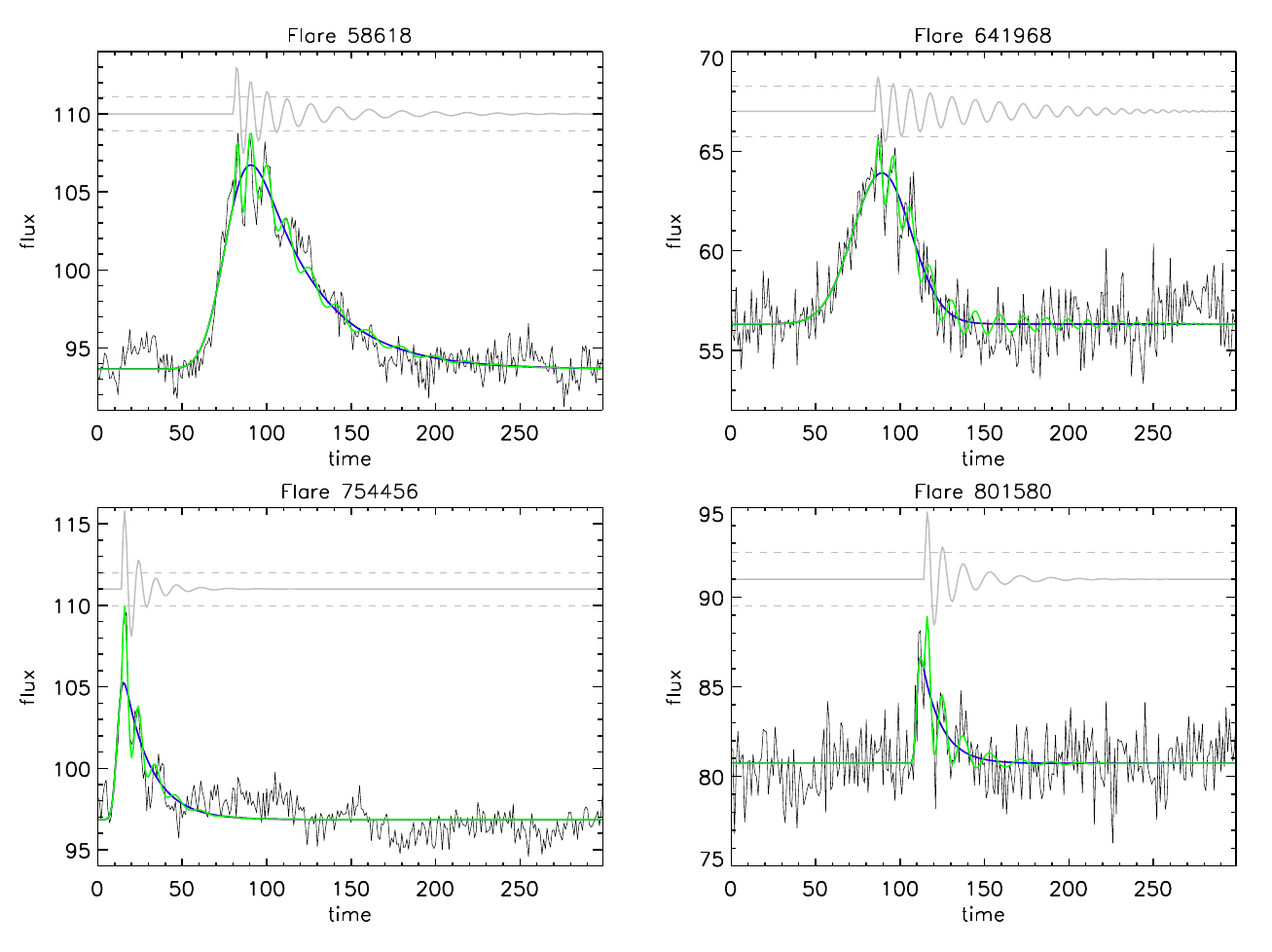}\\
\caption{Method of forward-modeling QPP signals based on the Bayesian inference and MCMC sampling used in \citet{2017A&A...600A..78P}.
Black lines show the simulated flare data; green lines represent the model fit based on the MA$p$ values of model parameters.
Blue and gray lines correspond to the background trend and detrended signal (shifted for visibility), respectively.
The gray dashed horizontal lines denote the estimated level of (white) noise in the signal.}
\label{figure[djp2]}
\end{figure*}

\begin{figure*}[htp]
\centering
\includegraphics[width=0.9\textwidth]{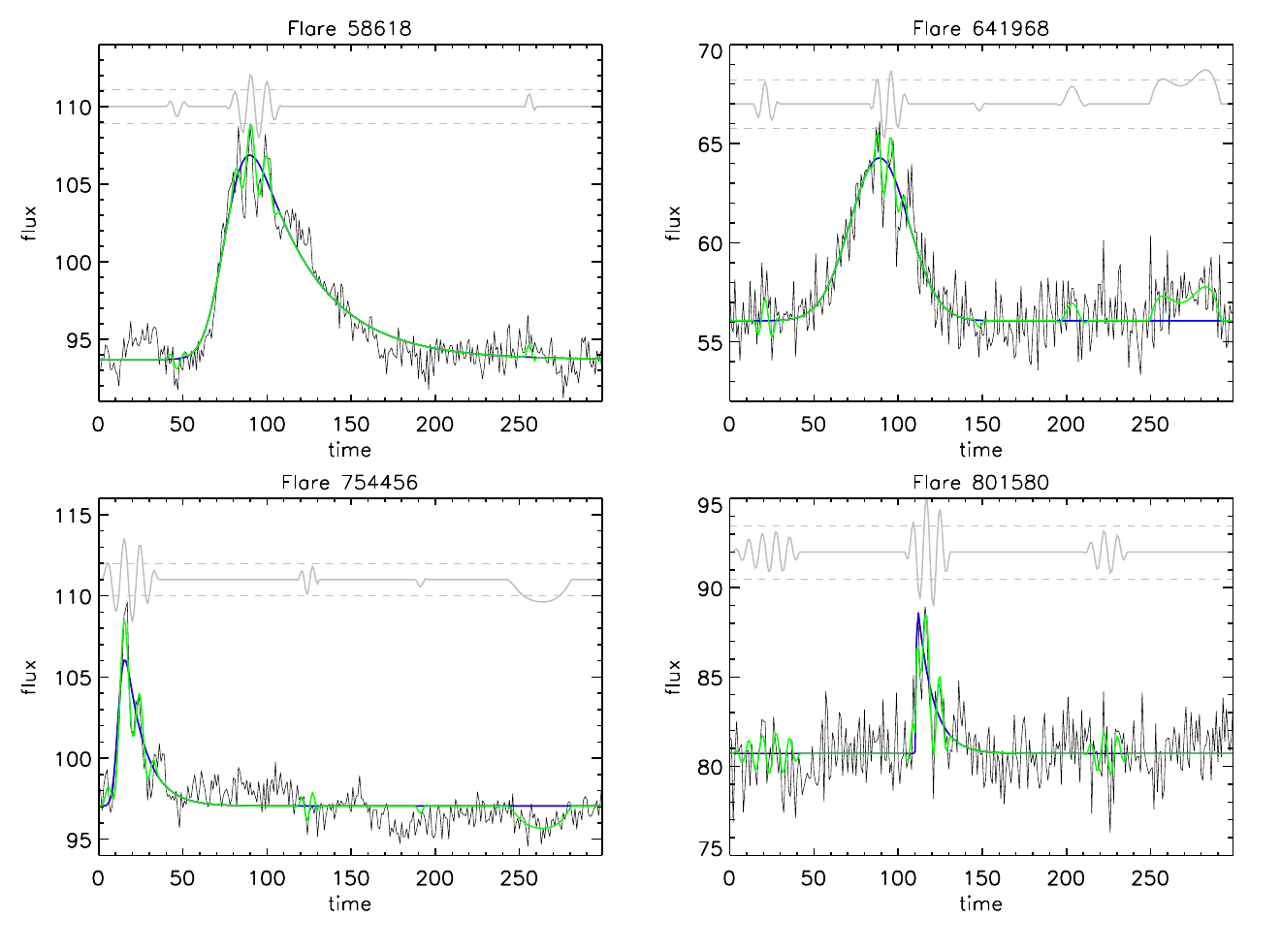}\\
\caption{Same as in Figure~\ref{figure[djp2]} but for a model based on continuous amplitude modulation rather than a damping profile.}
\label{figure[djp3]}
\end{figure*}

The nonstationary flares were also analyzed by the forward-modeling method of DJP. Figure \ref{figure[djp2]} shows the results of forward-modeling the four nonstationary QPP flares based on an exponentially decaying sinusoid (with the potential for a nonstationary period). Figure~\ref{figure[djp3]} shows the corresponding results for a method based on a signal with continuous amplitude modulation rather than defined start and decay times. This method is motivated by the characteristic shape of QPPs formed by dispersive evolution of fast wave trains, i.e. having both period and amplitude modulation. (This is more general than the actual form of the QPPs used in this study, which only have exponentially decreasing amplitudes.) As can be seen, in both Figure \ref{figure[djp2]} and \ref{figure[djp3]}, the model appears to fit the data well. Figure \ref{figure[non_stat]} shows that the average periods extracted from the method (based on Figure \ref{figure[djp2]}) agree well with the input periods once the variation in period over the lifetime of the QPPs is accounted for. However, we remind the reader that reliable extraction of the QPPs relies on correct specification of the model used to fit the data. Furthermore, the false-alarm rate for this method was not tested. However, we note that the forward-modeling method was able to extract the periodicity of the two simple sinusoidal QPPs that were analyzed (Flare 106440 had an input period of 12.4 and DJP found a periodicity of $12.4_{-0.9}^{+0.6}$; Flare 220365 had an input period of 14.5 and DJP found a periodicity of $14.3\pm0.2$). All of these results indicate that MCMC is a good way of obtaining QPPs' parameters and could perhaps be implemented once detections have been made with one of the robust methodologies (e.g. AFINO--ARI, wavelet--LAH, periodogram--CEP, TVD--manual).

\subsection{HH3 and the impact of smoothing}\label{ssec:smooth}

Some of the techniques employed by the hounds (TVD and JAM) rely on detrending the data before using periodograms to assess the significance of a signal. In both cases detrending was performed by removing a smoothed component from the data. However, as we saw with HH1, this must be done carefully, such as in the non-automated manner used by TVD and as described in Section \ref{ssec:tvd}, to obtain robust results: when TVD manually chose an appropriate smoothing window individually for each flare, the results were found to be robust, but, choosing a single smoothing width for all flares, as done by JAM, produced a large number of false and imprecise detections (see Table \ref{table[HH1]}). To investigate this further, TVD analyzed a third set of flares, HH3, which contained 18 flares, using a range of different smoothing widths on each flare. 

\begin{figure*}[htp]
  \centering   \includegraphics[width=0.45\textwidth]{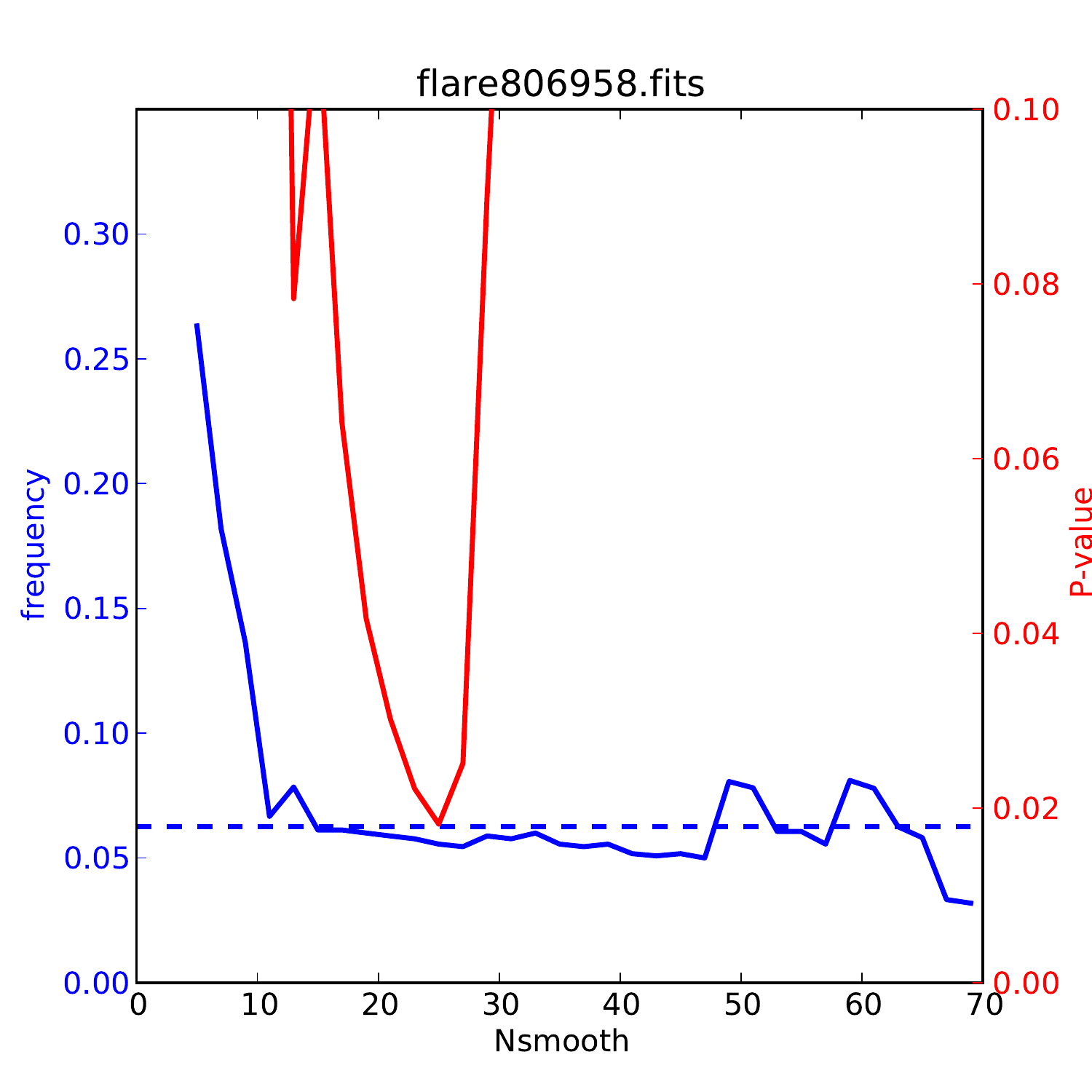}
  \includegraphics[width=0.45\textwidth]{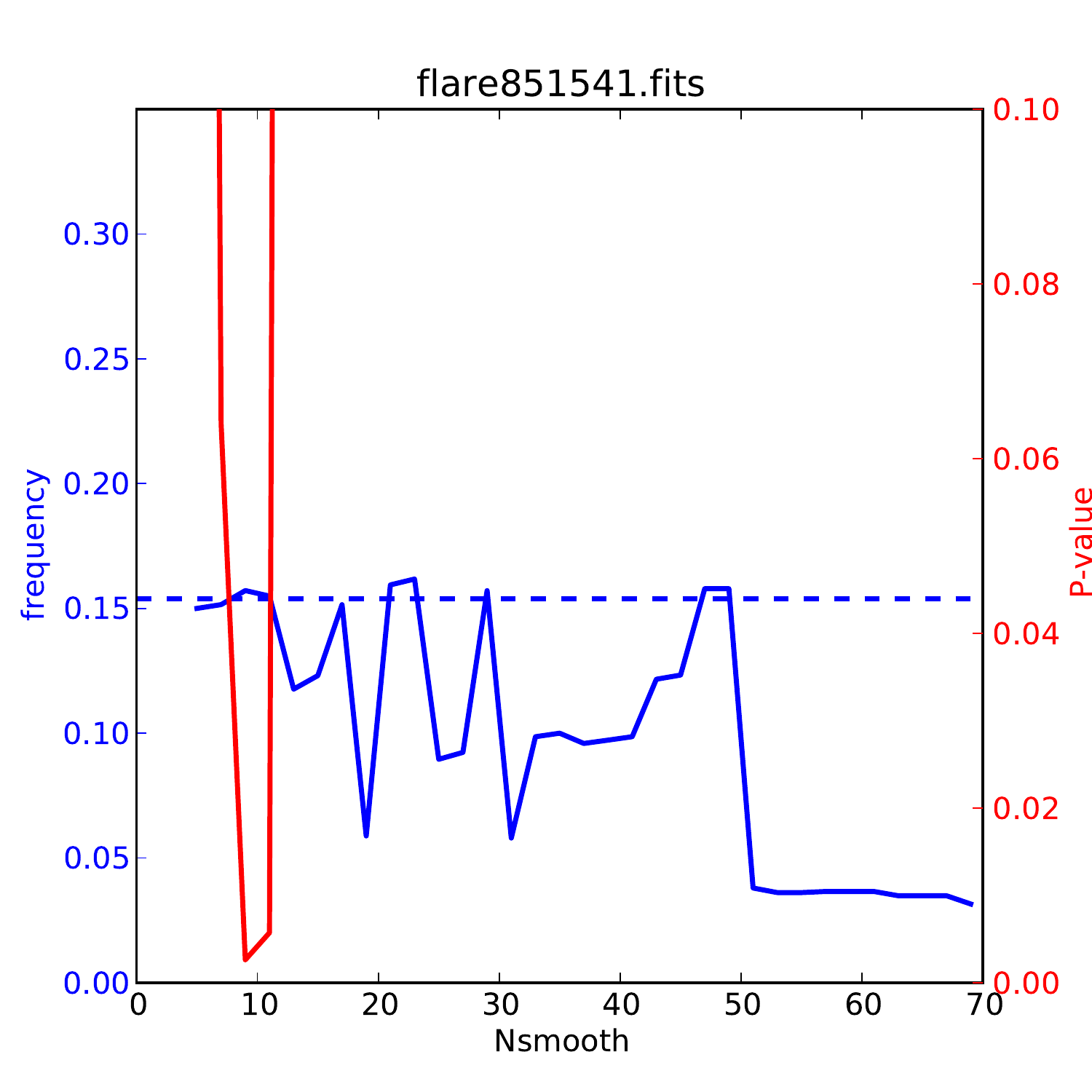}\\
  \includegraphics[width=0.45\textwidth]{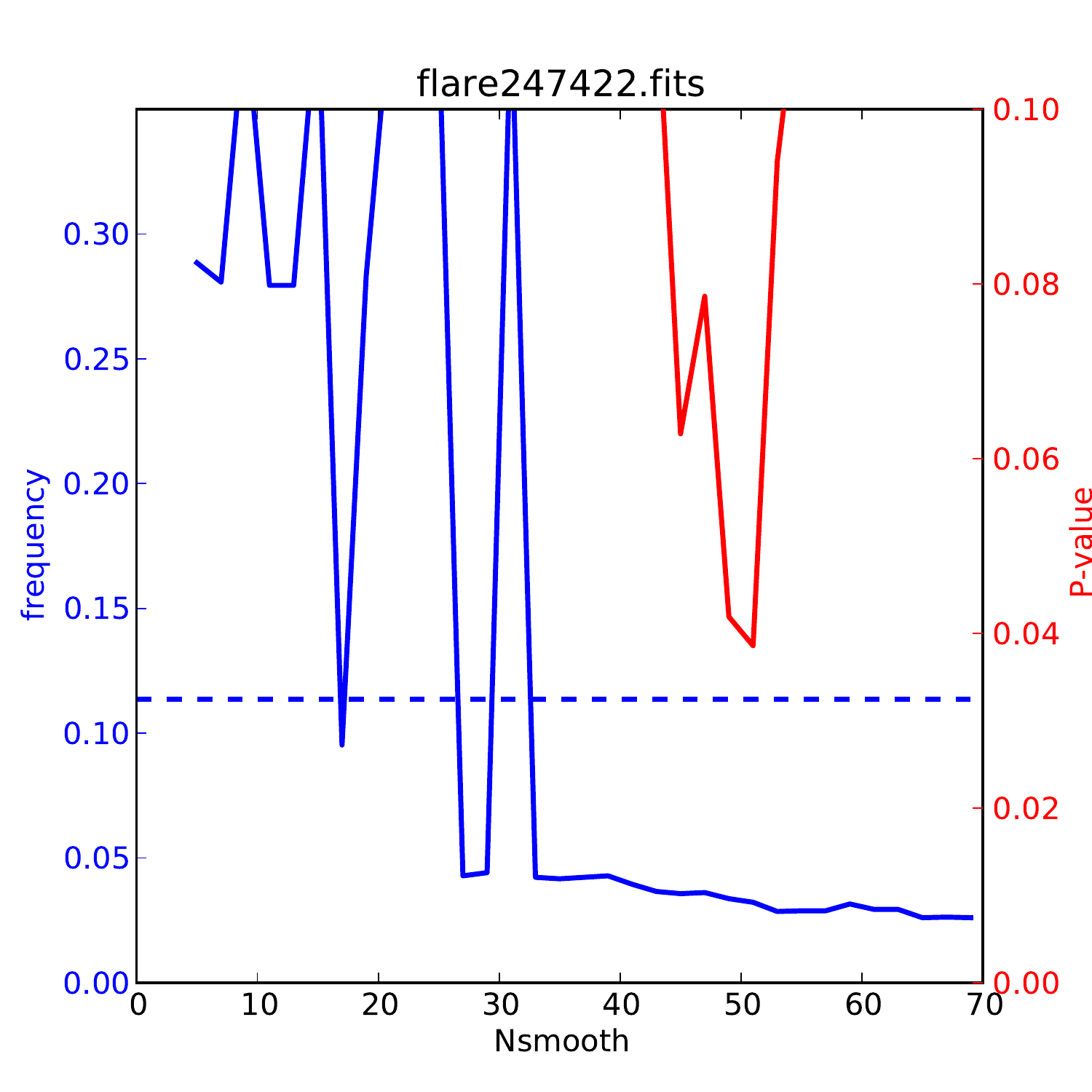}
  \includegraphics[width=0.45\textwidth]{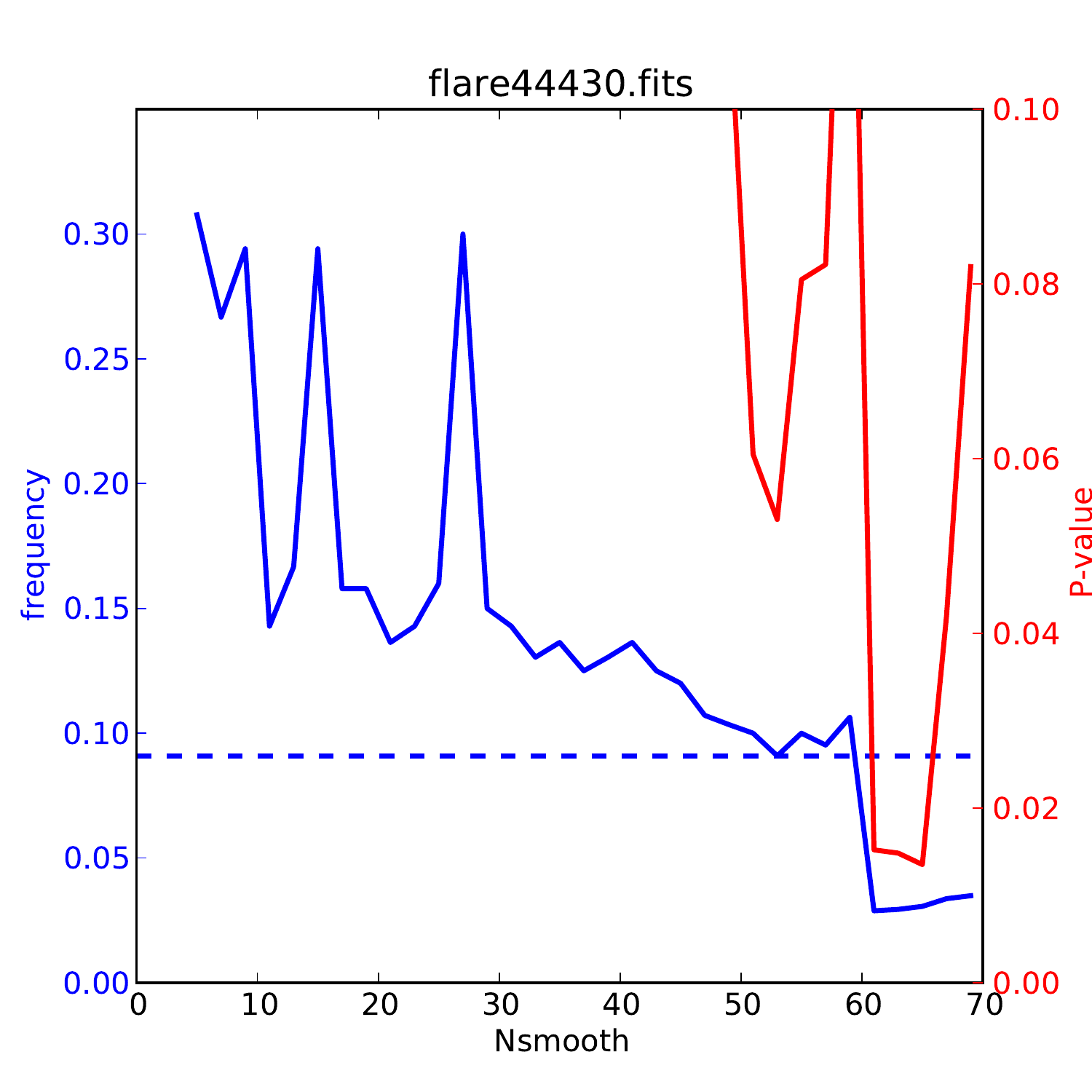}\\
   \caption{Frequency of the highest peak in the detrended periodogram is plotted as a function of the width of the smoothing window used to detrend the data (plotted in blue and corresponding to the left-hand ordinate). Also plotted is the false-alarm probability (or $p$ value) as a function of the width of the smoothing window (in red and corresponding to the right-hand ordinate). The horizontal dashed line gives the input frequency of the QPPs. Top left: Flare 806958 had an input period period of 16.0 and a detected period of 18.0 (or a frequency of 0.056) and so is an example of a precise detection. Top right: Flare 851541 had an input period of 6.5 and a detected period of 6.4 (or a frequency of 0.156) and so is an example of a precise detection. Bottom left: Flare 247422 had an input period of 8.8 and a detected period of 31.0 (or a frequency of 0.032) and so is an example of an imprecise detection. This detection was not flagged as untrustworthy by TVD. Bottom right: Flare 44430 had an input period of 11.0 and a detected period of 32.7 (or a frequency of 0.031) and so is an example of an imprecise detection. This detection was flagged as untrustworthy by TVD.}
   \label{figure[smooth_range]}
\end{figure*}

In this test TVD cycled through using different smoothing windows, $N_\textrm{smooth}$, to remove the background trend, from 5 to 63 in steps of two. For each detrended time series, a periodogram was found and the false-alarm probability (or $p$-value) and frequency of the largest peak recorded. Examples of the recorded frequency and $p$ value as a function of $N_\textrm{smooth}$ for four flares are shown in Figure \ref{figure[smooth_range]}. Here the $p$ value is the probability of observing a peak in the power spectrum at least as high as that of the largest observed peak if the data contained white noise only. In the method employed by TVD, as described in Section \ref{ssec:tvd}, detections were claimed if the false-alarm probability was below 5\% i.e. if the minimum $p$ value in Figure \ref{figure[smooth_range]} was below 0.05. The top two panels show examples where precise detections were made. In both cases a clear minimum in the $p$ value was observed. For Flare 806958 the observed frequency is relatively flat once the smoothing window is above approximately 11. This appears to account for the relatively broad range of potential smoothing windows with low $p$ values. This could be related to the fact that the input period of this QPP was relatively long (16.0). There is more variation in the frequency with the lowest $p$ value in the analysis of Flare 851541, which had an input period of 6.5. Here a much narrower range of smoothing windows produced low $p$ values. We notice also the drop in the frequency with the lowest $p$ value at high smoothing widths. This is a common feature of this analysis and can be seen in the bottom two panels of Figure \ref{figure[smooth_range]}. It is possible that this drop-off would also have been observed in Flare 806958 if the analysis had been extended to higher smoothing widths relative to the input period. The bottom two panels of Figure \ref{figure[smooth_range]} show examples of flares where detections were claimed but these detections were imprecise. The $N_\textrm{smooth}$ coinciding with the minima in the $p$ values correspond to frequencies beyond the drop-off. For Flare 44430 there is a secondary minimum in the false-alarm probability that would have produced a frequency of approximately 0.1, which is close to the input frequency of 0.09 (or a period of 11.0).

\begin{figure*}[htp]
  \centering   \includegraphics[width=0.45\textwidth]{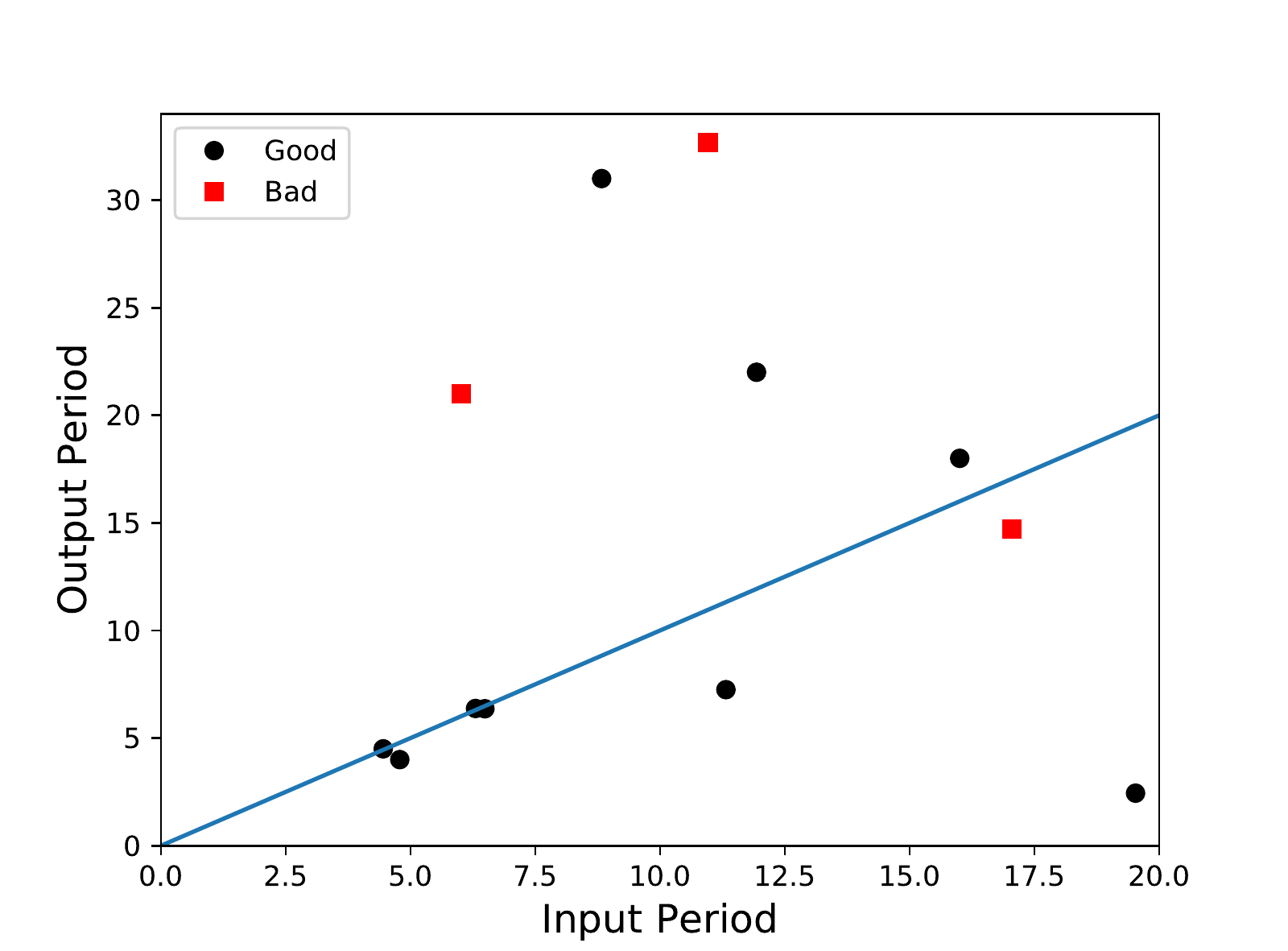}
   \includegraphics[width=0.45\textwidth]{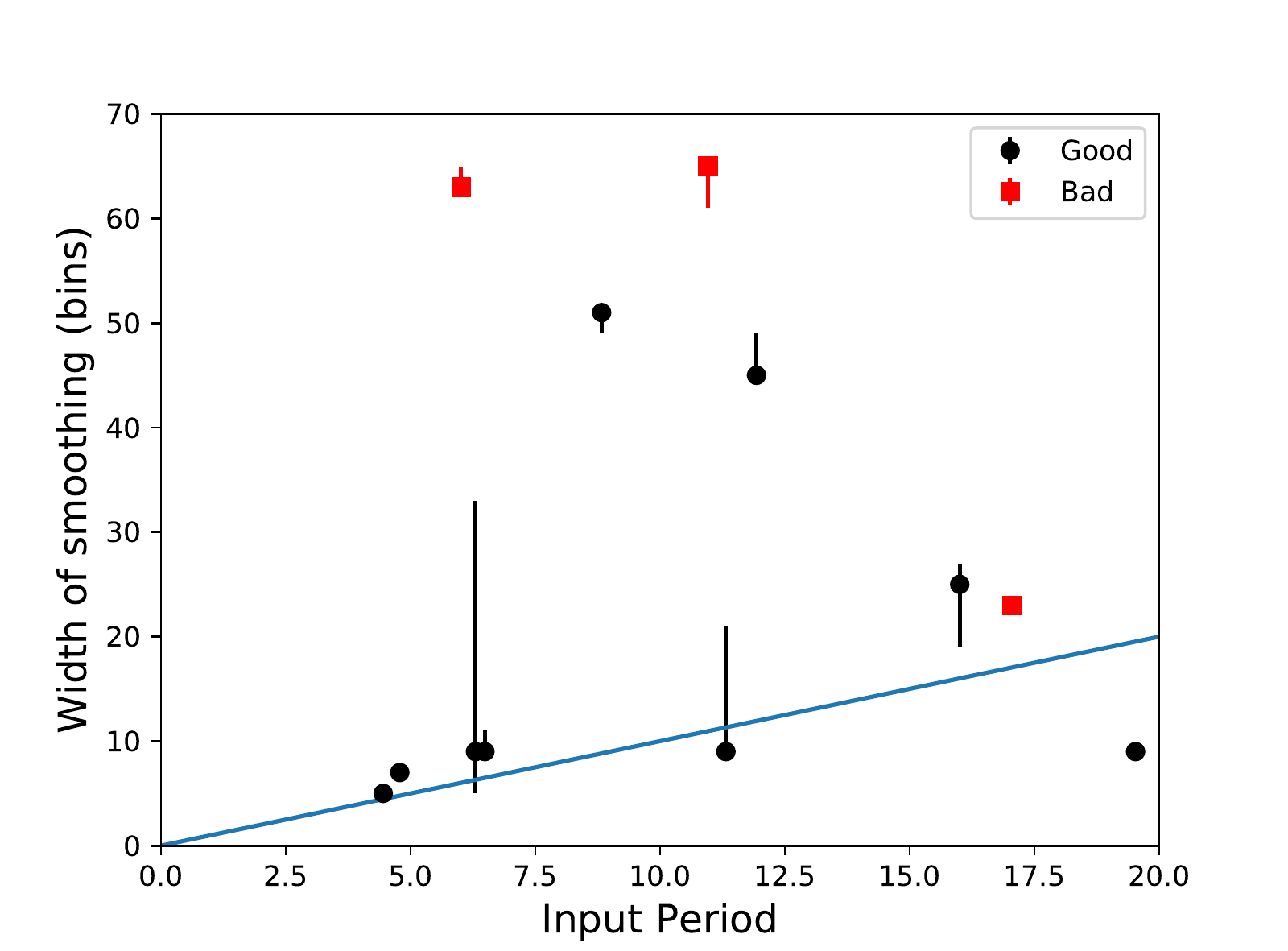}\\
   \caption{Left: Comparison of output and input periods obtained by TVD in HH3. The data points have been split into ``Good,'' where TVD was satisfied with the results, and ``Bad,'' where TVD was unconvinced by the output. Right: Width of the smoothing window that produced the lowest false-alarm probability, and the periodicities plotted in the right panel. The uncertainties represent the range of values for which the false-alarm probability was below 5\% and within 10\% of the periodicity with the lowest false-alarm probability.}
   \label{figure[HH3_smooth]}
\end{figure*}

Figure \ref{figure[HH3_smooth]} shows how precise the detections made by TVD in HH3 were. The data have been separated out into ``Good,'' where TVD was satisfied with the extracted period, and ``Bad,'' where TVD was unconvinced by the output. The ``Good'' or ``Bad'' assessment was based on TVD's previous experience in analysing QPP light curves. A result was taken as ``Bad'' when the trend did not fit well the ``visible'' trend (matching the expectations from experience), or if the obtained period did not match the ``visible'' period (once again as measured using experience). For example, ``Bad'' detections were often highlighted when $N_{\textrm{smooth}}$ was sufficiently long that the background trend was not removed properly, leading to spurious periodicities in the power spectrum that dominated over the true QPP periodicity. These were identified by visual inspection of the figures produced for each flare, examples of which are shown in Figure \ref{figure[smooth_range]}, and the residual time series obtained once the smoothed time series had been subtracted. The left panel of Figure \ref{figure[HH3_smooth]} supports the earlier finding of HH2, that the automated process for determining the appropriate smoothing window is less robust than the manual one. Precise detections (where the difference in the input and output periodicities was less than three) were made in only six out of 12 claimed detections, with only two of the remaining 6 imprecise detections being highlighted by TVD as unreliable. This can be compared to 91\% precise detections obtained in HH1 (see Table \ref{table[HH1]}). The right panel of Figure \ref{figure[HH3_smooth]} shows that precise detections tend to be made when the smoothing window is close to the periodicity that you are trying to detect. This was found by TVD when manually selecting the best smoothing window while analysing the flares in HH1 (see Section \ref{ssec:tvd}).

\section{Best Practice Blueprint for the Detection of QPPs}\label{sec:blueprint}

The short-lived and often nonstationary nature of QPPs means that they are difficult to detect robustly. Therefore, when attempting to find evidence for QPPs, it is extremely important to minimize the number of type I errors, where the null hypothesis is wrongly rejected. In this paper that would mean making false QPP detections. This paper demonstrates that there is more than one way to robustly search for QPP signatures (e.g. Table \ref{table[HH2]}), with the AFINO (ARI and LAH), wavelet (LAH), and periodogram method of CEP producing particularly low numbers of false detections (by which we mean both false claims and imprecise detections). Furthermore, these methods have already been used in a number of studies to detect QPPs \citep[e.g.][]{2016ApJ...827L..30H, 2016ApJ...833..284I, 2017A&A...608A.101P} and this article demonstrates that we can be confident in the detections previously made. All these methods make relatively large numbers of type II errors, i.e. a large number of QPPs were missed (see Table \ref{table[HH1]}). However, type II errors are preferential to type I errors: it is better to use an approach with a low false-alarm rate and a high precision rate, so you can be more confident about what you find in real data where the answer is not known, even if this is at the expense of missing detections. To further improve confidence in detections, it would be preferential to employ more than one detection method. As an aside we note that the the AFINO method and CEP's periodogram method both make detections in only $25-35$\% of flares containing QPPs. This detection rate is similar to that found by both authors in recent surveys \citep{2016ApJ...833..284I, 2017A&A...608A.101P}, implying that the number of real flares containing QPPs may be substantially higher than implied by these surveys.

\textit{Recommendation 1:} Minimize type I errors, using simulations to test robustness of detection methods. AFINO (ARI and LAH), wavelet (LAH) and periodgram (CEP) methods were the most robust methods identified here.

The three methods mentioned above, which produced the lowest false-alarm rates, all incorporated statistics pertinent to red noise in their detection methods. It is worth keeping in mind that the simulated flares always included red noise, although the tests were performed blind so the hounds did not know this for definite when performing their searches. Real data will contain colored noise but it is possible that the structure of the noise could differ from that included here e.g. the relative contributions of red and white noise could differ, or the correlation between successive data points may differ from the range prescribed here. 

\textit{Recommendation 2:} Take red noise into account in detection methods.

This paper also shows that care needs to be taken when detrending. Both TVD and JAM detrended by smoothing. JAM used a constant value for the width of the smoothing window. This method produced lots of false detections in both HH1 and HH2, despite attempts to improve the detection procedure between the two exercises (see Tables \ref{table[HH2]} and \ref{table[HH1]}). TVD varied the width of the optimal smoothing window on a flare-by-flare basis, which substantially reduced the number of false detections. In HH2, the process by which the optimal width was determined was automated (see Section \ref{ssec:smooth}). However, this automation detrending also led to a relatively large number of false detections (see Table \ref{table[HH2]}) and it was only through human intervention that the number of false detections was reduced. On the other hand, in HH1, TVD manually selected the optimal smoothing window and produced a low number of false detections, comparable with the AFINO, wavelet (LAH), and periodogram method of CEP. The nonautomated detrending method outlined in Section \ref{ssec:tvd} is a good blueprint to follow when detrending. However, we note that alternative methods of detrending, such as through EMD or spline interpolation, may also produce robust results. We therefore advise users to test their detrending methods using simulated flares, as is done here, to test reliability before use on real data. We also point the reader in the direction of \citet{2018SoPh..293...61D}, who propose a set of criteria to help identify real periodicities and discard artifacts when detrending. These criteria include, for example, excluding periodicities inside the cone of influence and only considering detections with periods less than the smoothing window used to detrend the data. This paper demonstrates that when performed with due care and attention and by an experienced user detrending by removing a smooth component can produce reliable and robust results. 

\textit{Recommendation 3:} If you are going to detrend, do it carefully and manually, treating each timeseries individually and being wary of automated methods. Use simulations to test methods and become familiar with potential pitfalls.

The impact of trimming the data around the QPPs on the likelihood and robustness of detection was considered in Section \ref{ssec:trim}. Whether or not trimming is advantageous appears to depend on the detection method employed: trimming increased the likelihood of CEP making a detection, with no detrimental effect on the robustness of these detections. However, trimming reduced the robustness of detections made with AFINO and LAH's wavelet. We therefore recommend stringent testing of the impact of trimming on a particular method before use on real data. 

\textit{Recommendation 4:} Only trim the data around the QPPs if you are sure it benefits detection. Use simulations where necessary to test this. Of the methods employed here CEP's periodogram benefitted but AFINO and wavelet did not.

Although only a small sample was considered in this study, it is reasonable to conclude that the periodogram-based methods are not ideally suited to detecting nonstationary QPPs. However, EMD and MCMC fitting were able to produce precise detections of these QPPs. Therefore, if aiming to specifically detect nonstationary QPPs, it would be worth employing these methodologies. It is interesting to note that the wavelet method did not detect the nonstationary QPPs when the whole raw time series was considered (LAH's method). This is potentially because LAH used the statistical significance of peaks in the global power spectrum to determine whether a detection was significant. Period drifts are likely to broaden peaks in the global spectrum at the expense of absolute power, meaning the broad peaks are not statistically significant. Statistical tests for peaks covering multiple period bins, such as those described in \citet{2017A&A...602A..47P}, may resolve this issue. However, if detrending is performed, for example, using the EMD technique, the nonstationary QPPs are revealed with the wavelet, including the drift in period (see Figure \ref{fig:EMDmethodcomp}). Therefore, a combination of EMD and wavelet techniques could also improve the robustness of the EMD detections. It is important to stress that if EMD is employed, it is necessary that the user has a good grasp on how to make appropriate choices for the value of the shift factor. It is possible that GPs (JRAD) could prove to be a useful analysis mechanism for nonstationary QPPs. However, substantial work is still required to ensure robustness. 

\textit{Recommendation 5:} For nonstationary signals use EMD, wavelet on a detrended EMD signal and MCMC fitting.

We should note here that EMD produced a relatively large number of false detections, raising questions over the robustness of the method. Further examination reveals that the majority of these false detections arise from the red noise and are composed of signals of the order of, or less than, one period in length. Although in the hare-and-hounds exercises performed here it is easy to distinguish between red noise and signal, in real data the distinction may not be so clear-cut. Any ``red noise'' observed in real data may contain interesting information about the system being observed. For example, the underlying shape of the flare can contribute to the red-noise signal in a periodogram spectrum. This raises the question of how we define QPPs in the first place and demonstrates the importance of a classification system for such quasi-periodic events as suggested by \citet{2019PPCF...61a4024N}. 

\textit{Recommendation 6:} Decide a priori on your definition of QPPs, including the number of periods required for detection of QPPs. For example, in these simulations, including an \textit{a priori} selection criterion that any detections contain at least three full periods would have substantially improved the robustness of the EMD detections.

Some of the methods were far more time-consuming than others, and so when deciding which method to employ, the number of time series being considered should be kept in mind. AFINO, LAH's wavelet, and CEP's periodogram are all relatively quick methods that require little user input and so are suitable for large-scale statistical studies. The requirement for user input when using smoothing to detrend the data means that TVD's method was relatively time-consuming but this method could be employed for specific case studies. EMD was also user intensive and therefore better suited to case studies. The MCMC method employed by DJP is currently user intensive and better suited to case studies. However, there is the potential for improvements in this regard. Bayesian analysis requires prior information with reasonable boundaries to be defined. However, limits on the parameters in the Bayesian model could potentially be constructed either by combining with other methods or based on the results of previous large statistical surveys. Similarly, there are multiple models that could be tested with Bayesian analysis. A priori decisions on this, based on theoretical models and QPP classification, or potentially machine learning mechanisms, could allow more automation. It is also worth noting that MCMC statistical studies are not unprecedented in solar physics \citep[e.g.][]{2017A&A...605A..65G}. 

\textit{Recommendation 7:} Consider the number of time series to be examined: If performing a large statistical study (containing, for example, more than 50 time series) AFINO (ARI \& LAH), wavelet (LAH) and periodogram (CEP) are good tools. These methods can also be used to ensure robustness in studies containing fewer time series, but you could consider also using alternative methods, such as periodogram (TVD - manual), EMD and MCMC fitting, which may reveal different features of the QPPs e.g. non-stationarity.

Multiple harmonics were included in some of the HH1 simulated flares, although not enough for a statistical study. While some of the hounds did highlight the fact that they thought there might be multiple QPPs included in certain simulations, a detailed study was not conducted, as the hounds predominantly concentrated on the most prominent detection. Constructive interference means that multiple harmonics are difficult to identify and a more in-depth study is required to determine how effective each of the methods are at identifying multiple signals. A logical way to proceed would be to use the robust methods to identify flares containing statistically significant QPPs and then perform a more detailed case study to determine how many QPPs are present. 

\textit{Recommendation 8:} To determine whether multiple harmonics are present,  more detailed case studies are required. Ensure a time series warrants further investigation using one of the robust methods to identify the dominant statistically significant QPPs. Then look for further harmonics with a more detailed analysis.

\section{Future prospects}\label{sec:future}
In this study, the investigations of stellar QPPs have been based primarily on observations from {\it Kepler} data. The {\it TESS} satellite \citep{2014SPIE.9143E..20R}
and {\it PLATO} \citep{2014ExA....38..249R} are now expected to bring us more stellar flare data. The {\it TESS} satellite, which was successfully launched in 2018 April, has a 2-minute time cadence mode, which is similar to {\it Kepler}'s 1-minute time cadence mode. 

\citet{2019arXiv190100443G} recently reported 763 flaring stars, including 632 M dwarfs, from the first 2 months of {\it TESS} 2-minute cadence data. The amplitudes (relative fluxes) of their detected flares are from (2-3) $\times$ 10$^{-2}$
to 10$^{1}$ and durations are from 10$^{-1}$\,h to 10$^{1}$\,hr. The bolometric energies of the detected flares are typically  10$^{34}$ -- 10$^{36}$\,erg on FGK dwarfs and 10$^{32}$ -- 10$^{34}$\,erg on late M dwarfs. As shown in Figure 5 of \citet{2019arXiv190100443G}, the number of late-M dwarfs is particularly increased compared with the sample from the {\it Kepler} data, and their {\it TESS} magnitudes are 10-15 mag. These values suggest that we can also conduct QPP analyzes with \textit{TESS} data, and in particular, potential QPP data from late M dwarf flares are increased compared with the previous studies. The data of \citet{2019arXiv190100443G} only use the first two months of {\it TESS} data, and so the number of flare stars increased by more than a factor of 10 after the analyzes of the whole {\it TESS} dataset (2 yr and the almost the whole sky). 

To best examine the synergies between solar and stellar flares, we would want to compare data that are as similar as possible. For example, data should ideally be observed in the same waveband. Similar to \textit{Kepler}, \textit{TESS} makes white-light observations. QPPs have also been detected in a flare observed by the Next Generation Transit Survey (NGTS) \citep{2019MNRAS.482.5553J}, which observes in white light, like \textit{Kepler} and \textit{TESS} but with a much faster cadence of 10\,s, which allows much shorter period QPPs to be detected. However, white-light flares are rarely studied in solar physics because they are difficult to observe. This issue can be tackled in two manners: First, we can attempt to make observations of solar flares that are as similar to the white-light observations as possible. These are likely to be resolved observations but may provide a hint toward the commonality of QPPs in solar and stellar flares. Second, we can attempt to make multiwavelength observations of stellar QPP flares. For example, there are flares that were observed by both \textit{XMM-Newton} and \textit{Kepler} \citep{2019A&A...622A.210G}. The number of detected stellar QPP flares is still relatively low and overlaps between \textit{Kepler}, \textit{K2} and \textit{TESS} and other wavelength observations remain understudied. Such simultaneous observations may enable us to determine whether the drivers of white-light QPPs are the same as the drivers of, for example, X-ray QPPs. 

There is now evidence that QPPs are a common feature of solar flares \citep{2010SoPh..267..329K, 2015SoPh..290.3625S, 2016ApJ...833..284I, 2017A&A...608A.101P}. However, these QPPs come in many different forms and so could require several different mechanisms to explain them all. Studies of solar QPP would, therefore, benefit from a classification system, as suggested in \citet{2019PPCF...61a4024N}. For the physics of each classification to be distinguished, we need to accrue enough QPPs of each classification to be able to perform statistical studies on their properties. The robust methods described in this paper should, therefore, be utilized to identify as many QPPs as possible. 
Finally, this study has shown that we can now reliably detect solar and stellar QPPs with a number of different methods. However, the majority (although not all) of the methods provide only limited to no information on the properties of those QPPs other than their period. Now that we can be confident in our detections, we can attempt to develop techniques, such as MCMC and forward modeling, that are capable of robustly extracting additional physical properties. Given sufficiently detailed theoretical models, studies along these lines could then potentially be used to distinguish between the different QPP excitation mechanisms. This, combined with the classification of QPPs mentioned above, which may well rely on these techniques, will enable us to take studies of QPPs to the next level.

\acknowledgments
We acknowledge support from the International Space Science Institute for the team \lq{Quasi-periodic Pulsations in Stellar Flares: a Tool for Studying the Solar-Stellar Connection}\rq{} and the International Space Science Institute Beijing for the team `MHD Seismology of the Solar Corona'. A.M.B., D.Y.K., V.M.N. and T.M. acknowledge the support of the Royal Society International Exchanges grant IEC/R2/170056. D.Y.K. and V.M.N. are supported by the STFC grant ST/P000320/1. L.A.H. was supported by an Enterprise Partnership
Scheme studentship from the Irish Research Council (IRC) between Trinity College Dublin and Adnet System Inc. J.A.M. acknowledges generous support from the Leverhulme Trust and this work was funded by a Leverhulme Trust Research Project Grant: RPG-2015-075. J.A.M. acknowledges IDL support provided by STFC and for support via grant number ST/L006243/1. V.M.N. acknowledges support by the Russian Foundation for Basic Research grant No. 18-29-21016. Y.N. was supported by JSPS KAKENHI Grant Numbers JP16J00320.  T.V.D. and D.J.P. were supported by GOA-2015-014 (KU~Leuven). This project has received funding from the European Research Council (ERC) under the European Union's Horizon 2020 research and innovation programme (grant agreement No 724326).

\textbf{Astropy \citep{2013A&A...558A..33A, 2018AJ....156..123A}, Celerite \citep{foreman-mackey2017}, Matplotlib \citep{Hunter:2007}, NumPy (\url{http://www.numpy.org/}), Python (\url{http://www.python.org}), SciPy \citep{scipy_ref}}
\appendix

\section{Appendix information}\label{sec:appendix} Table \ref{table[hh1_detections]} contains a detailed breakdown of which types of QPPs the various methods detected. The majority of QPPs detected were single sinusoidal QPPs, which is expected because the majority of methods were based on some from of transform to the frequency domain, based on the assumption that any signals are sinusoidal in nature. 

\begin{sidewaystable}\caption{Breakdown of Number of Flare Time Series in which a ``Detection'' was Claimed for HH1.}\label{table[hh1_detections]}
\centering
\begin{tabular}{c|cc|cc|cc|cc|cc|cc|cc|cc|cc|c|c}
  \hline
  & \multicolumn{2}{|c|}{1 QPP} & \multicolumn{2}{|c|}{2 QPPs} & \multicolumn{2}{|c|}{Lin. bgd} & \multicolumn{2}{|c|}{Quad. bgd} & \multicolumn{2}{|c|}{Non-stat.} & \multicolumn{2}{|c|}{1 flare} & \multicolumn{2}{|c|}{2 flares} & \multicolumn{2}{|c|}{Non-P. Multi.} & \multicolumn{2}{|c|}{P. multi.} & Stellar & Solar\\
 & G & E & G & E & G & E & G & E & G & E & G & E & G & E & G & E & G & E & & \\
 & (25) & (25) & (2) & (2) & (2) & (1) & (1) & (2) & (2) & (2) & (0) & (1) & (0) & (1) & (3) & (3) & (4) & (4) &(15)& (6)\\
 \hline
 LAH - W & 7 & 3 & 1 & 1 & 0 & 0 & 0 & 0 & 0 & 0 & 0 & 0 & 0 & 0 & 0 & 0 & 0 & 0 & 0 & 1 \\
 LAH - A & 6 & 8 & 2 & 1 & 0 & 0 & 0 & 0 & 0 & 0 & 0 & 0 & 0 & 0 & 1 & 1 & 1 & 0 & 2 & 4 \\
 ARI & 11 & 3 & 2 & 1 & 1 & 0 & 0 & 0 & 0 & 0 & 0 & 0 & 0 & 0 & 0 & 0 & 0 & 0 & 1 & 2 \\
 CEP & 14 & 5 & 2 & 1 & 1 & 0 & 1 & 0 & 0 & 0 & 0 & 0 & 0 & 0 & 0 & 0 & 0 & 0 & 0 & 2 \\
 TVD\tablenotemark{a} & 9 & 9 & 0 & 1 & 1 & 0 & 1 & 0 & 1 & 0 & 0 & 0 & 0 & 0 & 1 & 0 & 1 & 0 & 3 & 4 \\
 JAM & 8 & 6 & 1 & 1 & 0 & 0 & 1 & 1 & 1 & 0 & 0 & 0 & 0 & 1 & 0 & 1 & 0 & 1 & 4 & 2 \\
 JRAD & 24 & 15 & 2 & 2 & 2 & 1 & 1 & 1 & 0 & 2 & 0 & 1 & 0 & 0 & 2 & 2 & 4 & 2 & 10 & 5 \\
 \hline
 TM - F\tablenotemark{b} & 9 (10) & 0 & 2 (3) & 0 & 1 & 0 & 0 & 0 & 0 & 0 & 0 & 0 & 0 & 0 & 0 & 0 & 0 & 3 & 0 & 0 \\
 TM - EMD\tablenotemark{b} & 12 (15) & 2 (2) & 2 (3) & 0 & 1 (2) & 0 & 0 & 0 & 2 (3) & 2 (2) & 0 & 0 & 0 & 0 & 0 & 0 & 3 & 0 & 0 & 1 \\
 \hline
 CEP ``flare'' & 2 & 2 & 0 & 1 & 0 & 0 & 0 & 0 & 0 & 0 & 0 & 0 & 0 & 0 & 0 & 0 & 0 & 0 & 0 & 0 \\
 CEP ``whole'' & 2 & 2 & 0 & 0 & 0 & 0 & 1 & 0 & 0 & 0 & 0 & 0 & 0 & 0 & 1 & 0 & 0 & 0 & 0 & 0 \\
 AFINO trim & 8 & 8 & 0 & 1 & 0 & 0 & 0 & 0 & 0 & 0 & 0 & 0 & 0 & 0 & 0 & 0 & 0 & 1 & 2 & 1 \\
 LAH - W trim & 12 & 12 & 0 & 0 & 1 & 0 & 0 & 0 & 1 & 0 & 0 & 0 & 0 & 0 & 0 & 0 & 1 & 0 & 3 & 5 \\
 \hline
\end{tabular}
\tablecomments{Here ``G'' stands for flares constructed from two half-Gaussian curves, while ``E'' stands for flares with a two-stage exponential decay, as described in Section \ref{sec:sims}.}
\tablenotetext{a}{TVD only analyzed 58 flares.}
\tablenotetext{b}{TM only analyzed 26 flares. Since, on occasion, more than one ``detection'' was made per flare the number in brackets contains the total claimed detections.}
\end{sidewaystable}

\bibliographystyle{aasjournal}
\bibliography{HandH_bib} 

\end{document}